\documentclass[a4paper,10pt]{article}

\usepackage[utf8]{inputenc}
\usepackage{authblk}

\usepackage{ifthen,ifpdf}

\ifpdf
  \usepackage{hyperref}
  \hypersetup{colorlinks=false,allcolors=blue}
  \usepackage{hypcap}
  \pdfpagewidth=\paperwidth
  \pdfpageheight=\paperheight
\fi

\usepackage[utf8]{inputenc}
\usepackage[T1]{fontenc}

\usepackage{amsmath, amsthm, amssymb}
\usepackage{mathtools}
\usepackage{enumerate}
\usepackage{afterpage}
\usepackage{tabularx}
\usepackage{dsfont}

\usepackage{graphicx}

\usepackage{algorithm}
\usepackage{algpseudocode}
\usepackage{caption}
\usepackage[labelformat=simple]{subcaption}

\usepackage{color}
\usepackage{units}
\usepackage{array, multirow}
\usepackage{booktabs}

\newcolumntype{M}[1]{>{\vspace{3pt}\raggedleft\arraybackslash}m{#1}}

\usepackage{siunitx}[=v2]
\usepackage{tikz}
\usetikzlibrary{calc}%to move common legend between two graphs
\usetikzlibrary{matrix}
\usepackage[export]{adjustbox}
\usepackage{pgfplots}
\usepackage{pgfplotstable}
\usepgfplotslibrary{units}
\pgfplotsset{compat=1.9,samples=1000,every axis/.append style={
font=\large,
line width=1pt,
tick style={line width=0.8pt}}}

\usetikzlibrary{arrows}
\usetikzlibrary{hobby}
\usetikzlibrary{calc}
\usetikzlibrary{arrows}
\usetikzlibrary{shapes}
\usetikzlibrary{positioning}
\usetikzlibrary{intersections}
\usetikzlibrary{decorations.pathreplacing}
\usetikzlibrary{patterns}
\usetikzlibrary{decorations.pathmorphing}
\usetikzlibrary{decorations.markings}
\usetikzlibrary{3d}
\usetikzlibrary{matrix}
\usetikzlibrary{spy}

\usepgfplotslibrary{polar}
\usepgfplotslibrary{external}
\usepgfplotslibrary{fillbetween}
\usepgfplotslibrary{groupplots}
\usepgfplotslibrary{patchplots}

\usepackage{cite}

\usepackage{anysize} 
\marginsize{2.5cm}{2.5cm}{2cm}{2cm}

\usepackage{wrapfig}

\newcommand{\eff}{\textrm{eff}}
\newcommand{\app}{\textrm{app}}
\newcommand{\tot}{\textrm{tot}}
\newcommand{\ran}{\textrm{ran}}
\newcommand{\sys}{\textrm{sys}}
\newcommand{\err}{\texttt{err}}
\newcommand{\rect}{\textrm{rect}}
\newcommand{\cub}{\textrm{cub}}
\newcommand{\sph}{\textrm{sph}}

\renewcommand{\div}{\textrm{div }}

\newcommand{\cref}{c^{\textrm{ref}}}

\usepackage{scalerel,stackengine,amsmath}
\newcommand\equalhat{\mathrel{\stackon[1.5pt]{=}{\stretchto{%
    \scalerel*[\widthof{=}]{\wedge}{\rule{1ex}{3ex}}}{0.5ex}}}}
    
\def\Xint#1{\mathchoice
   {\XXint\displaystyle\textstyle{#1}}%
   {\XXint\textstyle\scriptstyle{#1}}%
   {\XXint\scriptstyle\scriptscriptstyle{#1}}%
   {\XXint\scriptscriptstyle\scriptscriptstyle{#1}}%
   \!\int}
\def\XXint#1#2#3{{\setbox0=\hbox{$#1{#2#3}{\int}$}
     \vcenter{\hbox{$#2#3$}}\kern-.5\wd0}}

\def\dashint{\Xint-}

\usepackage{epstopdf}
\usepackage{physics}

\usepackage{printlen}
\everymath=\expandafter{\the\everymath\displaystyle}
\usepackage{layouts}
\newcommand\inputpgf[2]{{
 		\let\pgfimageWithoutPath\pgfimage
 		\renewcommand{\pgfimage}[2][]{\pgfimageWithoutPath[##1]{#1/##2}}
 		\input{#1/#2}
}}
\usepackage{import}
\usepgfplotslibrary{external}
\usepackage{longtable}
\usepackage{tikz-cd}
\usepackage{lmodern}

\makeatletter
\DeclareRobustCommand{\pdot}{\mathbin{\mathpalette\pdot@\relax}}
\newcommand{\pdot@}[2]{%
	\ooalign{%
		$\m@th#1\circ$\cr
		\hidewidth$\m@th#1\cdot$\hidewidth\cr
	}%
}
\makeatother
\theoremstyle{plain}

\theoremstyle{remark}

\numberwithin{equation}{section}

\usepackage{bm}

\newcommand{\R}{\mathds{R}}

\newcommand{\Q}{\mathds{Q}}

\newcommand{\Sym}[1]{\textrm{Sym}(#1)}

\DeclareMathOperator{\Id}{\textrm{Id}}

\newcommand{\mean}[1]{\left\langle {#1} \right\rangle}

\renewcommand{\div}{\textrm{div }}

\newcommand{\review}[1]{{\color{black}{#1}}}

\newcommand{\sty}[1]{\bm{#1}}

\newcommand{\fa}{\sty{ a}}

\newcommand{\fe}{\sty{ e}}

\newcommand{\fu}{\sty{ u}}
\newcommand{\fv}{\sty{ v}}
\newcommand{\fw}{\sty{ w}}
\newcommand{\fx}{\sty{ x}}
\newcommand{\fy}{\sty{ y}}
\newcommand{\fz}{\sty{ z}}

\newcommand{\fA}{\bm{ A}}
\newcommand{\fB}{\sty{ B}}
\newcommand{\fC}{\sty{ C}}

\newcommand{\fI}{\sty{ I}}

\newcommand{\fL}{\sty{ L}}
\newcommand{\fM}{\sty{ M}}
\newcommand{\fN}{\sty{ N}}

\newcommand{\fP}{\sty{ P}}
\newcommand{\fQ}{\bm{ Q}}
\newcommand{\fR}{\sty{ R}}

\newcommand{\fV}{\sty{ V}}

\newcommand{\fxi}{\bm{\xi}}
\newcommand{\fpsi}{\bm{\psi}}
\newcommand{\feta}{\mbox{\boldmath $\eta$}}

\newcommand{\Qtensor}{$\mu$Q-tensor}

\makeatletter
\def\omino#1#2{{%   % #2 will be the scaling factor
		\unitlength10\p@
		\@tempcnta\z@
		\@tempcntb\@ne
		\count@\z@
		\xomino#1\relax
		\scalebox{#2}{\begin{picture}(\@tempcnta,\@tempcntb)(0,-\@tempcntb)%
					\@tempcnta\z@
					\@tempcntb\@ne
					\count@\z@
					\xxomino#1\relax
		\end{picture}}%
}}%

\def\xomino#1{%
	\ifx\relax#1%
	\else
	\ifx\\#1%
	\ifnum\count@>\@tempcnta \@tempcnta\count@\fi
	\advance\@tempcntb\@ne
	\count@\z@
	\else
	\advance\count@\@ne
	\fi
	\expandafter\xomino
	\fi}

\def\xxomino#1{%
	\ifx\relax#1%
	\else
	\ifx\\#1%
	\advance\@tempcntb\@ne
	\count@\z@
	\else
	\advance\count@\@ne
	\ifx*#1%
	\put(\count@,-\@tempcntb){\kern-10pt\rule{10pt}{10pt}}%
	\fi
	\fi
	\expandafter\xxomino
	\fi}

\makeatother

\title{Symmetries in stochastic homogenization\\ and \review{adjustments} for the RVE method}

\author[1]{Binh Huy Nguyen}
\author[1,2,3,*]{Matti Schneider}

\affil[1]{University of Duisburg-Essen, Institute of Engineering Mathematics}
\affil[2]{Center for Nanointegration Duisburg-Essen (CENIDE)}
\affil[3]{Fraunhofer Institute for Industrial Mathematics ITWM, Kaiserslautern}
\affil[*]{correspondence to: \texttt{matti.schneider@uni-due.de}}

\date{\today}

\begin{document}
\maketitle 
%\printinunitsof{in}\prntlen{\textwidth}\prntlen{\textheight}

\begin{abstract}
\noindent We investigate the implications of a given symmetry of a random microstructure on the obtained effective tensor and its fluctuation in the context of thermal conductivity, and study strategies for enforcing these symmetries in postprocessing via orthogonal projectors. Within the framework of the representative volume element (RVE) method, we establish the invariance conditions for the effective tensor and its fluctuation under different symmetry groups of the microstructure. Interestingly, the symmetry of the considered cell type in the RVE method may break the ensemble symmetry and compromise the approximation of the effective properties.\\
To rectify this issue, we introduce dedicated techniques which permit to enforce the expected symmetries in postprocessing and study the implications on the bounds for the effective properties as well as the total, the random and the systematic errors. We provide theoretical arguments that suitable projections lead to unbiased variance-reduction strategies which furthermore enforce the expected symmetries exactly.
\\
Through large-scale FFT-based homogenization simulations, we study the symmetry structure of the estimated effective conductivities and their fluctuations. Moreover, we demonstrate the power of the symmetry-projection techniques for fiber-reinforced composite microstructures of industrial scale.\\
%identifying such measures of other classes of material responses.\\
%\noindent We work out the implications of a given symmetry of a random microstructure on the obtained effective tensor and its fluctuations in the context of thermal conductivity. Interestingly, the symmetry of the considered cell type in the representative-volume element method has an influence on the estimated fluctuation tensor.
\end{abstract}

%\todo{
%\begin{enumerate}
%	\item @BHN: read and critisize
%	\item @BHN: update numerics
%	\item then back to MS
%\end{enumerate}
%}

\newpage

%\tableofcontents
%
%\newpage

%%%%%%%%%%%%%%%%%%%%%%%%%%%%%%%%%%%%%%%%%%%%%%%%%%%%%%%%%%%%%%%%%%%%%%%%%%%%%%%%%%%%%%%%%%%%%%%%%%%%%%%%%%%%%%%%%%%%%%%%%%%%%%%%%%%%%%%%%%%%%%%%%%%%%%%%%%%%%%%%%%%%%%%%%%%%%%%%%%%%%%%%%%%%

\section{Introduction}
\label{sec:intro}

\subsection{State of the art}
\label{sec:intro_state}

%\todo{update}\\
%Weber2019Symmetry~\cite{Weber2019Symmetry}\\
In the modern age of digital material design, computational homogenization has became an indispensable tool for various engineering applications where the material behavior at the macroscopic scale can be linked to their microscopic features. Irrespective of the intricacies of the microstructure involved, i.e., multiple phases or sophisticated inclusion geometries, the effective material behavior turns out to be quite simple, in general, provided the microstructure scale is sufficiently small compared to the considered component scale. Identifying such emergent behavior is the goal of homogenization method, which resolves a microscale boundary value problem and employs the resulting solutions to extract the macroscopic behavior~\cite{zohdi2008introduction}. Homogenization methods can be helpful in various ways. For architected materials, periodic homogenization methods employing periodic unit cells turn out to be appropriate. For materials with random microstructure such as metal matrix composite or reinforced fibers composite, stochastic homogenization is more favorable. \\
While the effective properties of architected material are obtained by considering the periodic unit cell, material with random microstructure are more challenging. For a start, the aleatoric uncertainty of the microstructure necessitates a stochastic quantification. These approaches often rely on numerical simulations to study the influence of microstructure descriptors such as geometric shapes, grain size distributions, spatial arrangements and aspect ratio on the overall macroscopic material properties. For instance, in Hiriyur et al.~\cite{hiriyur2011uncertainty}, the extended finite element method (XFEM) in combination with the Monte Carlo method was employed to quantify the uncertainty of the effective elastic Young's modulus of heterogeneous microstructures consisting of circular and elliptic inclusions in a two dimensional (2D) unit cell of fixed size. Stefanou et al.~\cite{stefanou2017stochastic} performed a similar study to extract statistical information, including the probability distribution and the correlation function of the so-called mesoscale tensor random field in 2D elasticity. While it is important to study the influence of microstructure in a unit cell of fixed size, these numerical studies are not sufficient to incorporate  the central object in stochastic homogenization, namely the representative volume element (RVE). Introduced by Hill~\cite{hill1963elastic}, the RVE should be statistically typical for the mixture and sufficiently large such that the apparent property is independent of the boundary conditions. This definition, however, is too strict for practical considerations, often leading to large volume elements, and rather serves as a guiding principle than a law to be rigidly enforced. Relaxing on Hill's RVE definition, Drugan and Willis~\cite{drugan1996micromechanics} proposed RVE to be a volume element whose apparent properties represent the effective constitutive response with sufficient accuracy. During the last three decades, several alternative definitions of RVEs were proposed. For instance, Evesque~\cite{evesque2005fluctuations} postulated that the determination of the minimum volume of the RVE should take into account the local fluctuations from one RVE to the other. In a series of works on the statistical representative volume element (SVE), Ostoja-Starzewski and co-workers~\cite{khisaeva2006size,ostoja2007microstructural,ostoja2016scaling} determined the RVE size based on the statistical homogeneity and ergodicity, Hill's homogeneity condition~\cite{hill1963elastic} and the underlying variational principle. In the case of a carbon-reinforced fiber composite, Trias et al.~\cite{trias2006determination} estimated the minimum characteristic length of an SVE to be 50 times the fiber radius. Other alternative RVE definitions were surveyed by Gitman et al.~\cite{gitman2007representative} and El Moumen et al.~\cite{el2021numerical}. To assert Drugan and Willis's RVE criteria~\cite{drugan1996micromechanics}, Kanit et al.~\cite{kanit2003determination} investigated the RVE size via a statistical approach by combining FEM and Monte-Carlo method, where the apparent properties of several random realizations of microstructures with different sizes were monitored against its empirical mean for a fixed volume size as well as against the effective property of a very large volume element size. These errors are coined as dispersion (or random) and bias (or systematic) errors, respectively. Additionally, the authors studied the impact of three different boundary conditions, namely Dirichlet, Neumann and periodic boundary conditions. Their work finds that periodic boundary conditions yields the closest approximation of the effective properties.\\
Parallel to the findings from the engineering community, the mathematical community was also active in this area. The mathematical theory of stochastic homogenization of a stationary and ergodic coefficient field defines the (deterministic) effective properties abstractly, but their evaluation requires computations on the whole space. To put it differently, a single infinitely large volume element almost surely provides the effective material property. The mathematical approach permits to study the RVE method~\cite{bourgeat2004approximations,sab2005periodization,bystrom2006periodic,le2010some} more closely, which approximates the effective properties via an ensemble of possibly periodized finite-size realization of the microstructure. Mathematical results do not only show convergence of the apparent properties obtained by the RVE method to the deterministic effective property as the volume element size approaches infinity but may also be used to quantify the rate of convergence. Specifically, in the works of Gloria and Otto~\cite{gloria2012optimal} and Gloria et al.~\cite{gloria2015quantification}, the authors quantified the optimal convergence rates of the random and systematic errors, demonstrating a guideline to identify RVE based on desired accuracy of the effective property approximation, at least for simple type of microstructure. To numerically verify the convergence rate for 2D conductivity coefficients as a function of the volume element size, Khoromskaia et al.~\cite{khoromskaia2020numerical} employed an efficient FEM discretization in combination with a Monte Carlo method. In a comprehensive work with elaborated microstructure generators and advanced FFT-based computational homogenization methods, Schneider et al.~\cite{schneider2022representative} not only attested the optimal convergence rate of the RVE method but also shed light on the superior asymptotic convergence of the periodized ensemble over the snap-shot ensemble which naturally arises in digital imaging. Furthermore, in addition to the microscopic oscillations, the RVE method in stochastic homogenization also characterizes the macroscopic fluctuation. By recognizing the H-convergence criterion that relates the microscopic oscillatory flux to the product of homogenized tensor and oscillatory field, Duerinckx et al.~\cite{Duerinckx2020muQ,duerinckx2020robustness} proposed the homogenization commutator and its associated two-scale expansion, namely the standard homogenization commutator, to capture and quantify the macroscopic fluctuations. The leading-order term of such fluctuations is determined by the variance of the homogenization commutator, which may also be represented as the scaled covariance of the apparent tensors in the RVE method. For the scalar-valued elliptic problem, this fluctuation is represented by a fourth-order \Qtensor{}, which may be obtained at no extra cost within the RVE method as shown in~\cite{khoromskaia2020numerical} for the 2D cubic setting. A similar concept of the spatial correlation function of random field apparent tensor was studied in the SVE approach~\cite{ostoja1993micromechanics,ostoja1998random,ostoja2016scaling}. In this method, by virtue of a moving window on a realized microstructure and Hill's macrohomogeneity condition, the apparent material tensor arises as a random field and is therefore scale-dependent. Numerical studies of the correlation structure of elastic random field in 2D were investigated in different works~\cite{sena2013stiffness,savvas2016determination,stefanou2017stochastic,Jetti2024TRFs}. \\
In homogenization theory, the random heterogeneous microstructures determine the macroscopic properties and thus encode the inherent material symmetries via the statistics of their representatives. Leveraging symmetry in homogenization may be advantageous, especially for periodic homogenization. For instance, as studied in Barbarosie et al.~\cite{Barbarosie2017DomainSymmetry}, a smaller unit cell with proper symmetry-constrained periodic boundary conditions, determined from the group representation of the domain symmetry, was employed in a finite element program to reduce the computational expense from a larger domain for determining effective material tensor of two-phase planar composite. The concepts of plane and space group symmetry of the unit cell were also utilized to construct a large database of periodic micro-architectures in combination with topology optimization for the inverse homogenization by Podest\'a et al.~\cite{podesta2019symmetry} and M\'endez et al.~\cite{mendez2019making}. \review{By establishing the relation between lattice and material symmetries, Giusteri and Penta~\cite{giusteri2022periodic,giusteri2024corrigendum} successfully identified a unit cell symmetry of a locally isotropic two-phase composite with spherical inclusions such that the homogenized elastic tensor exhibited isotropy under the vanishing condition of one elastic coefficient}. Prior to these works, in a more general setting, Ptashnyk and Seguin~\cite{Ptashnyk2016PeriodicSymmetry} formulated the invariance condition of the effective elastic tensor in periodic homogenization under a volume-preserving affine transformation. On the other hand, in stochastic homogenization, based on the probability space setting from Papanicolaou~\cite{papanicolaou1979boundary}, Alexanderian~\cite{Alexanderian2012} showed how ensemble symmetries lead to invariance properties of the effective second-order diffusion tensor, also performing numerical experiments to verify isotropy invariance of tile-based 2D random media. A similar invariance condition was also described in Zhikov et al.~\cite{zhikov1994homogenization} and demonstrated explicitly for some simple 2D problems. It should be noted that the invariance conditions in these works are prescribed on the deterministic effective tensor. In practice, the apparent material tensor, often obtained from a cubic volume element, is not only tensor-value random variable but also non-invariant with respect to its associated ensemble symmetry. In elasticity, it is well known that the symmetry class of a fourth-order elastic tensor is approximated by the its distance to a known symmetry class~\cite{Gazis1963,Moakher2006,Weber2019Symmetry}. However, the ramification of such approximation in the context of stochastic homogenization for the apparent material tensor is not discussed at length, although it is recognized that the shape of the computational volume element has an effect on the convergence and symmetry of the apparent tensors~\cite{Gluege2012SphericalCell,Gluege2013SphericalCell}. As the cubic volume element unintentionally breaks the ensemble symmetry, the use of spherical volume element may remedy such undesired consequences by preserving the ensemble symmetry. This feature was numerically demonstrated in references~\cite{Gluege2012SphericalCell,Gluege2013SphericalCell} within the SVE framework. On the related works, the invariance of the one-point expectation and two-point covariance function of wide-sense stationary tensor random field such as conductivity and linear elasticity under different symmetry class was provided by Malyarenko and Ostoja-Starzewski~\cite{malyarenko2017random,malyarenko2017arandom,malyarenko2019tensor} based on group representation theory. In case the microheterogeneity information is unavailable, whether from digital generation or experimental acquisition, the material tensor random field is assumed to follow a specific probabilistic model. Soize~\cite{soize2006non} proposed to generate non-Gaussian, positive definite elastic tensor random fields depending on four parameters, i.e., three correlation lengths and a level of random fluctuation, and employed the results as an a priori mesoscale material tensor in stochastic homogenization~\cite{soize2008tensor}. Meanwhile, Guilleminot and Soize~\cite{guilleminot2012generalized} constructed a probabilistic model of a tensor random field with underlying symmetry from the maximum entropy principle. Thanks to the small number of parameters, these probabilistic tensor random field modeling approaches were also incorporated with stochastic finite elements to identify multi-scale elastic material parameters~\cite{nguyen2015multiscale}. More recently, Shivanand et al.~\cite{shivanand2024stochastic} employed a combination of spectral decomposition and the exponential map to model different class of invariant stochastic positive definite second- and fourth-order elastic tensors. \\
Despite of different paths to translate the uncertainties from the micro-heterogeneity into the macroscopic-scale properties, the common goal of stochastic homogenization is to provide an uncertainty-quantified apparent material tensor, i.e., the associated statistical means and covariances, for the macroscopic analysis whose solutions and the deduced physical quantities are inherently stochastic. As mentioned earlier, the RVE method provides a theoretical probe on the fluctuation of macroscopic observables, namely the \Qtensor~\cite{Duerinckx2020muQ,duerinckx2020robustness}.
It appears clear that the symmetries of the ensemble are not only inherited by the homogenized tensor but also its fluctuation. Such interplay between symmetry and randomness also has practical implication. For instance, in case the variance of one component of the homogenized material tensor is characterized, the variance of other components may be deduced without the need for additional experiments. Thus, the objective of this paper is to investigate the ramifications of symmetries in stochastic homogenization using RVE method, particularly the apparent material tensor and its corresponding \Qtensor{}.   

%The interplay between microstructure configurations, i.e. material properties and random arrangements, and macroscopic properties points to the possibility of structure of fluctuation, which is the purpose of this paper.  

%The next compelling question is what the error looks like. This question also has practical implication, as we will show how the errors are related to one another. That means, if one component of the material tensor is determined (mean and variance), the variance of other parameters can be extrapolated without the need for additional experiment.\\
%The aim of this paper is multi-fold. We will first establish a theoretical derivation to identify the structure of the \Qtensor{} under some specific symmetry group. Furthermore, we will demonstrate through large-scale numerical examples different symmetry of the \Qtensor

\newpage
 
\subsection{Contributions}
\label{sec:intro_contributions}
%\todo{update}\\
In this work, we aim to investigate the implications of symmetries in stochastic homogenization, specifically on the apparent material tensor and its associated \Qtensor{}. From a practical point of view, the \Qtensor{} not only quantifies the fluctuation of the macroscopic tensor but also characterizes the interrelations between the random errors of the apparent coefficients. Put differently, the structure of the \Qtensor{} allows the uncertainty quantification of one material coefficient from that of the other. \review{Although computing effective elastic properties by the RVE method is of primary engineering interest~\cite{sena2013stiffness,Jetti2024TRFs,ravichandran2025quantifying}, we consider the case of static heat conduction with its associated quartic \Qtensor{}, for simplicity. In this way, the intricacies of the eighth-order \Qtensor{} associated to linear elasticity are avoided.} Concretely, we make the following contributions:
\begin{itemize}
	\item We establish the symmetries of the effective tensor and its associated fluctuation \Qtensor{} as consequences of invariances of the ensemble.
	\item We clarify the symmetry breaking induced by the shape of the volume element.
	\item We introduce and study symmetry-enforcing projectors to improve the RVE method.
\end{itemize}
For ease of construction, we recall the analytic setting of the stochastic homogenization from a stationary and ergodic ensemble in section \ref{sec:theory_setup} with an emphasis on the emergence of the effective conductivity tensor. Section \ref{sec:theory_unitCells} lays out the practical RVE method to approximate the effective conductivity tensor by an apparent quantity via a proxy problem, defined on a general computational volume element. The apparent conductivity properties are quantified in terms of the random and the systematic error, decaying at different rates, in addition to their complimentary fluctuation tensor. As long as symmetry is concerned, the first driving question of the paper at hand is:
\begin{center}
	\fbox{%
		\parbox{.9\textwidth}{\emph{What is the impact of a given symmetry of the underlying ensemble on the apparent conductivity tensor and its associated \Qtensor{}?}}
	}
\end{center}
In section \ref{sec:theory_symmetryProperties}, we rely on the probability measure and propose the invariance conditions of the effective and apparent conductivity tensors and their corresponding \Qtensor{s}, which is stricter than the condition proposed by Zhikov et al.\cite{zhikov1994homogenization} and Alexandrian~\cite{Alexanderian2012}. More importantly, the focal point is placed on the breaking of the ensemble symmetry by the geometry of the computational volume element in the RVE method. To put it in a different perspective, the discrepancy between the apparent and the effective conductivity tensors are caused by the lack of symmetry of the unit cell. This leads us to the second mission:
\begin{center}
	\fbox{%
		\parbox{.9\textwidth}{\emph{How to remedy the lack of symmetry of the apparent conductivity tensor and its associated \Qtensor{} induced by cubic cells?}}
	}
\end{center}
We utilize the equivalent interpretation of an orthogonal projector as an averaging operator to extract the invariant part of the apparent conductivity tensor and its associated \Qtensor{} in section \ref{sec:SymmetryInformedStrats_projectors}. Consequently, an a posteriori symmetry-informed strategy is proposed to improve the approximation errors in section \ref{sec:SymmetryInformedStrats_postprocessing}. In addition to the explicit demonstrations of invariance condition and the projection, we further demonstrate our claims through computational homogenization of various ensemble symmetry of short fiber-reinforced composite in section \ref{sec:computations}. Discussions on the results are given in section \ref{sec:conclusion}.

\newpage

\section{Stochastic homogenization with symmetries}
\label{sec:theory}

\subsection{Stochastic homogenization of thermal conductivity}
\label{sec:theory_setup}

In this section, we recall some parts of the theory of stochastic homogenization~\cite{papanicolaou1979boundary,zhikov1994homogenization}, and refer to Neukamm~\cite[§2.2]{Neukamm2017LectureNotes} for an accessible introduction.\\
We consider the phase space of conductivity tensors defined on the whole space,
\begin{equation}\label{eq:theory_setup_space}
	\mathcal{A} = \left\{ \fA\in L^\infty( \R^d ; \Sym{d}) \,\middle|\,  \alpha_-\,\|\fxi\|^2 \leq \fxi\cdot \fA(\fx) \fxi \leq \alpha_+\,\|\fxi\|^2, \quad \fxi \in \R^d, \quad \text{a.e. }\fx \in \R^d\right\},
\end{equation}
in $d$ spatial dimensions, where $\Sym{d}$ denotes the vector space of symmetric $d\times d$ tensors and $\alpha_{\pm}$ are given positive constants. Stochastic homogenization is encoded by a probability measure $\mu$ on the space \eqref{eq:theory_setup_space} which we assume to be stationary and ergodic. To express these notions mathematically, we denote by
\begin{equation}\label{eq:theory_setup_expectation}
	\mean{F} \equiv \int_{\mathcal{A}} F(\fA) \, d\mu(\fA)
\end{equation}
the ensemble average of an integrable function $F$ on $\mathcal{A}$, i.e., $F\in L^1(\mathcal{A},\mu)$. We assume that the shift operator
\begin{equation}\label{eq:theory_setup_shift_operator}
	\tau:\R^d \times \mathcal{A} \rightarrow \mathcal{A}, \quad (\fz, \fA) \mapsto \fA(\cdot + \fz) \equiv \tau_{\fz}(\fA),
\end{equation}
is measurable. Then, \emph{stationarity} means that the identity
\begin{equation}\label{eq:theory_setup_stationarity}
	\mean{F} = \mean{F \circ \tau_{\fz}}
\end{equation}
holds for all shift vectors $\fz \in \R^d$ and any integrable function $F\in L^1(\mathcal{A},\mu)$. The condition \eqref{eq:theory_setup_stationarity} encodes the fact that the shifting operation \eqref{eq:theory_setup_shift_operator} does not influence the outcome of reasonable observables.\\
In contrast, \emph{ergodicity} ensures the validity of the equation
\begin{equation}\label{eq:theory_setup_ergodicity}
	\mean{F} = \lim_{L \rightarrow \infty} \dashint_{Y^{\cub}_L} F(\tau_{\fx}(\fA))\, d\fx
\end{equation}
for $\mu$-almost every element $\fA \in \mathcal{A}$, every integrable function $F\in L^1(\mathcal{A},\mu)$ and where we write
\begin{equation}\label{eq:theory_setup_cube_defn}
	Y^{\cub}_L = \left\{ \fx \in \R^d \, \middle| \, - \frac{L}{2} \leq x_i \leq \frac{L}{2} \text{ for } i=1,2,\ldots,d \right\}
\end{equation}
for the cube with edge length $L$ centered at the origin. The ergodicity property \eqref{eq:theory_setup_ergodicity} ensures that an ensemble average \eqref{eq:theory_setup_expectation} may be approximated by a spatial average on \emph{a single realization} (almost surely). Loosely speaking, a single realization contains the variations encoded by all other realizations, at least almost surely.\\
The assumed properties permit to define the effective conductivity tensor
\begin{equation}\label{eq:theory_setup_Aeff}
	\fA^{\eff} \in \Sym{d},
\end{equation}
which also obeys the bounds
\begin{equation}\label{eq:theory_setup_Aeff_bounds}
	\alpha_-\,\|\fxi\|^2 \leq \fxi\cdot \fA^{\eff} \fxi \leq \alpha_+\,\|\fxi\|^2 \quad \text{for all} \quad \fxi \in \R^d.
\end{equation}
These effective properties \eqref{eq:theory_setup_Aeff} are useful for approximating, for almost every coefficient field $\fA \in \mathcal{A}$, the solution $\phi_{\epsilon} \in H^1_0(\Omega)$ to the problem
\begin{equation}\label{eq:theory_setup_phi_eps_problem}
	-\div \left( \fA_{\epsilon} \nabla \phi_{\epsilon} \right) = f - \div g \quad \text{with} \quad \fA_{\epsilon}(\fx) = \fA\left(\frac{\fx}{\epsilon}\right),
\end{equation}
where $\Omega \subseteq \R^d$ is an arbitrary bounded domain, $\epsilon > 0$ is a small positive parameter and $f \in L^2(\Omega)$ as well as $g \in L^2(\Omega)^d$ are deterministic. More precisely, the unique solution $\phi_{0} \in H^1_0(\Omega)$ to the problem
\begin{equation}\label{eq:theory_setup_phi_0_problem}
	-\div \left( \fA^{\eff} \nabla \phi_{0} \right) = f - \div g
\end{equation}
with homogeneous coefficients arises as the limit of the problems \eqref{eq:theory_setup_phi_eps_problem} as $\epsilon \rightarrow 0$ in the sense:
\begin{enumerate}
	\item The solutions $\phi_{\epsilon}$ converge to $\phi_{0}$ weakly in $H^1(\Omega)$, i.e., 
	\begin{equation}\label{eq:theory_setup_weak_convergence_temperature}
		\|\phi_{\epsilon} - \phi_{0}\|_{L^2(\Omega)} \rightarrow 0 \quad \text{and} \quad \int_\Omega \psi \, \partial_i \phi_{\epsilon} \, d\fx \rightarrow \int_\Omega \psi \, \partial_i \phi_{0} \, d\fx
	\end{equation}
	as $\epsilon \rightarrow 0$ for all smooth fields $\psi \in C_0^\infty(\Omega)$ and all indices $i=1,2,\ldots,d$.
	\item The associated fluxes $\fA_{\epsilon} \nabla\phi_{\epsilon}$ converge to the flux $\fA^{\eff}\nabla\phi_{0}$ weakly in $L^2(\Omega)$, i.e.,
	\begin{equation}\label{eq:theory_setup_weak_convergence_flux}
		\int_\Omega \fpsi \cdot \fA_{\epsilon} \nabla\phi_{\epsilon} \, d\fx \rightarrow \int_\Omega \fpsi \cdot \fA^{\eff}\nabla\phi_{0} \, d\fx
	\end{equation}
	as $\epsilon \rightarrow 0$ for all smooth fields $\fpsi \in C^\infty(\Omega)^d$.
\end{enumerate}
Loosely speaking, the effective properties \eqref{eq:theory_setup_Aeff} permit to approximate solutions to highly heterogeneous and stochastic problems \eqref{eq:theory_setup_phi_eps_problem} by solutions to the homogeneous and deterministic problem \eqref{eq:theory_setup_weak_convergence_flux} \emph{on average}, at least in case the parameter $\epsilon$ is small.

\subsection{Apparent tensors on unit cells}
\label{sec:theory_unitCells}

To compute the effective conductivity \eqref{eq:theory_setup_Aeff} in practice, the conventional strategy proceeds via suitable proxy problems on cells of finite size. In principle, \emph{any}  bounded domain 
\begin{equation}\label{eq:theory_unitCells_generalCell}
	Y \subseteq \R^d
\end{equation} 
can be used. In computational practice, rectangular cells are typically preferred, i.e.,
\begin{equation}\label{eq:theory_unitCells_rectangularCell}
	Y^{\rect}_{L_1,\ldots,L_d} = \left\{ \fx \in \R^d \, \middle| \, - \frac{L_i}{2} \leq x_i \leq \frac{L_i}{2} \text{ for } i=1,2,\ldots,d \right\}
\end{equation}
for positive real numbers $L_i$ ($i=1,2,\ldots,d$), and the most popular unit cell is the cube \eqref{eq:theory_setup_cube_defn}
\begin{equation}\label{eq:theory_unitCells_cubicCell}
	Y^{\cub}_L = \left\{ \fx \in \R^d \, \middle| \, - \frac{L}{2} \leq x_i \leq \frac{L}{2} \text{ for } i=1,2,\ldots,d \right\}
\end{equation}
for $L>0$. It is also possible to consider \emph{spherical} unit cells~\cite{Gluege2012SphericalCell,Gluege2013SphericalCell}, i.e., those which have the form
\begin{equation}\label{eq:theory_unitCells_sphericalCell}
	Y^{\sph}_R \equiv B_R \equiv \left\{ \fx \in \R^d \,\middle|\, \|\fx\| \leq R \right\}
\end{equation}
with a positive radius $R$.\\
Let us return to a \emph{general cell} \eqref{eq:theory_unitCells_generalCell} and discuss the approach to compute approximations of the effective properties \eqref{eq:theory_setup_Aeff}. For a prescribed vector $\bar{\fxi}$ and a given realization $\fA \in \mathcal{A}$, the unit-cell corrector $\phi_{Y,\fA,\bar{\fxi}}$ is defined as the unique element $\phi \in H^1_0(Y)$ which solves the cell problem
\begin{equation}\label{eq:theory_unitCells_correctorEqStrong}
	\div \fA(\bar{\fxi} + \nabla \phi) = 0.
\end{equation}
Thus, we consider vanishing Dirichlet boundary conditions on the cell $Y$. The results which we discuss also hold for other boundary conditions like Neumann or periodic boundary conditions, which are only sensible for rectangular cells \eqref{eq:theory_unitCells_rectangularCell}.\\
As the solutions $\phi_{Y,\fA,\bar{\fxi}}$ are linear in the vector $\bar{\fxi} \in \R^d$, the associated apparent conductivity is defined implicitly by flux averaging
\begin{equation}\label{eq:theory_unitCells_Aapp}
	\fA^{\app}_Y \bar{\fxi} = \dashint_{Y} \fA(\bar{\fxi} + \nabla \phi_{Y,\fA,\bar{\fxi}}) \, d\fx.
\end{equation}
The apparent conductivity \eqref{eq:theory_unitCells_Aapp} is symmetric, i.e., $\fA^{\app}_Y \in \Sym{d}$, and satisfies the bounds 
\begin{equation}\label{eq:theory_unitCells_Aapp_bounds}
	\alpha_-\,\|\fxi\|^2 \leq \fxi\cdot \fA^{\app}_Y \fxi \leq \alpha_+\,\|\fxi\|^2 \quad \text{for all} \quad \fxi \in \R^d
\end{equation}
for almost every realization $\fA \in \mathcal{A}$, i.e., identical estimates as the effective properties \eqref{eq:theory_setup_Aeff_bounds}. Due to the finitude of the cell $Y$, the apparent conductivity is a random variable, in contrast to the effective conductivity \eqref{eq:theory_setup_Aeff}.\\
The apparent properties \eqref{eq:theory_unitCells_Aapp} may be used to approximate the effective conductivity \eqref{eq:theory_setup_Aeff}. More precisely, it holds that
\begin{equation}\label{eq:theory_unitCells_Aapp_convergence}
	\fA^{\app}_{L Y} \rightarrow \fA^{\eff} \quad \text{as} \quad L \rightarrow \infty,
\end{equation}
almost surely, where $LY$ refers to the unit cell \eqref{eq:theory_unitCells_generalCell}, rescaled by a positive factor $L>0$. We refer to Bourgeat-Piatnitskii~\cite{bourgeat2004approximations} for proofs. Their arguments are restricted to the rectangular cell \eqref{eq:theory_unitCells_rectangularCell}, but carry over seamlessly to the general case \eqref{eq:theory_unitCells_generalCell}, at least for pure Dirichlet and pure Neumann boundary conditions. In a nutshell, the key argument is to recognize the unit-cell problem \eqref{eq:theory_unitCells_correctorEqStrong} for the corrector $\phi_{L Y,\fA,\bar{\fxi}}$ as a problem of the form \eqref{eq:theory_setup_phi_eps_problem} with parameter $\epsilon = 1/L$ and pass to the limit \eqref{eq:theory_setup_phi_0_problem}. Due to the homogeneity of the limit problem 
\begin{equation}\label{eq:theory_unitCells_correctorEqStrongAeff}
	\div \fA^{\eff}(\bar{\fxi} + \nabla \phi_0) = 0,
\end{equation}
the fluctuation $\phi_0 \in H^1_0(Y)$ needs to vanish, and we obtain the convergence statement \eqref{eq:theory_unitCells_Aapp_convergence} directly as a consequence of the weak convergence \eqref{eq:theory_setup_weak_convergence_flux} of the fluxes.\\
In case the error 
\begin{equation}\label{eq:theory_unitCells_total_error}
	\err^{\tot}_Y := \|\fA^{\app}_Y - \fA^{\eff}\|
\end{equation}
is sufficiently small (almost surely), the cell \eqref{eq:theory_unitCells_generalCell} is called a \emph{representative volume element} (RVE).\\
Kanit et al.~\cite{kanit2003determination}, and, independently, Gloria-Otto~\cite{gloria2012optimal} noticed the following decomposition of the average squared total error
\begin{equation}\label{eq:theory_unitCells_error_decomposition}
	\mean{\left\|\fA^{\app}_Y - \fA^{\eff} \right\|^2} = \mean{\left\|\fA^{\app}_Y - \mean{\fA^{\app}_Y} \right\|^2} + \left\|\mean{\fA^{\app}_Y} - \fA^{\eff} \right\|^2
\end{equation}
into the random error
\begin{equation}\label{eq:theory_unitCells_random_error}
	\err^{\ran}_Y := \sqrt{\mean{\left\|\fA^{\app}_Y - \mean{\fA^{\app}_Y} \right\|^2}},
\end{equation}
also called dispersion, and the systematic error
\begin{equation}\label{eq:theory_unitCells_systematic_error}
	\err^{\sys}_Y := \left\|\mean{\fA^{\app}_Y} - \fA^{\eff} \right\|,
\end{equation}
also called bias. The systematic error \eqref{eq:theory_unitCells_systematic_error} is purely deterministic, and quantifies the shortcoming of the unit cell $Y$ to represent the random material at large. In contrast, the random error \eqref{eq:theory_unitCells_random_error} measures the standard deviation of the apparent conductivity \eqref{eq:theory_unitCells_Aapp}.\\
In practical simulations, it was observed that the two errors \eqref{eq:theory_unitCells_random_error} and \eqref{eq:theory_unitCells_systematic_error} converge with a different order. Theoretical considerations, e.g., Gloria et al.~\cite{gloria2015quantification}, however, pointed out that no explicit rates for the individual errors \eqref{eq:theory_unitCells_random_error} and \eqref{eq:theory_unitCells_systematic_error} may be determined for the general stationary and ergodic case. Rather, specific more restrictive ergodicity hypotheses are required to show explicit convergence rates for the errors \eqref{eq:theory_unitCells_random_error} and \eqref{eq:theory_unitCells_systematic_error}, like finite range of dependence~\cite{ArmstrongSmart}, a mixing condition with algebraic decorrelation rate~\cite{ArmstrongMourrat} or more sophisticated weighted functional inequalities in probability~\cite{GloriaNeukammOttoLSI2019,WeightedFunctionalInequalities,WeightedFunctionalInequalities2}. Moreover, the imposed type of boundary condition plays a pivotal role. It turns out that both Dirichlet and Neumann boundary conditions only lead to a convergence behavior
\begin{equation}\label{eq:theory_unitCells_resonance_error}
	\err^{\sys}_{LY} \sim \frac{1}{L}
\end{equation}
of the systematic error~\cite{Clozeau2024}, whereas periodic boundary conditions, together with a periodization of the ensemble~\cite{sab1992homogenization,schneider2022representative} lead to the strongly improved decay
\begin{equation}\label{eq:theory_unitCells_systematic_error_decay}
	\err^{\sys}_{LY} \sim \frac{1}{L^d}
\end{equation}
in $d$ spatial dimensions, at least under restrictive ergodicity hypotheses, see Clozeau et al.~\cite{Clozeau2024}. In contrast, the random error \eqref{eq:theory_unitCells_random_error} typically decays with the rate
\begin{equation}\label{eq:theory_unitCells_random_error_decay}
	\err^{\ran}_{LY} \sim \frac{1}{L^{\frac{d}{2}}}
\end{equation}
expected by the central limit theorem, although exceptions are known~\cite{DirrenbergerForestJeulin2014}.\\
To quantify the fluctuations of the apparent conductivity~\eqref{eq:theory_unitCells_Aapp}, we follow Duerinckx et al.~\cite{Duerinckx2020muQ}, and define the cell-associated \Qtensor{} 
\begin{equation}\label{eq:theory_unitCells_muQ}
	\Q_Y = \text{vol}(Y) \mean{ \left( \fA^{\app}_Y - \mean{\fA^{\app}_Y} \right) \otimes \left( \fA^{\app}_Y - \mean{\fA^{\app}_Y} \right)}
\end{equation}
for the scenario \eqref{eq:theory_unitCells_random_error_decay} as the rescaled covariance tensor of the apparent properties~\eqref{eq:theory_unitCells_Aapp}. For almost every realization $\fA$, the fourth-order tensor $\Q_Y$ defines an element of the space $\Sym{\Sym{d}}$, i.e., it possesses the major symmetry 
\begin{equation}\label{eq:theory_unitCells_muQapp_major_symmetry}
	\fB : \Q_Y : \fC = \fC : \Q_Y : \fB, \quad \fB, \fC \in \Sym{d},
\end{equation}
due to its definition \eqref{eq:theory_unitCells_muQ} as a covariance, and the minor symmetries inherited by the apparent properties $\fA^{\app}_Y$. Moreover, the apparent \Qtensor{} \eqref{eq:theory_unitCells_muQ} is positive semidefinite
\begin{equation}\label{eq:theory_unitCells_muQapp_positive}
	\fB : \Q_Y : \fB \geq 0, \quad \fB \in \Sym{d},
\end{equation}
due to its definition \eqref{eq:theory_unitCells_muQ}, more precisely
\begin{equation}\label{eq:theory_unitCells_muQapp_positive2}
	\fB : \Q_Y : \fB = \text{vol}(Y) \mean{ \left[\left( \fA^{\app}_Y - \mean{\fA^{\app}_Y} \right):\fB \right]^2} \geq 0, \quad \fB \in \Sym{d}.
\end{equation}
Sometimes it is convenient to re-write the formula \eqref{eq:theory_unitCells_muQ} in the form
\begin{equation}\label{eq:theory_unitCells_muQ2}
	\Q_Y = \text{vol}(Y) \left( \mean{\fA^{\app}_Y \otimes \fA^{\app}_Y} - \mean{\fA^{\app}_Y} \otimes \mean{\fA^{\app}_Y} \right).
\end{equation}
For a specific random conductance ensemble, Duerinckx et al.~\cite{Duerinckx2020muQ} showed that the apparent \Qtensor{} converges to a deterministic limit $\Q \in \Sym{\Sym{d}}$ 
\begin{equation}\label{eq:theory_unitCells_muQ_convergence}
	\Q_{LY} \rightarrow \Q \quad \text{as} \quad L \rightarrow \infty,
\end{equation}
the ensemble's \Qtensor{}, which inherits the symmetry property \eqref{eq:theory_unitCells_muQapp_major_symmetry}
\begin{equation}\label{eq:theory_unitCells_muQ_major_symmetry}
	\fB : \Q : \fC = \fC : \Q : \fB, \quad \fB, \fC \in \Sym{d},
\end{equation}
and the positive semidefiniteness property \eqref{eq:theory_unitCells_muQapp_positive}
\begin{equation}\label{eq:theory_unitCells_muQ_positive}
	\fB : \Q : \fB \geq 0, \quad \fB \in \Sym{d}.
\end{equation}
from the apparent tensor \eqref{eq:theory_unitCells_muQ} via the limit procedure \eqref{eq:theory_unitCells_muQ_convergence}. Moreover, Duerinckx et al.~\cite{Duerinckx2020muQ} established the quantitative decay estimate
\begin{equation}\label{eq:theory_unitCells_muQ_convergence_quantitative}
	\left\| \Q_{LY} - \Q \right\| \sim \left(\frac{\log L}{L}\right)^{\frac{d}{2}}  \quad \text{as} \quad L \rightarrow \infty
\end{equation}
for periodized ensembles on cubic cells.

\subsection{Symmetry properties of effective and apparent tensors}
\label{sec:theory_symmetryProperties}

To encode the symmetry of our ensemble \eqref{eq:theory_setup_expectation}, we consider a given closed subgroup $\mathcal{G}$ of the orthogonal group
\begin{equation}\label{eq:theory_setup_orthogonal_group}
	O(d) = \left\{ \fQ \in \R^{d \times d} \,\middle|\, \fQ^T \fQ = \fI \right\}
\end{equation}
of matrices which preserve the Euclidean norm. Then, we study the group action
\begin{equation}\label{eq:theory_setup_group_action}
	\mathcal{G} \times \mathcal{A} \rightarrow \mathcal{A}, \quad (\fQ, \fA) \mapsto \fA^{\fQ},
\end{equation}
defined point-wise via the formula
\begin{equation}\label{eq:theory_setup_group_action_definition}
	\fA^{\fQ}(\fx) = \fQ\fA(\fQ^T \fx) \fQ^T, \quad \fx \in \R^d,
\end{equation}
for $\fA \in \mathcal{A}$ and $\fQ \in \mathcal{G}$. It is not difficult to see that this prescription is well-defined, i.e., preserves the eigenvalue bounds \eqref{eq:theory_setup_space}, and does indeed lead to a group action, i.e., the identity
\begin{equation}\label{eq:theory_setup_group_action_well_defined}
	\left(\fA^{\fQ_1}\right)^{\fQ_2}(\fx) = \fA^{\fQ_2\fQ_1}(\fx), \quad \fx \in \R^d,
\end{equation}
holds for all elements $\fQ_1,\fQ_2 \in \mathcal{G}$. We assume that the action \eqref{eq:theory_setup_group_action} is $\mu$-measurable for an ensemble \eqref{eq:theory_setup_expectation} of conductivity tensors.\\
We say that the ensemble $\mu$ is $\mathcal{G}$-invariant if, for any $\mu$-measurable set $\mathcal{S} \subseteq \mathcal{A}$ and every group element $\fQ \in \mathcal{G}$, the transformed set
\begin{equation}\label{eq:theory_setup_transformed_set_measurable}
	\mathcal{S}^{\fQ} = \left\{ \fA^{\fQ} \in \mathcal{A} \,\middle| \, \fA \in \mathcal{S} \right\}
\end{equation}
is also $\mu$-measurable and has identical measure
\begin{equation}\label{eq:theory_setup_transformed_set_same_measure}
	\mu(\mathcal{S}^{\fQ}) = \mu(\mathcal{S}).
\end{equation}
As a consequence, any $\mu$-integrable function $F:\mathcal{A} \rightarrow \R$ satisfies the equation
\begin{equation}\label{eq:theory_setup_invariance_assumption}
	\int_{\mathcal{A}} F(\fA)\,d\mu(\fA) = \int_{\mathcal{A}} F\left(\fA^{\fQ}\right)\,d\mu(\fA)
\end{equation}
for all group elements $\fQ \in \mathcal{G}$.\\
The following properties hold:
\begin{enumerate}
	\item For a $\mathcal{G}$-invariant ensemble, the associated effective conductivity \eqref{eq:theory_setup_Aeff} is $\mathcal{G}$-invariant, i.e., the identity
	\begin{equation}\label{eq:theory_setup_Aeff_invariance}
		\fQ^T \fA^{\eff} \fQ = \fA^{\eff}
	\end{equation}
	holds for all $\fQ \in \mathcal{G}$.
	\item For a $\mathcal{G}$-invariant ensemble, the associated $\mu$Q-tensor \eqref{eq:theory_unitCells_muQ_convergence} is $\mathcal{G}$-invariant, i.e., the identity
	\begin{equation}\label{eq:theory_setup_muQ_invariance}
		\Q^{\fQ} = \Q
	\end{equation}
	holds for all $\fQ \in \mathcal{G}$ and the implicitly defined action
	\begin{equation}\label{eq:theory_setup_action_on_muQ}
		\fxi\otimes_s \feta : \Q^{\fQ} : \fxi\otimes_s \feta = \left[(\fQ \fxi) \otimes_s (\fQ \feta)\right] : \Q : \left[(\fQ \fxi) \otimes_s (\fQ \feta)\right], \quad \fxi, \feta \in \R^d.
	\end{equation}
\end{enumerate}
Item 1 is known in the literature, see Alexanderian~\cite{Alexanderian2012} and Zhikov-Kozlov~\cite{zhikov1994homogenization} who use a less stringent invariance condition than described by eq.~\eqref{eq:theory_setup_transformed_set_same_measure}. We proceed differently -- the results \eqref{eq:theory_setup_Aeff_invariance} and \eqref{eq:theory_setup_muQ_invariance} are a byproduct of their unit-cell based counterparts \eqref{eq:theory_unitCells_Aapp_invariance} and \eqref{eq:theory_unitCells_muQ_invariance} discussed below. For a spherical unit cell \eqref{eq:theory_unitCells_sphericalCell}, the limiting procedures \eqref{eq:theory_unitCells_Aapp_convergence} and the \Qtensor{} \eqref{eq:theory_unitCells_muQ_convergence} imply the statements \eqref{eq:theory_setup_Aeff_invariance} and \eqref{eq:theory_setup_muQ_invariance} directly.
\\
The effective properties \eqref{eq:theory_setup_Aeff} satisfy the invariance property \eqref{eq:theory_setup_Aeff_invariance}, together with the associated \Qtensor{} \eqref{eq:theory_setup_muQ_invariance}. However, as these quantities cannot be inferred directly, we are more interested in the ramifications for the RVE method described in section \ref{sec:theory_unitCells}. To this end, we define the symmetry group of a given unit cell $Y$ via
\begin{equation}\label{eq:theory_unitCells_symmetryGroupCell_orthogonal}
	O_Y(d) = \left\{ \fQ \in O(d) \, \middle| \, \fQ Y = Y\right\},
\end{equation}
comprising those orthogonal transformations which preserve the unit cell \eqref{eq:theory_unitCells_generalCell}. Apparently, the spherical unit cell \eqref{eq:theory_unitCells_sphericalCell} is the most symmetric, i.e., it holds
\begin{equation}\label{eq:theory_unitCells_symmetryGroupCell_spherical}
	O_{Y^{\sph}_R}(d) = O(d).
\end{equation}
Other unit cells may reduce the symmetry, e.g., we have
\begin{equation}\label{eq:theory_unitCells_symmetryGroupCell_rectangular}
	O_{Y^{\rect}_{L_1,\ldots,L_d}}(d) = \left\{ \textrm{diag}(\alpha_1, \alpha_2, \ldots, \alpha_d) \,\middle|\, \alpha_i \in \{-1,+1\}, \quad i=1,2,\ldots,d \right\},
\end{equation}
for a rectangular unit cell \eqref{eq:theory_unitCells_rectangularCell} with \emph{distinct} edge-lengths, i.e., $L_i \neq L_j$ $(i\neq j)$.\\
The symmetry group \eqref{eq:theory_unitCells_symmetryGroupCell_orthogonal} of the cube \eqref{eq:theory_unitCells_cubicCell} in three spatial dimensions, $d=3$, equals the full octahedral group~\cite[§3.10.D]{Lomont1987book}
\begin{equation}\label{eq:theory_unitCells_symmetryGroupCell_cubic}
	O_{Y^{\cub}_L}(3) \equiv O_h(3),
\end{equation}
which has $48$ elements and is, for instance, generated by the three elements
\begin{equation}\label{eq:theory_unitCells_symmetryGroupCell_cubic_generator}
	\left[
	\begin{array}{rrr}
		1 &0 &0\\
		0 &1 &0\\
		0 &0 &-1
	\end{array}
	\right],
	\quad
	\left[
	\begin{array}{rrr}
		1 &0 &0\\
		0 &0 &1\\
		0 &1 &0
	\end{array}
	\right]
	\quad \text{and} \quad
	\left[
	\begin{array}{rrr}
		0 &1 &0\\
		1 &0 &0\\
		0 &0 &1
	\end{array}
	\right].
\end{equation}
As the action of any element $\fQ$ of the group $O_Y(d)$ preserves the unit cell $Y$, the identity
\begin{equation}\label{eq:theory_unitCells_covariance_corrector}
	\phi_{Y,\fA^{\fQ},\fQ\bar{\fxi}}(\fQ\fx) = \phi_{Y,\fA,\bar{\fxi}}(\fx)
\end{equation}
holds for any realization $\fA \in \mathcal{A}$, any vector $\bar{\fxi} \in \R^d$ and almost every point $\fx \in Y$, see Appendix \ref{apx:theory_covariantCorrectors}. With the result \eqref{eq:theory_unitCells_covariance_corrector} at hand, we may directly infer the invariance results
\begin{equation}\label{eq:theory_unitCells_Aapp_invariance}
	\fQ \mean{\fA^{\app}_Y} \fQ^T = \mean{\fA^{\app}_Y}
\end{equation}
and
\begin{equation}\label{eq:theory_unitCells_muQ_invariance}
	\left(\Q_Y\right)^{\fQ} = \Q_Y
\end{equation}
for all elements $\fQ$ of the cell-symmetry group
\begin{equation}\label{eq:theory_unitCells_symmetryGroupCell}
	\mathcal{G}_Y = \mathcal{G} \cap O_Y(d).
\end{equation}
The derivation of these results is discussed in the Appendices \ref{apx:theory_eff} and \ref{apx:theory_muQ}.\\
Before discussing the consequences of the statements \eqref{eq:theory_unitCells_Aapp_invariance} and \eqref{eq:theory_unitCells_muQ_invariance}, we take a look at an illustrative computational example. We consider a cubic volume element \eqref{eq:theory_setup_cube_defn} which comprises orthotropically distributed short fibers, as illustrated in the second column of Tab.\ref{tab:ramification-ortho}. The details of such microstructure generation and conductivity coefficients are given in section \ref{sec:computation_orthotropy} and not relevant for the current discussion. The symmetry group of the ensemble is the orthotropic group, which is not broken by the cubic cell:
\begin{equation}\label{eq:ramification_G_Y}
	\mathcal{G}_Y = \left\{ \fQ \in O(3) \, \middle| \, \fQ = \text{diag} \qty(\pm 1, \pm 1,\pm 1) \right\}.
\end{equation}
This group consists of eight group elements: the identity $\Id$, the inversion $-\Id$, reflections with respect to the $x-,y-,z-$planes, $M_{yz},M_{xz},M_{xy}$, respectively, and $180^\circ$-rotations around the $x-,y-,z-$axes, $R_x,R_y,R_z$, respectively, as demonstrated in the second column of Tab.~\ref{tab:ramification-ortho}.
\begin{longtable}{ccccc}
	\caption{Illustrating the reason for the invariance condition \eqref{eq:theory_unitCells_Aapp_invariance}. The transformed microstructures (red) have the same probability as the original microstructure (blue)}
	\label{tab:ramification-ortho}\\
	\toprule
	\endfirsthead
	\caption*{Apparent of symmetry versus symmetry of apparent (cont.)}\\
	\toprule
	\endhead 
	\endfoot 
	\bottomrule
	\endlastfoot
	$\fQ \in \mathcal{G}_Y$ & $\fA^{\fQ}$ \eqref{eq:theory_setup_group_action_definition} & $ \left(\fA^{\fQ}\right)^{\app}_Y $  & $\left(\fA^{\app}_Y\right)^{\fQ}$\\ \noalign{\medskip}
	\midrule
	$\Id=\text{diag}\qty(1,1,1)$ & \begin{minipage}{.15\textwidth}
		\includegraphics[width=.8\linewidth, trim = 2in  0in  13.5in  3in ,clip]{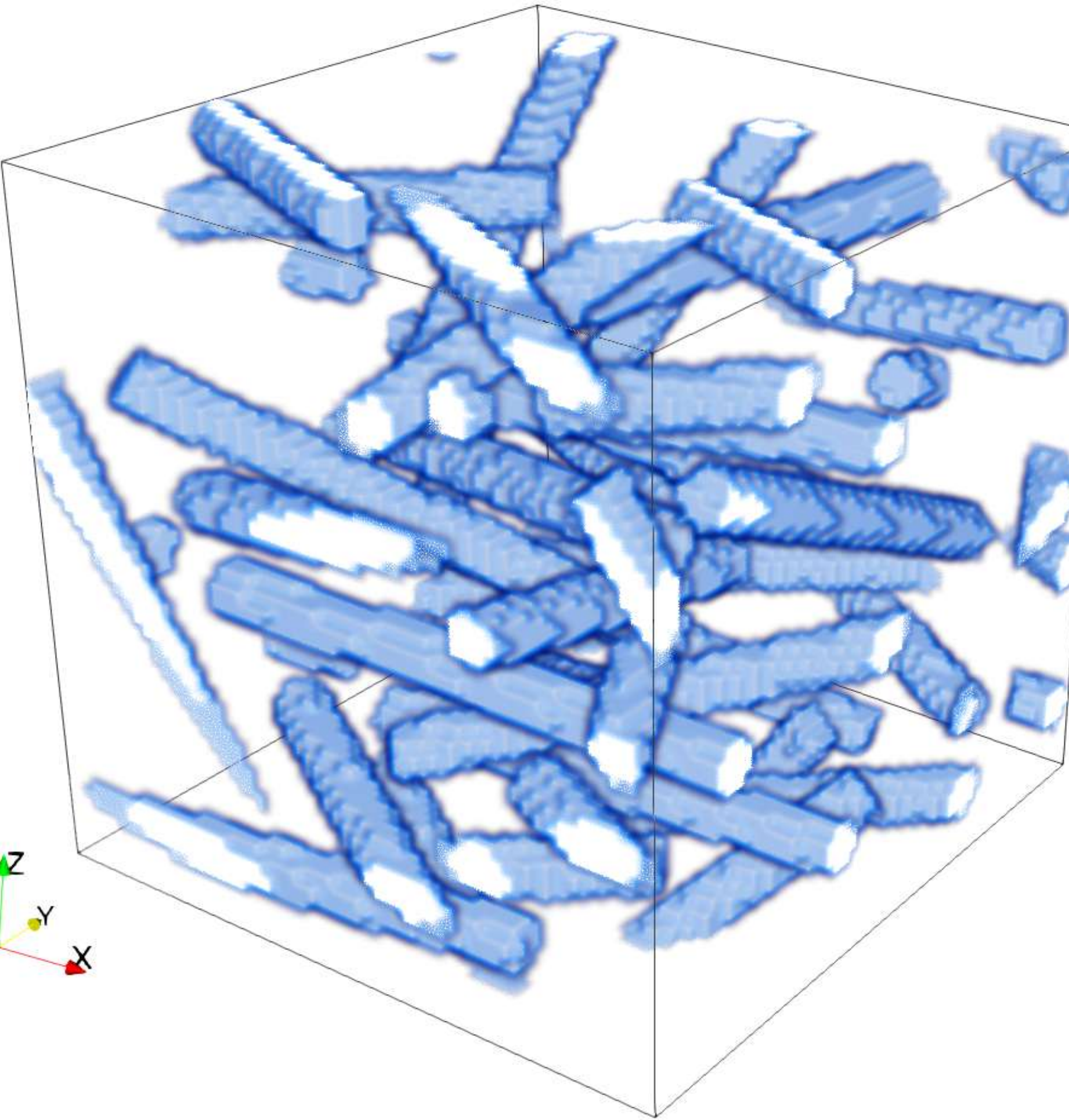}
	\end{minipage} & \resizebox{.25\textwidth}{!}{$\mqty[0.3032 & 0.0001 & -0.0001 \\
		0.0001 & 0.2801  & 0.0004 \\
		-0.0001 & 0.0004 & 0.2675]$}  & \resizebox{.25\textwidth}{!}{$\mqty[0.3032 & 0.0001 & -0.0001 \\
		0.0001 & 0.2801  & 0.0004 \\
		-0.0001 & 0.0004 & 0.2675]$} \\ \noalign{\medskip}
	$-\Id=\text{diag}\qty(-1,-1,-1)$ & \begin{minipage}{.15\textwidth}
		\includegraphics[width=.8\linewidth, trim = 2in  0in  13.5in  3in ,clip]{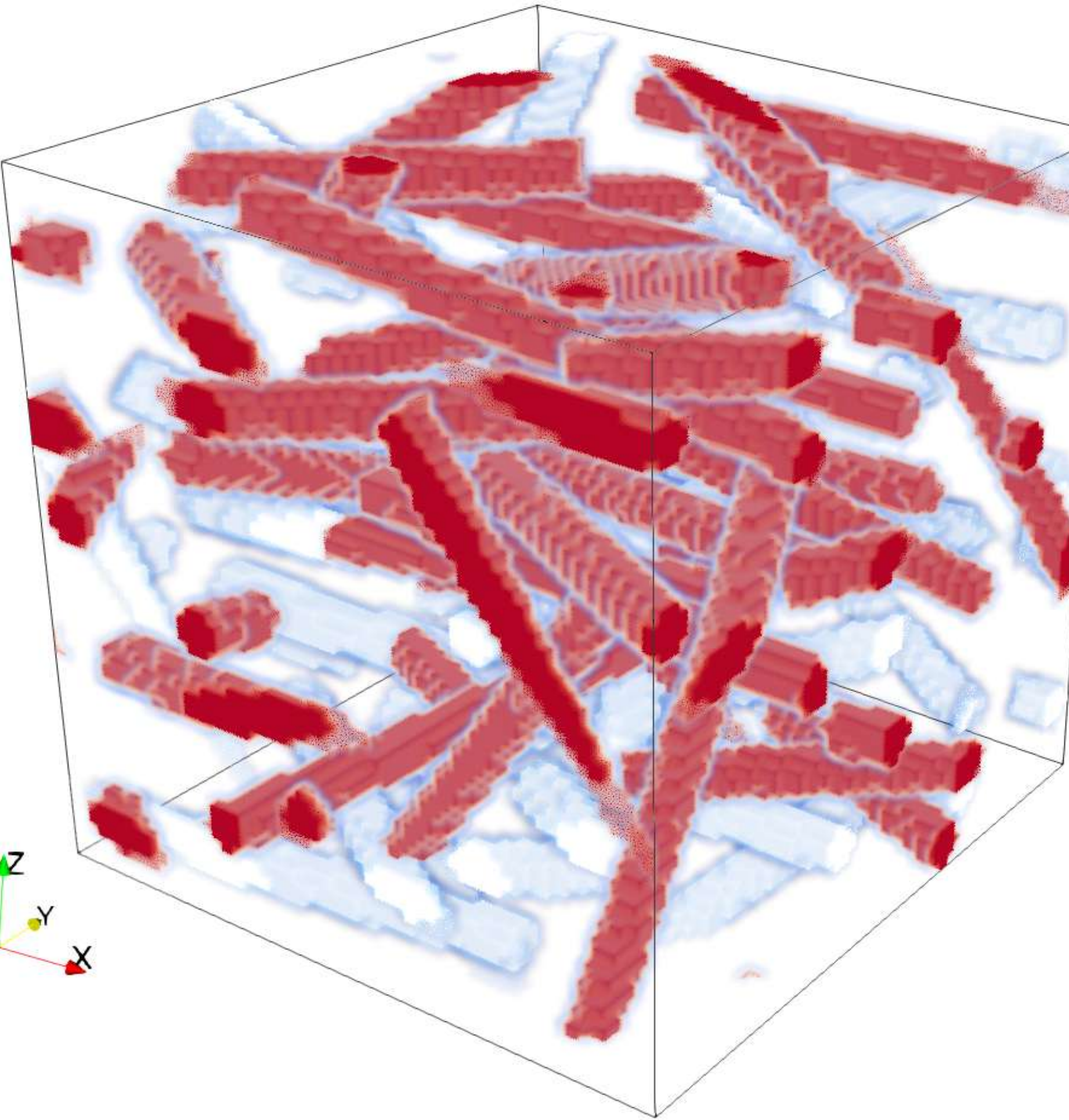}
	\end{minipage} & \resizebox{.25\textwidth}{!}{$\mqty[0.3032 & 0.0001 & -0.0001 \\
		0.0001 & 0.2801  & 0.0004 \\
		-0.0001 & 0.0004 & 0.2675]$} & \resizebox{.25\textwidth}{!}{$\mqty[0.3032 & 0.0001 & -0.0001 \\
		0.0001 & 0.2801  & 0.0004 \\
		-0.0001 & 0.0004 & 0.2675]$} \\ \noalign{\medskip}
	$M_{yz}=\text{diag}\qty(-1,1,1)$ & \begin{minipage}{.15\textwidth}
		\includegraphics[width=.8\linewidth, trim =2in  0in  13.5in  3in ,clip]{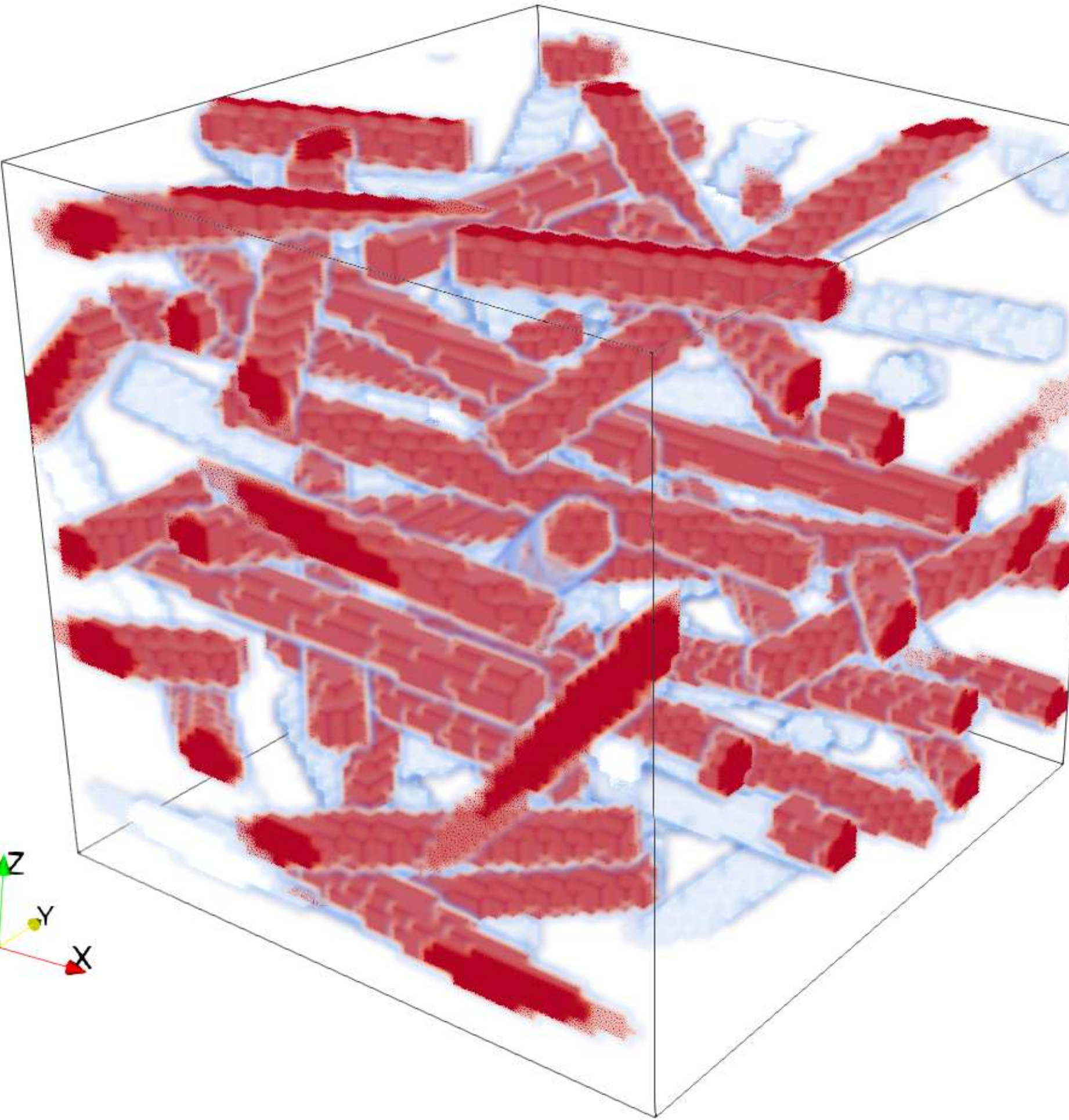}
	\end{minipage} & \resizebox{.25\textwidth}{!}{$\mqty[0.3032 & -0.0001 & 0.0001 \\
		-0.0001 & 0.2801  & 0.0004 \\
		0.0001 & 0.0004 & 0.2675]$} & \resizebox{.25\textwidth}{!}{$\mqty[0.3032 & -0.0001 & 0.0001 \\
		-0.0001 & 0.2801  & 0.0004 \\
		0.0001 & 0.0004 & 0.2675]$} \\ \noalign{\medskip}
	$M_{xz}=\text{diag}\qty(1,-1,1)$ & \begin{minipage}{.15\textwidth}
		\includegraphics[width=.8\linewidth, trim =2in  0in  13.5in  3in ,clip]{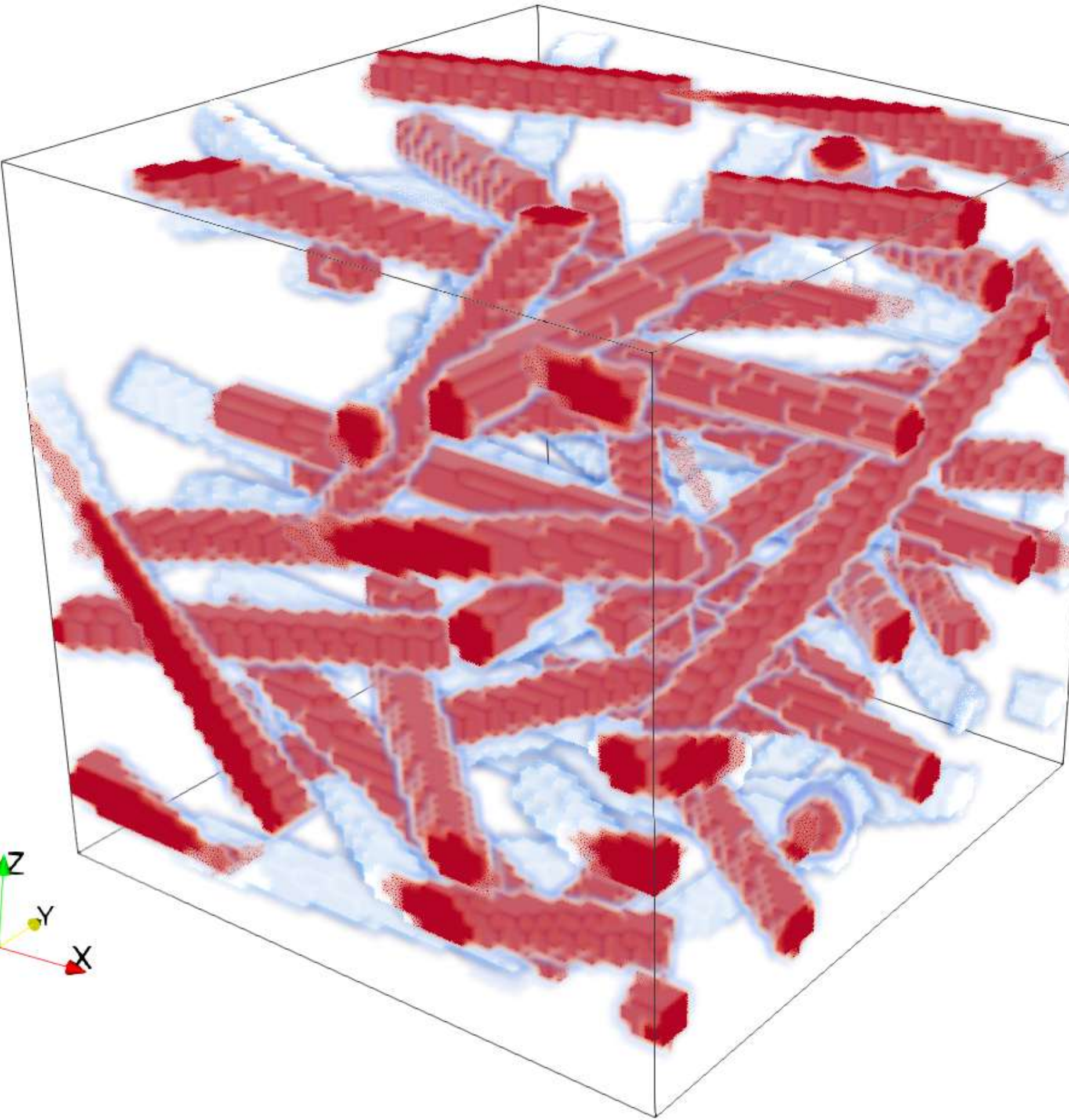}
	\end{minipage} & \resizebox{.25\textwidth}{!}{\textbf{$\mqty[0.3032 & -0.0001 & -0.0001 \\
			-0.0001 & 0.2801  & -0.0004 \\
			-0.0001 & -0.0004 & 0.2675]$}}  & \resizebox{.25\textwidth}{!}{$\mqty[0.3032 & -0.0001 & -0.0001 \\
		-0.0001 & 0.2801  & -0.0004 \\
		-0.0001 & -0.0004 & 0.2675]$} \\ \noalign{\medskip}
	$M_{xy}=\text{diag}\qty(1,1,-1)$ & \begin{minipage}{.15\textwidth}
		\includegraphics[width=.8\linewidth, trim = 2in  0in  13.5in  3in ,clip]{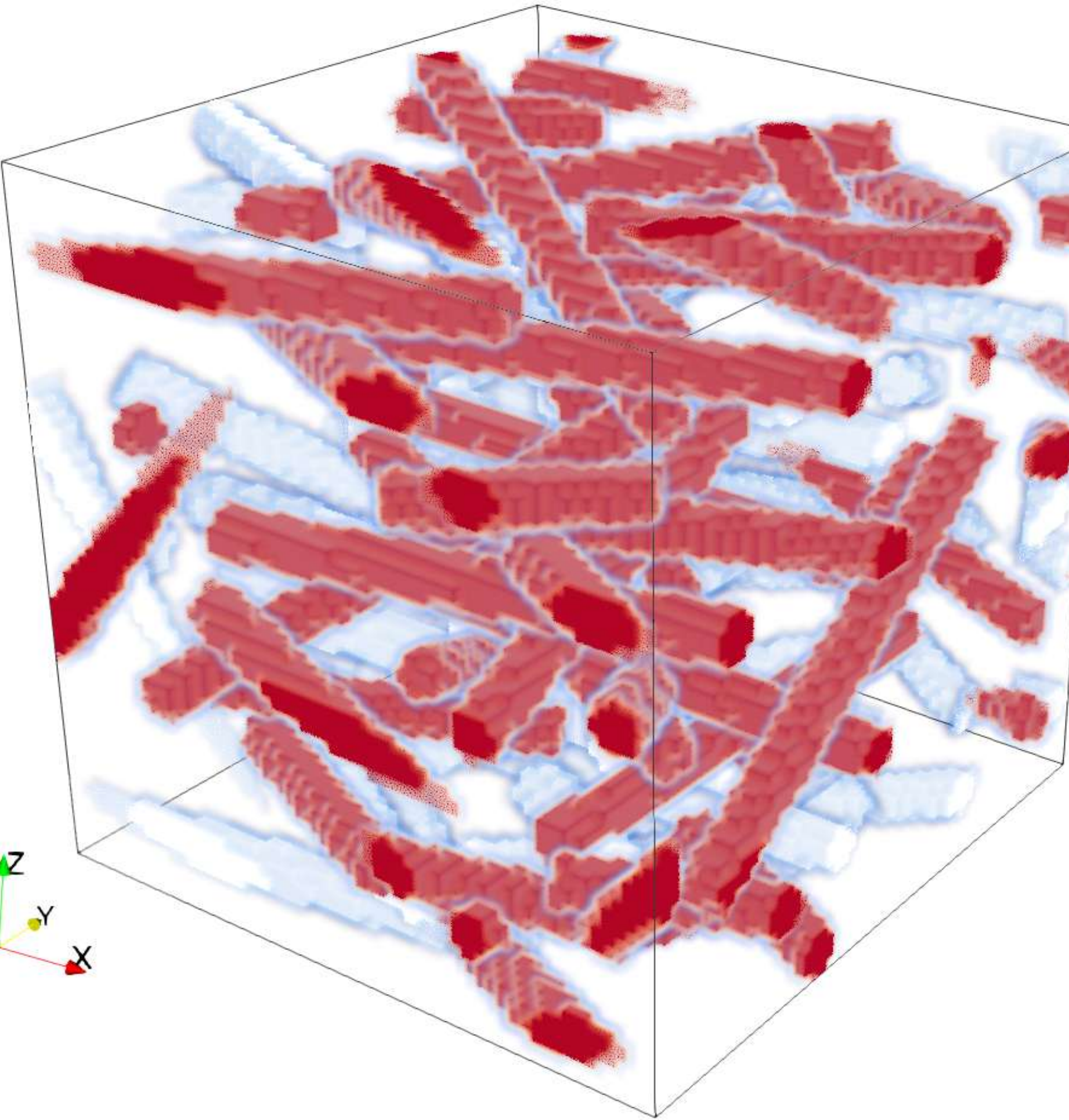}
	\end{minipage} &  \resizebox{.25\textwidth}{!}{$\mqty[0.3032 & 0.0001 & 0.0001 \\
		0.0001 & 0.2801  & -0.0004 \\
		0.0001 & -0.0004 & 0.2675]$} & \resizebox{.25\textwidth}{!}{$\mqty[0.3032 & 0.0001 & 0.0001 \\
		0.0001 & 0.2801  & -0.0004 \\
		0.0001 & -0.0004 & 0.2675]$} \\ \noalign{\medskip}
	$R_x=\text{diag}\qty(1,-1,-1)$ & \begin{minipage}{.15\textwidth}
		\includegraphics[width=.8\linewidth, trim = 2in  0in  13.5in  3in,clip]{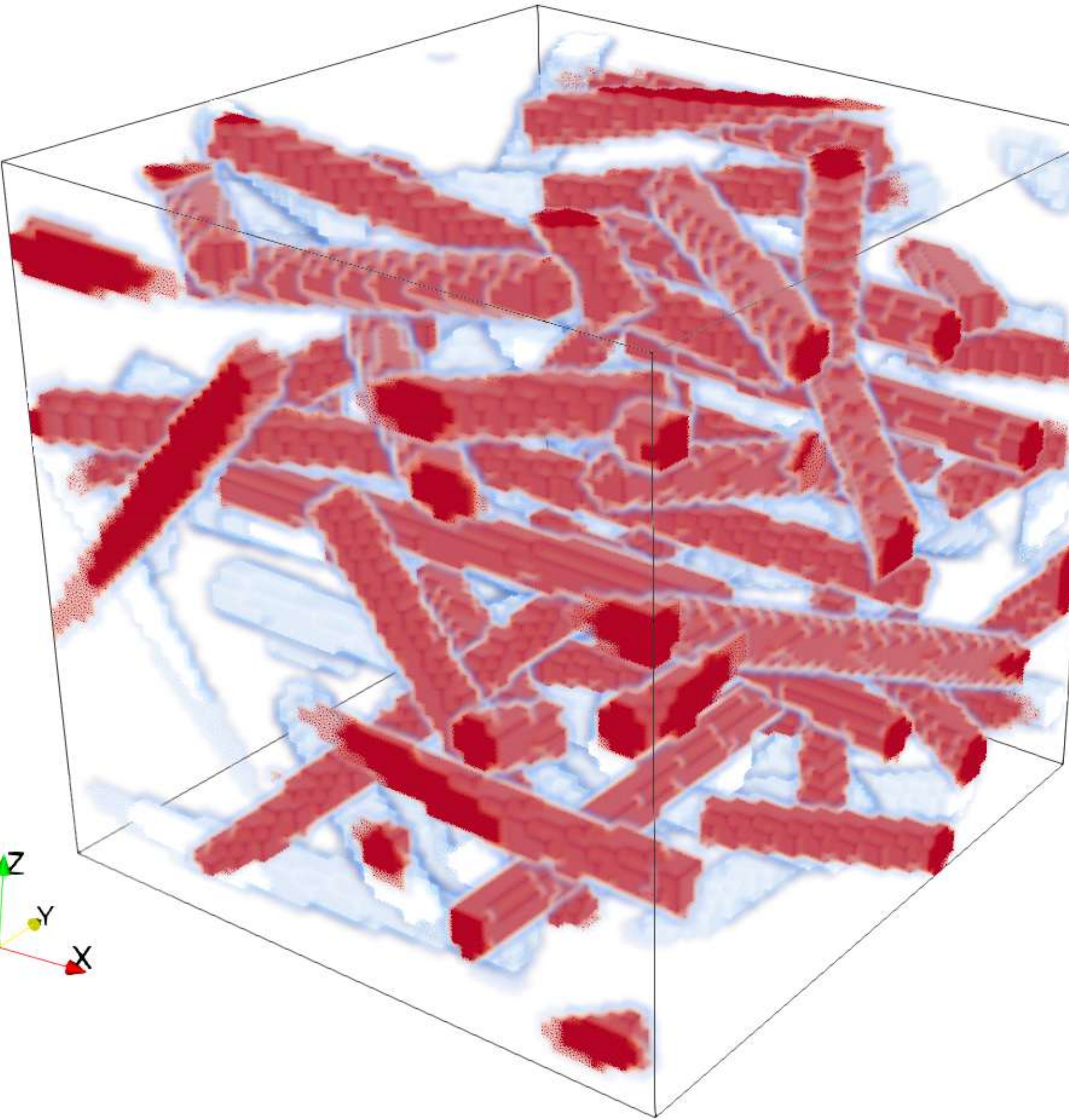}
	\end{minipage} & \resizebox{.25\textwidth}{!}{$\mqty[0.3032 & -0.0001 & 0.0001 \\
		-0.0001 & 0.2801  & 0.0004 \\
		0.0001 & 0.0004 & 0.2675]$} &  \resizebox{.25\textwidth}{!}{$\mqty[0.3032 & -0.0001 & 0.0001 \\
		-0.0001 & 0.2801  & 0.0004 \\
		0.0001 & 0.0004 & 0.2675]$} \\ \noalign{\medskip}
	$R_y=\text{diag}\qty(-1,1,-1)$ & \begin{minipage}{.15\textwidth}
		\includegraphics[width=.8\linewidth, trim =2in  0in  13.5in  3in,clip]{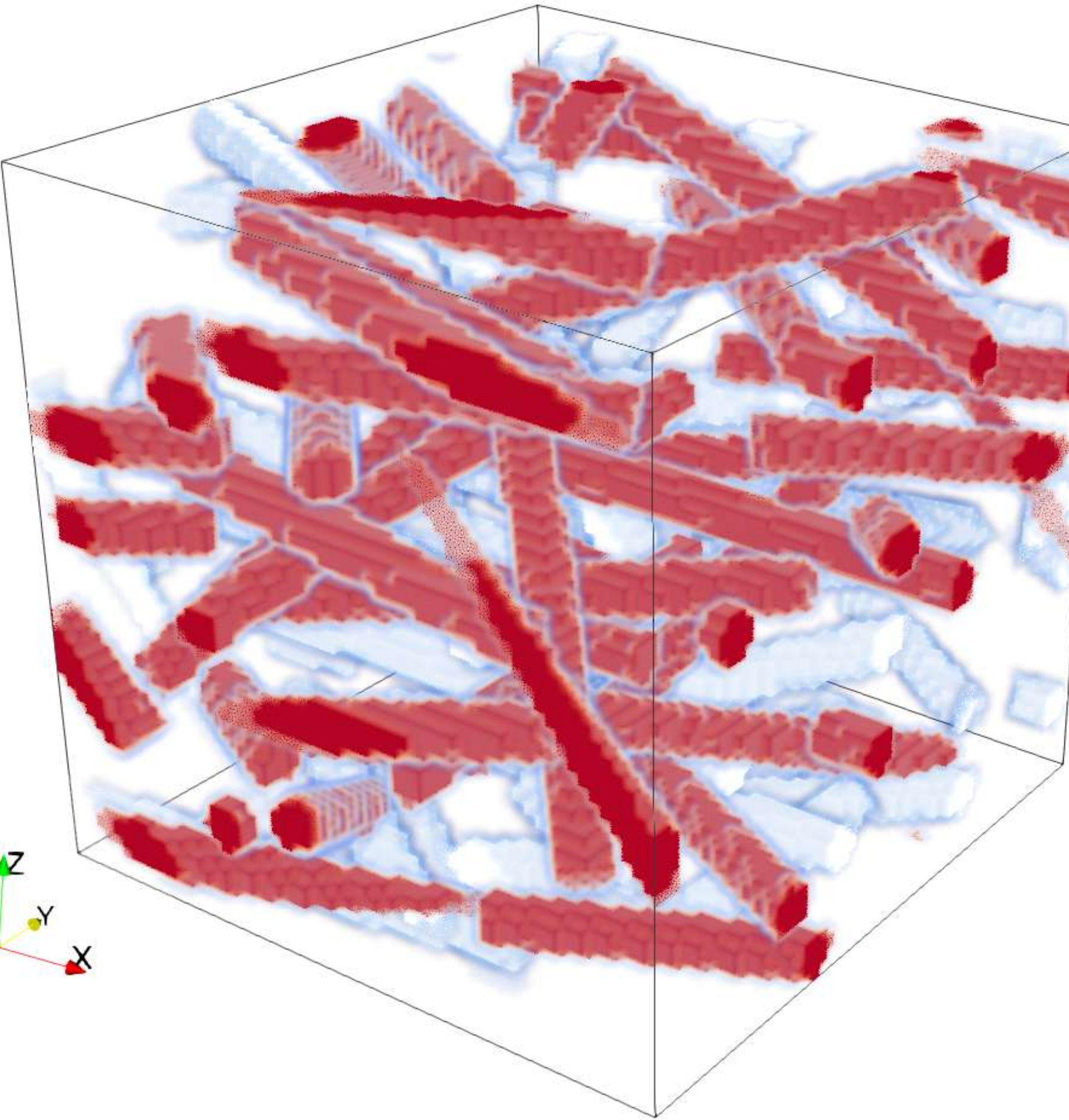}
	\end{minipage} &  \resizebox{.25\textwidth}{!}{$\mqty[0.3032 & -0.0001 & -0.0001 \\
		-0.0001 & 0.2801  & -0.0004 \\
		-0.0001 & -0.0004 & 0.2675]$} & \resizebox{.25\textwidth}{!}{$\mqty[0.3032 & -0.0001 & -0.0001 \\
		-0.0001 & 0.2801  & -0.0004 \\
		-0.0001 & -0.0004 & 0.2675]$} \\ \noalign{\medskip}
	$R_z=\text{diag}\qty(-1,-1,1)$ & \begin{minipage}{.15\textwidth}
		\includegraphics[width=.8\linewidth, trim = 2in  0in  13.5in  3in,clip]{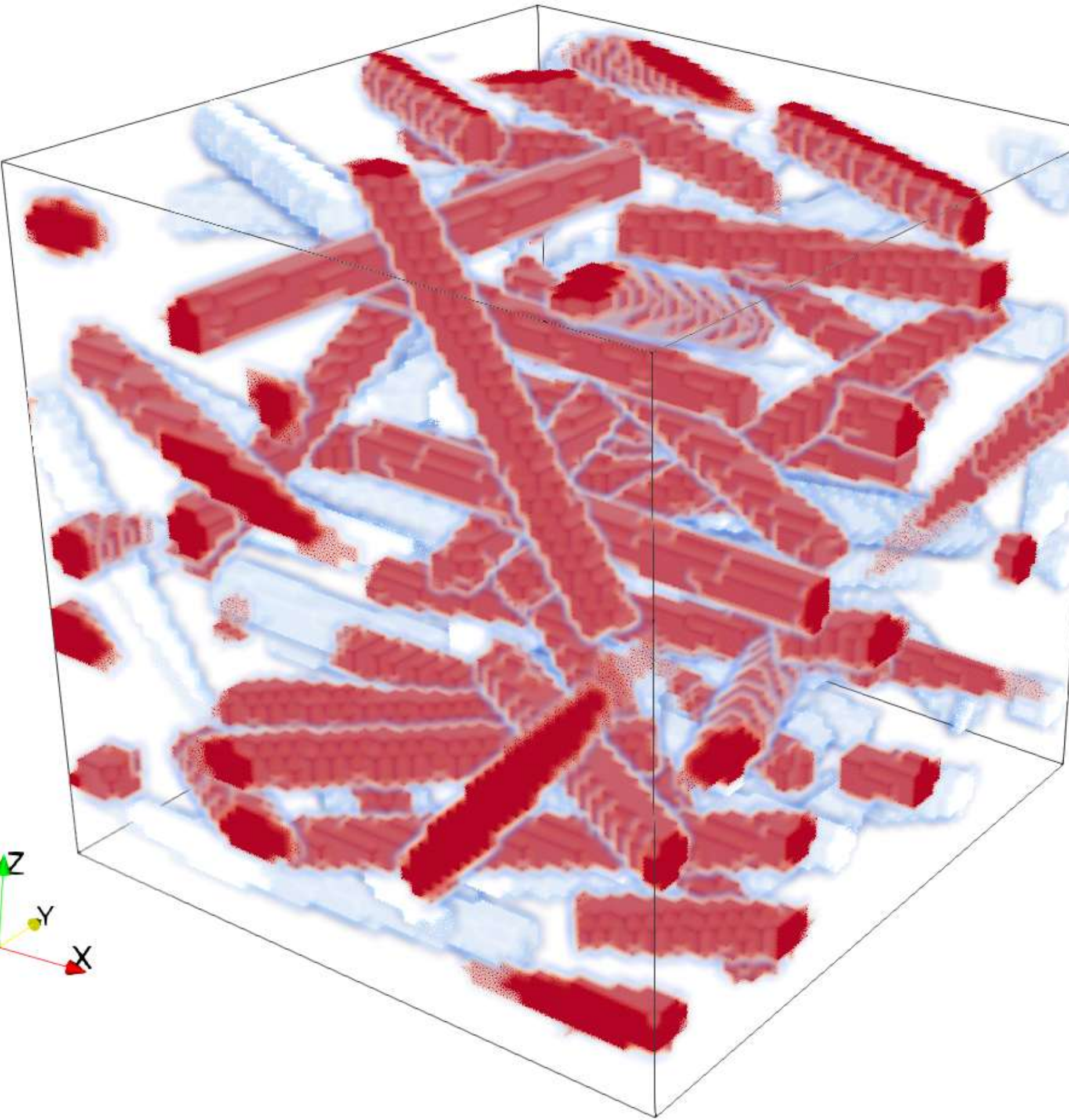}
	\end{minipage} &\resizebox{.25\textwidth}{!}{$\mqty[0.3032 & 0.0001 & 0.0001 \\
		0.0001 & 0.2801  & -0.0004 \\
		0.0001 & -0.0004 & 0.2675]$} & \resizebox{.25\textwidth}{!}{$\mqty[0.3032 & 0.0001 & 0.0001 \\
		0.0001 & 0.2801  & -0.0004 \\
		0.0001 & -0.0004 & 0.2675]$}  \\ \noalign{\medskip}
	Average &  & \resizebox{.25\textwidth}{!}{$\mqty[0.3032 & 0.0000 & 0.0000 \\
		0.0000 & 0.2801  & 0.0000 \\
		0.0000 & 0.0000 & 0.2675]$}  &  \resizebox{.25\textwidth}{!}{$\mqty[0.3032 & 0.0000 & 0.0000 \\
		0.0000 & 0.2801  & 0.0000 \\
		0.0000 & 0.0000 & 0.2675]$} \\ \noalign{\medskip}
\end{longtable}
Actually, all eight microstructures shown in Tab.~\ref{tab:ramification-ortho} have the same probability. Thus, the average of the associated apparent conductivity is exactly orthotropic, as expected.\\
After this illustrative example we return to discussing the ramifications of the invariance properties \eqref{eq:theory_unitCells_Aapp_invariance} and \eqref{eq:theory_unitCells_symmetryGroupCell}. As a first consequence, we consider a spherical unit cell \eqref{eq:theory_unitCells_sphericalCell} which comes with the \emph{full} symmetry group \eqref{eq:theory_unitCells_symmetryGroupCell_spherical}. In particular, the statements \eqref{eq:theory_unitCells_Aapp_convergence} and \eqref{eq:theory_unitCells_muQ_convergence} imply the results \eqref{eq:theory_setup_Aeff_invariance} and \eqref{eq:theory_setup_muQ_invariance} for the effective conductivity and the \Qtensor{}.\\
In general, however, the symmetry group $\mathcal{G}_Y$ of the cell $Y$, defined in eq.~\eqref{eq:theory_unitCells_symmetryGroupCell} is \emph{strictly smaller} than the original symmetry group $\mathcal{G}$ of the considered ensemble.\\
A central purpose of this article is to repair this defect, see section \ref{sec:SymmetryInformedStrats}.

\subsection{Relevant ensemble- and cell-symmetry groups}
\label{sec:theory_examples}

\subsubsection{Isotropic ensembles}
\label{sec:theory_examples_isotropic}

We consider the full rotation group~\cite{brannon2018rotation}
\begin{equation}\label{eq:special_isotropic_SO}
	SO(d) = \left\{ \fQ \in \R^{d \times d} \,\middle| \, \fQ^T \fQ = \fI \quad \text{and} \quad \det \fQ = 1\right\}.
\end{equation}
In case the ensemble \eqref{eq:theory_setup_expectation} satisfies the invariance condition \eqref{eq:theory_setup_transformed_set_same_measure} for the group \eqref{eq:special_isotropic_SO}, the implied invariance property \eqref{eq:theory_setup_Aeff_invariance} of the effective conductivity tensor \eqref{eq:theory_setup_Aeff} shows that it is proportional to the identity tensor $\fI$, i.e., in three spatial dimensions, $d=3$, the representation~\cite{brannon2018rotation}
\begin{equation}\label{eq:special_isotropic_A}
	\fA^{\eff} = \alpha \fI  \equiv  \mqty[\alpha & 0 & 0 \\ 0 & \alpha & 0 \\ 0 & 0 & \alpha]
\end{equation} 
is valid for some scalar conductivity $\alpha \in \R$. Similarly, the associated $SO(3)$-invariant \Qtensor{}, see eq.~\eqref{eq:theory_setup_muQ_invariance}, is an isotropic fourth-order \Qtensor{}, admitting the form~\cite{d2024representation}
\begin{equation}\label{eq:special_isotropic_Qtensor}
	\Q \equalhat \mqty[Q_{1111} & Q_{1122} & Q_{1122} & 0 & 0 & 0 \\
						Q_{1122} & Q_{1111} & Q_{1122} & 0 & 0 & 0 \\
						Q_{1122} & Q_{1122} & Q_{1111} & 0 & 0 & 0 \\	
					0 & 0 & 0 &  \frac{Q_{1111}-Q_{1122}}{2} & 0 & 0   \\
					0 & 0 & 0 & 0 &  \frac{Q_{1111}-Q_{1122}}{2} & 0   \\
					0 & 0 & 0 & 0 & 0 &    \frac{Q_{1111}-Q_{1122}}{2} 					
					]
\end{equation}
in Voigt's notation with two independent coefficients $Q_{1111}$ and $Q_{1122}$. 

\subsubsection{Transversely isotropic ensembles}
\label{sec:theory_examples_transversely_isotropic}

We consider the case of rotations which fix a specific axis, described by a unit vector $\fa \in \R^d$, 
\begin{equation}\label{eq:special_trans_isotropic_symmetry_group_general}
	\mathcal{G} = \left\{ \fQ \in SO(d) \,\middle|\, \fQ \fa = \fa\right\}.
\end{equation}
Put differently, ensembles invariant w.r.t. the symmetry group \eqref{eq:special_trans_isotropic_symmetry_group_general} are transversely isotropic w.r.t. the axis $\fa$. Apparently, the group \eqref{eq:special_trans_isotropic_symmetry_group_general} is isomorphic to the group $SO(d-1)$.\\
To be more specific, we restrict to dimension three, $d=3$, and consider the axis $\fa = \fe_3$. Then, the group \eqref{eq:special_trans_isotropic_symmetry_group_general} becomes
\begin{equation}\label{eq:special_trans_isotropic_symmetry_group_special}
	\mathcal{G} = \left\{ \fQ = \mqty[ a & b & 0 \\ c & d & 0 \\ 0& 0& 1] \,\middle|\, \mqty[ a & b \\ c & d] \in SO(2) \right\}.
\end{equation}
The associated transversely isotropic effective conductivity has the form~\cite{brannon2018rotation}
\begin{equation}\label{eq:special_trans_apparent}
	\fA^{\eff} = \mqty[A_{11} & 0 & 0 \\
								0 & A_{11} & 0 \\
								0 & 0 & A_{33}].
\end{equation}
with two independent constants $A_{11}$ und $A_{33}$.
Similarly, the associated \Qtensor{} is also transversely isotropic and attains the form~\cite{zohdi2008introduction}
\begin{equation}\label{eq:special_trans_isotropic_Qtensor}
	\Q \equalhat \mqty[Q_{1111} & Q_{1122} & Q_{1133} & 0 & 0 & 0 \\
	Q_{1122} & Q_{1111} & Q_{1133} & 0 & 0 & 0 \\
	Q_{1133} & Q_{1122} & Q_{3333} & 0 & 0 & 0 \\	
	0 & 0 & 0 & Q_{2323} & 0 & 0  \\
	0 & 0 & 0 & 0 & Q_{2323} & 0  \\
	0 & 0 & 0 & 0 & 0  & \frac{Q_{1111}-Q_{1122}}{2}					
	]
\end{equation}
in Voigt's notation, which is characterized by five independent coefficients $Q_{1111}$, $Q_{1122}$, $Q_{1133}$, $Q_{3333}$ and $Q_{2323}$. 

\subsubsection{Orthotropic ensembles}
\label{sec:theory_examples_orthotropic}

We consider the group
\begin{equation}\label{eq:special_ortho_group}
	\mathcal{G} = \left\{ \textrm{diag}(\alpha_1, \alpha_2, \ldots, \alpha_d) \,\middle|\, \alpha_i \in \{-1,+1\}, \quad i=1,2,\ldots,d \right\},
\end{equation}
generated by all reflections with respect to the coordinate planes in the Euclidean space $\R^d$. Formally, the group \eqref{eq:special_ortho_group} coincides with the cell-symmetry group \eqref{eq:theory_unitCells_symmetryGroupCell_rectangular} associated to a non-equiaxed rectangular cell.\\
For an orthotropic ensemblein three spatial dimensions, the effective conductivity tensor has the form
\begin{equation}\label{eq:special_ortho_apparent}
	\fA^{\eff} = \mqty[A_{11} & 0 & 0 \\
	0 & A_{22} & 0 \\
	0 & 0 & A_{33}],
\end{equation}
involving three potentially different conductivities $A_{11}$, $A_{22}$ and $A_{33}$. Moreover, the corresponding \Qtensor{} is characterized by nine independent coefficients~\cite{zohdi2008introduction}, and may be expressed in the following matrix form
\begin{equation}\label{eq:special_ortho_isotropic_Qtensor}
	\Q \equalhat \mqty[Q_{1111} & Q_{1122} & Q_{1133} & 0 & 0 & 0 \\
	Q_{1122} & Q_{2222} & Q_{2233} & 0 & 0 & 0 \\
	Q_{1133} & Q_{2233} & Q_{3333} & 0 & 0 & 0 \\	
	0 & 0 & 0 & Q_{2323} & 0 & 0 \\
	0 & 0 & 0 & 0 & Q_{1313} & 0  \\
	0 & 0 & 0 & 0 & 0  & Q_{1212}					
	]
\end{equation}
using Voigt's notation.

\subsubsection{Spherical cell symmetry}
\label{sec:theory_examples_spherical}

The symmetry group \eqref{eq:theory_unitCells_symmetryGroupCell_orthogonal} associated to spherical cells \eqref{eq:theory_unitCells_sphericalCell} coincides with the full symmetry of the ensemble, i.e., the ensemble symmetry is not broken by the cell. In particular, the mean apparent properties $\mean{\fA^{\app}_{Y^{\sph}_R}}$ share the same symmetries as the effective properties $\fA^{\eff}$. Also, the \Qtensor{} of the cell \eqref{eq:theory_unitCells_sphericalCell} has the full symmetries expected from the ensemble's \Qtensor{}.

\subsubsection{Cubic cell symmetry}
\label{sec:theory_examples_cubic}

The predominant cell shape in computational homogenization is given by cubic cells \eqref{eq:theory_unitCells_cubicCell}. The associated cell-symmetry group \eqref{eq:theory_unitCells_symmetryGroupCell_orthogonal} is the octahedral group \eqref{eq:theory_unitCells_symmetryGroupCell_cubic}, generated by the three matrices \eqref{eq:theory_unitCells_symmetryGroupCell_cubic_generator} in three spatial dimensions. For isotropic ensemble symmetry \eqref{eq:special_isotropic_SO}, the associated mean apparent properties have the form
\begin{equation}\label{eq:special_cubic_apparent}
	\mean{\fA^{\app}_{Y^{\cub}_L}} = \alpha \fI  \equiv  \mqty[\alpha & 0 & 0 \\ 0 & \alpha & 0 \\ 0 & 0 & \alpha]
\end{equation} 
with a single independent constant $\alpha$. In particular, the expected apparent conductivity has the same form as the effective conductivity \eqref{eq:special_isotropic_A} in the isotropic case. However, this is merely a coincidence, as a quick glance at the properties of the \Qtensor{} reveal: For an isotropic ensemble, the mean \Qtensor{} has a cubic symmetry, i.e., admits the form
\begin{equation}\label{eq:special_cubic_Qtensor}
	\mean{\Q_{Y^{\cub}_L}} \equalhat \mqty[Q_{1111} & Q_{1122} & Q_{1122} & 0 & 0 & 0 \\
						Q_{1122} & Q_{1111} & Q_{1122} & 0 & 0 & 0 \\
						Q_{1122} & Q_{1122} & Q_{1111} & 0 & 0 & 0 \\	
					0 & 0 & 0 &  Q_{1212} & 0 & 0   \\
					0 & 0 & 0 & 0 &  Q_{1212} & 0   \\
					0 & 0 & 0 & 0 & 0 & Q_{1212}
					]
\end{equation}
in Voigt's notation with three independent constants $Q_{1111}$, $Q_{1122}$ and $Q_{1212}$. Thus, in contrast to the ensemble case \eqref{eq:special_isotropic_Qtensor}, an additional independent parameter appears.\\
Using cubic cells does not impose restrictions on transversely isotropic \eqref{eq:special_trans_isotropic_symmetry_group_general} and orthotropic ensembles \eqref{eq:special_ortho_group}, as long as the axes of transverse isotropy/orthotropy are aligned with the edges of the cube.

\subsubsection{Rectangular cell symmetry}
\label{sec:theory_examples_rectangular}

The symmetry group of rectangular cells \eqref{eq:theory_unitCells_rectangularCell} with distinct edge lengths coincides with the orthotropic group \eqref{eq:special_ortho_group}. Thus, using such non-equiaxed rectangular cells breaks the symmetry to orthotropic. In order to preserve the transversely isotropic symmetry \eqref{eq:special_trans_isotropic_symmetry_group_special}, it is sufficient to use a rectangular cell which are cylindrical in the third direction, i.e., the condition $L_1 = L_2$ should hold.

%%%%%%%%%%%%%%%%%%%%%%%%%%%%%%%%%%%%%%%%%%%%%%%%%%%%%%%%%%%%%%%%%%%%%%%%%%%%%%%%%%%%%%%%%%%%%%%%%%%%%%%%%%%%%%%%%%%%%%%%%%%%%%%%%%%%%%%%%%%%%%%%%%%%%%%%%%%%%%%%%%%%%%%%%%%%%%%%%%%%%%%%%%%%

%\newpage

\section{Symmetry-informed strategies for improving the RVE method}
\label{sec:SymmetryInformedStrats}

\subsection{Symmetry-enforcing projectors and their properties}
\label{sec:SymmetryInformedStrats_projectors}

Let $V$ be a (finite-dimensional) Hilbert space with inner product $\langle \cdot , \cdot \rangle_V$. A projector $\fP \in L(V)$ is a bounded linear operator on the space $V$ which is idempotent:
\begin{equation}\label{eq:SymmetryInformedStrats_projectors_idempotent}
	\fP^2 = \fP.
\end{equation}
Any projector $\fP$ gives rise to a decomposition of the Hilbert space $V$ into closed complementary subspaces,
\begin{equation}\label{eq:SymmetryInformedStrats_projectors_Z2grading}
	V = \textrm{im}(\fP) \oplus \textrm{ker}(\fP),
\end{equation}
represented by the image $\textrm{im}(\fP)$ and the kernel $\textrm{ker}(\fP)$ of the operator $\fP$. In terms of the complementary projector $\fP^c = \fI - \fP \in L(V)$, we may also write
\begin{equation}\label{eq:SymmetryInformedStrats_projectors_complementary_projector_properties}
	\textrm{im}(\fP) = \textrm{ker}(\fP^c) \quad \text{and} \quad \textrm{ker}(\fP) = \textrm{im}(\fP^c).
\end{equation}
We call a projector \eqref{eq:SymmetryInformedStrats_projectors_idempotent} orthogonal if it is self-adjoint, i.e., the equation
\begin{equation}\label{eq:SymmetryInformedStrats_projectors_self-adjoint_projector}
	\langle \fP \fu, \fv \rangle_V = \langle \fu, \fP \fv \rangle_V
\end{equation}
holds for all $\fu,\fv \in \fV$. Equivalent to the condition \eqref{eq:SymmetryInformedStrats_projectors_self-adjoint_projector} is the validity of the Pythagorean Theorem
\begin{equation}\label{eq:SymmetryInformedStrats_projectors_Pythagoras}
	\|\fv\|^2_V = \|\fP \fv\|_V^2 + \|\fP^c \fv\|_V^2, \quad \fv \in V.
\end{equation}
A trivial, yet quite significant consequence of the identity \eqref{eq:SymmetryInformedStrats_projectors_Pythagoras} is the non-expansiveness
\begin{equation}\label{eq:SymmetryInformedStrats_projectors_nonExpansive}
	\|\fP \fv\|_V \leq \|\fv\|_V, \quad \fv \in V,
\end{equation}
of an orthogonal projector $\fP$.\\
Let us consider a compact group $\mathcal{G}$, and suppose that an orthogonal representation
\begin{equation}\label{eq:SymmetryInformedStrats_projectors_representation_general}
	\fL:\mathcal{G} \rightarrow L(V), \quad \fQ \mapsto \fL_{\fQ},
\end{equation}
is given, i.e., a group homomorphism into the invertible linear operators on the space $V$, s.t. every element is length-preserving
\begin{equation}\label{eq:SymmetryInformedStrats_projectors_representation_norm-preserving}
	\| \fL_{\fQ} \fv\|_V = \| \fv \|_V \quad \text{for all} \quad \fv \in V, \quad \fQ \in \mathcal{G}.
\end{equation}
Denote by
\begin{equation}\label{eq:SymmetryInformedStrats_projectors_invariant_set}
	V^{\mathcal{G}} = \left\{ \fv \in V \, \middle| \, \fL_{\fQ}\fv = \fv \quad \text{for all} \quad \fQ \in \mathcal{G} \right\}
\end{equation}
the set of vectors which are invariant under the group action \eqref{eq:SymmetryInformedStrats_projectors_representation_general}. The set \eqref{eq:SymmetryInformedStrats_projectors_invariant_set} defines a closed subspace of the Hilbert space $V$.\\
Before coming to the main statement of relevance in this section, we need to discuss the concept of Haar measure first. A (Borel) measure $d\mu$ on a Lie group $\mathcal{G}$ is called left invariant if the condition
\begin{equation}\label{eq:SymmetryInformedStrats_projectors_left_invariant}
	\int_\mathcal{G} f(\fR \fQ) \, d\mu(\fQ) = \int_\mathcal{G} f(\fQ) \, d\mu(\fQ)
\end{equation}
holds for any integrable function $f \in L^1(\mathcal{G};d\mu)$ and every group element $\fR \in \mathcal{G}$. It is called right invariant provided the condition
\begin{equation}\label{eq:SymmetryInformedStrats_projectors_right_invariant}
	\int_\mathcal{G} f(\fQ\fR) \, d\mu(\fQ) = \int_\mathcal{G} f(\fQ) \, d\mu(\fQ)
\end{equation}
is satisfied for any integrable function $f \in L^1(\mathcal{G};d\mu)$ and every group element $\fR \in \mathcal{G}$. The group-inversion mapping
\begin{equation}\label{eq:SymmetryInformedStrats_projectors_inversion_mapping}
	\iota: \mathcal{G} \rightarrow \mathcal{G}, \quad \fQ \mapsto \fQ^{-1},
\end{equation}
permits to turn a left-invariant measure $d\mu$ \eqref{eq:SymmetryInformedStrats_projectors_left_invariant} into a right-invariant one \eqref{eq:SymmetryInformedStrats_projectors_right_invariant} via pull-back $\iota^* d\mu$, i.e.,
\begin{equation}\label{eq:SymmetryInformedStrats_projectors_pullback_inversion}
	\int_\mathcal{G} f(\fQ) \, d\left( \iota^* d\mu \right)(\fQ) := \int_\mathcal{G} f(\fQ^{-1}) \, d\mu(\fQ).
\end{equation}
On a compact Lie group, there is precisely one left-invariant measure \eqref{eq:SymmetryInformedStrats_projectors_left_invariant} which is moreover normalized
\begin{equation}\label{eq:SymmetryInformedStrats_projectors_invariant_Haar}
	\int_\mathcal{G} d\mu_{\mathcal{G}}(\fQ) = 1
\end{equation}
to be a probability measure. It is not difficult to see that the measure \eqref{eq:SymmetryInformedStrats_projectors_invariant_Haar}, called \emph{Haar measure}~\cite{ecker2024haar}, is also right-invariant \eqref{eq:SymmetryInformedStrats_projectors_right_invariant} and also inversion-invariant
\begin{equation}\label{eq:SymmetryInformedStrats_projectors_inversion_invariant}
	\int_\mathcal{G} f(\fQ) \, d\mu_{\mathcal{G}}(\fQ). := \int_\mathcal{G} f(\fQ^{-1}) \, d\mu_{\mathcal{G}}(\fQ), \quad f \in L^1(\mathcal{G};d\mu_{\mathcal{G}}).
\end{equation}
With this concept at hand, we may represent the orthogonal projector onto the closed subspace \eqref{eq:SymmetryInformedStrats_projectors_invariant_set} via averaging the orthogonal representation \eqref{eq:SymmetryInformedStrats_projectors_representation_general} over the group w.r.t. the normalized Haar measure \eqref{eq:SymmetryInformedStrats_projectors_invariant_Haar} on the group $\mathcal{G}$, i.e., the identity
\begin{equation}\label{eq:SymmetryInformedStrats_projectors_projection_via_averaging}
	\fP_{V^{\mathcal{G}}} \fv = \int_{\mathcal{G}} \fL_{\fQ}\fv \, d\mu_{\mathcal{G}}(\fQ)
\end{equation}
holds for all $\fv \in V$. The validity of this statement is ensured by showing that the right-hand side of the expression \eqref{eq:SymmetryInformedStrats_projectors_projection_via_averaging} defines an orthogonal projector onto the closed linear subspace \eqref{eq:SymmetryInformedStrats_projectors_invariant_set}, see Apx.~\ref{sec:apx_theory_projector}.\\
After these abstract considerations let us take a closer look at why the identity \eqref{eq:SymmetryInformedStrats_projectors_projection_via_averaging} is actually useful. The two hands of the identity \eqref{eq:SymmetryInformedStrats_projectors_projection_via_averaging} have a different spirit: For a finite-dimensional space, the left-hand side \eqref{eq:SymmetryInformedStrats_projectors_projection_via_averaging} may be expressed in the form
\begin{equation}\label{eq:SymmetryInformedStrats_projectors_projection_via_inner_products}
	\fP_{V^{\mathcal{G}}} \fv = \sum_{i=1}^{\dim V^{\mathcal{G}}} \langle \fv, \fe_i\rangle_V \, \fe_i,
\end{equation}
where $(\fe_1,\fe_2,\ldots,\fe_{\dim V})$ refers to an orthogonal basis of the space $V$ of a special form -- the first $\dim V^{\mathcal{G}}$ vectors span the subspace $V^{\mathcal{G}}$, and the remaining $\dim V - \dim V^{\mathcal{G}}$ vectors span its orthogonal complement. In particular, the expression \eqref{eq:SymmetryInformedStrats_projectors_projection_via_inner_products} is \emph{readily computable}. In contrast, the right-hand side \eqref{eq:SymmetryInformedStrats_projectors_projection_via_averaging} is less computable -- except for the case of finite groups -- but helps in establishing formal properties, as will become clear in the following.\\
For the article at hand, the first Hilbert space $V$ of interest is the space
\begin{equation}\label{eq:SymmetryInformedStrats_projectors_SymD}
	V = \Sym{d},
\end{equation}
equipped with the Frobenius inner product
\begin{equation}\label{eq:SymmetryInformedStrats_projectors_Frobenius}
	\langle \fA, \fB \rangle = \textrm{tr}(\fA \fB) \equiv \fA : \fB, \quad \fA,\fB \in \Sym{d}.
\end{equation}
The natural action \eqref{eq:theory_setup_group_action_definition}
\begin{equation}\label{eq:SymmetryInformedStrats_projectors_SymD_action}
	\mathds{L}_{\fQ}:\fA \equiv \fA^{\fQ} = \fQ \fA \fQ^T, \quad \fA \in \Sym{d}, \quad \fQ \in O(d),
\end{equation}
is orthogonal, as
\begin{equation}\label{eq:SymmetryInformedStrats_projectors_SymD_action_orthogonal}
	\langle \fA^{\fQ}, \fB^{\fQ} \rangle = \textrm{tr}(\fA^{\fQ} \fB^{\fQ}) = \textrm{tr}(\fQ\fA \fQ^T \fQ \fB \fQ^T) = \textrm{tr}(\fA \fB) = \langle \fA, \fB \rangle, \quad \fA,\fB \in \Sym{d},
\end{equation}
where we used the orthogonality of the matrix $\fQ$ and the fact that the trace vanishes on commutators, i.e., the identity
\begin{equation}\label{eq:SymmetryInformedStrats_projectors_trace_kills_commutators}
	\tr(\fM \fN) = \tr(\fN \fM), \quad \fM,\fN \in \R^{n \times n},
\end{equation}
is used with $\fM = \fQ\fA \fQ^T \fQ \fB$ and $\fN = \fQ^T$.\\
For the work at hand, the advantage of the averaging formulation \eqref{eq:SymmetryInformedStrats_projectors_projection_via_averaging} is its compatibility with eigenvalue bounds. More precisely, suppose that there are real constants $\alpha_\pm$, s.t. a tensor $\fA \in \Sym{d}$ satisfies the inequalities
\begin{equation}\label{eq:SymmetryInformedStrats_projectors_eigenvalue_bounds_before_projection}
	\alpha_- \, \|\fxi\|^2 \leq \fxi \cdot \fA \fxi \leq \alpha_+ \, \|\fxi\|^2
\end{equation}
for all vectors $\fxi \in \R^d$. Then, for any closed subgroup $\mathcal{G}$ of the orthogonal group $O(d)$, the projected tensor satisfies the same bounds
\begin{equation}\label{eq:SymmetryInformedStrats_projectors_eigenvalue_bounds_after_projection}
	\alpha_- \, \|\fxi\|^2 \leq \fxi \cdot \left( \mathds{P}^{\mathcal{G}}:\fA\right) \fxi \leq \alpha_+ \, \|\fxi\|^2,
\end{equation}
where we made use of the simplifying notation
\begin{equation}\label{eq:SymmetryInformedStrats_projectors_simplifying_notation}
	\mathds{P}^{\mathcal{G}} \equiv \mathds{P}_{\Sym{d}^{\mathcal{G}}}.
\end{equation}
For the argument to establish the bounds \eqref{eq:SymmetryInformedStrats_projectors_eigenvalue_bounds_after_projection}, we use the representation \eqref{eq:SymmetryInformedStrats_projectors_projection_via_averaging}
\begin{equation}\label{eq:SymmetryInformedStrats_projectors_projection_via_averaging_SymD}
	\mathds{P}^{\mathcal{G}}:\fA = \int_{\mathcal{G}} \mathds{L}_{\fQ}:\fA \, d\mu_{\mathcal{G}}(\fQ) \equiv \int_{\mathcal{G}} \fQ \fA \fQ^T \, d\mu_{\mathcal{G}}(\fQ).
\end{equation}
To derive the inequalities \eqref{eq:SymmetryInformedStrats_projectors_eigenvalue_bounds_after_projection}, we notice
\begin{equation}\label{eq:SymmetryInformedStrats_projectors_bounds_argument1}
	\fxi\cdot \left( \mathds{L}_{\fQ}:\fA \right) \fxi = \fxi\cdot \left( \fQ \fA \fQ^T\right) \fxi = \fQ^T\fxi \cdot \fA \fQ^T \fxi,
\end{equation}
valid for $\fxi \in \R^d$ and $\fQ \in O(3)$. Invoking the bounds \eqref{eq:SymmetryInformedStrats_projectors_eigenvalue_bounds_before_projection}, the two-sided inequality
\begin{equation}\label{eq:SymmetryInformedStrats_projectors_bounds_argument2}
	\alpha_- \, \| \fQ^T \fxi \|^2 \leq \fxi\cdot \left( \mathds{L}_{\fQ}:\fA \right) \fxi \leq \alpha_+ \, \| \fQ^T \fxi \|^2
\end{equation}
follows. As the element $\fQ \in O(3)$ is length-preserving -- together with its inverse $\fQ^T$ --  we obtain the estimate
\begin{equation}\label{eq:SymmetryInformedStrats_projectors_bounds_argument3}
	\alpha_- \, \| \fxi \|^2 \leq \fxi\cdot \left( \mathds{L}_{\fQ}:\fA \right) \fxi \leq \alpha_+ \, \| \fxi \|^2.
\end{equation}
Averaging the expression over the group $\mathcal{G}$ yields the claimed bounds \eqref{eq:SymmetryInformedStrats_projectors_eigenvalue_bounds_after_projection}.\\
The same form of arguments apply to tensors
\begin{equation}\label{eq:SymmetryInformedStrats_projectors_Qtensor}
	\mathds{Q} \in \Sym{\Sym{d}}
\end{equation}
which satisfy bounds
\begin{equation}\label{eq:SymmetryInformedStrats_projectors_Qtensor_bounds_before_projection}
	\alpha_- \, \|\fB\|^2 \leq \fB:\mathds{Q}:\fB \leq \alpha_+ \, \|\fB\|^2, \quad \fB \in \Sym{d},
\end{equation}
for real numbers $\alpha_{\pm}$, i.e., the estimates
\begin{equation}\label{eq:SymmetryInformedStrats_projectors_Qtensor_bounds_after_projection}
	\alpha_- \, \|\fB\|^2 \leq \fB:\left[\mathcal{P}^{\mathcal{G}}::\mathds{Q}\right]:\fB \leq \alpha_+ \, \|\fB\|^2, \quad \fB \in \Sym{d},
\end{equation}
follow for the orthogonal projector $\mathcal{P}^{\mathcal{G}}$ onto the closed subspace $\Sym{\Sym{d}}^{\mathcal{G}}$. For such tensors \eqref{eq:SymmetryInformedStrats_projectors_Qtensor}, we use Voigt's notation
\begin{equation}\label{eq:SymmetryInformedStrats_projectors_Qtensor_Voigt}
	\mathds{Q} \equalhat \underline{\underline{Q}} = \frac{1}{3}\left[
		\begin{array}{rrrrrr}
			Q_{11} & Q_{12} & Q_{13} & Q_{14} & Q_{15} & Q_{16}\\
			Q_{12} & Q_{22} & Q_{23} & Q_{24} & Q_{25} & Q_{26}\\
			Q_{13} & Q_{23} & Q_{33} & Q_{34} & Q_{35} & Q_{36}\\
			Q_{14} & Q_{24} & Q_{34} & Q_{44} & Q_{45} & Q_{46}\\
			Q_{15} & Q_{25} & Q_{35} & Q_{45} & Q_{55} & Q_{56}\\
			Q_{16} & Q_{26} & Q_{36} & Q_{46} & Q_{56} & Q_{66}\\
		\end{array}
	\right],
\end{equation}
and the inner product
\begin{equation}\label{eq:SymmetryInformedStrats_projectors_Qtensor_inner_product}
	\langle \mathds{Q}, \mathds{R} \rangle = \tr\left( \underline{\underline{S}}\,\underline{\underline{Q}}\,\underline{\underline{S}}\,\underline{\underline{R}}\right)
\end{equation}
with the scaling matrix
\begin{equation}\label{eq:SymmetryInformedStrats_projectors_Qtensor_inner_product_scaling}
	\underline{\underline{S}} = \textrm{diag}(1,1,1,2,2,2),
\end{equation}
which is invariant under the action \eqref{eq:theory_setup_action_on_muQ}.\\
We close this section by providing explicit expressions for the projectors associated to relevant ensemble-symmetry groups, see section \ref{sec:theory_examples}. Suppose a second-order tensor \eqref{eq:SymmetryInformedStrats_projectors_SymD}
\begin{equation}\label{eq:SymmetryInformedStrats_projectors_SymD_explicitProjectors_input}
	\fA = \mqty[ A_{11} & A_{12} & A_{13}\\
	A_{12} & A_{22} & A_{23}\\
	A_{13} & A_{23} & A_{33}]
\end{equation}
is given. Then, for isotropic symmetry \eqref{eq:special_isotropic_SO}, we obtain the identity
\begin{equation}\label{eq:SymmetryInformedStrats_projectors_SymD_explicitProjectors_iso}
	\mathds{P}^{\textrm{iso}}:\fA = \alpha \, \fI \quad \text{with} \quad \alpha = \frac{ A_{11} + A_{22} + A_{33}}{3},
\end{equation}
which coincides with cubic symmetry \eqref{eq:theory_unitCells_symmetryGroupCell_cubic}
\begin{equation}\label{eq:SymmetryInformedStrats_projectors_SymD_explicitProjectors_cubic}
	\mathds{P}^{\cub}:\fA = \mathds{P}^{\textrm{iso}}:\fA.
\end{equation}
For transverse isotropy \eqref{eq:special_trans_isotropic_symmetry_group_special}, the explicit expression
\begin{equation}\label{eq:SymmetryInformedStrats_projectors_SymD_explicitProjectors_trans_iso}
	\mathds{P}^{\textrm{ti}}:\fA = \mqty[\alpha_{\perp} & 0 & 0 \\
								0 & \alpha_{\perp} & 0 \\
								0 & 0 & \alpha_{\|}] \quad \text{with} \quad \alpha_{\perp} = \frac{ A_{11} + A_{22}}{2} \quad \text{and} \quad \alpha_{\|} = A_{33}
\end{equation}
follows, whereas the orthotropic group \eqref{eq:special_ortho_group} leads to the formula
\begin{equation}\label{eq:SymmetryInformedStrats_projectors_SymD_explicitProjectors_ortho}
	\mathds{P}^{\textrm{orth}}:\fA = \mqty[\alpha_1 & 0 & 0 \\
	0 & \alpha_2 & 0 \\
	0 & 0 & \alpha_3] \quad \text{with} \quad \alpha_i = A_{ii} \quad (i=1,2,3).
\end{equation}
For the \Qtensor{} in Voigt's notation \eqref{eq:SymmetryInformedStrats_projectors_Qtensor_Voigt}, the isotropic projection has the form~\cite{Gazis1963}
\begin{equation}\label{eq:SymmetryInformedStrats_projectors_Qtensor_projection_isotropic}
	\mathcal{P}^{\textrm{iso}}::\mathds{Q} = \left\langle \mathds{Q}, \mathds{E}_1^{\textrm{iso}} \right\rangle \, \mathds{E}_1^{\textrm{iso}} + \left\langle \mathds{Q}, \mathds{E}_2^{\textrm{iso}} \right\rangle \, \mathds{E}_2^{\textrm{iso}}
\end{equation}
with the orthonormal tensors
\begin{equation}\label{eq:SymmetryInformedStrats_projectors_Qtensor_projection_isotropic2}
	\mathds{E}_1^{\textrm{iso}} \equalhat \frac{1}{3}\left[
		\begin{array}{rrrrrr}
			1 &1 &1 &0 &0 &0\\
			1 &1 &1 &0 &0 &0\\
			1 &1 &1 &0 &0 &0\\
			0 &0 &0 &0 &0 &0\\
			0 &0 &0 &0 &0 &0\\
			0 &0 &0 &0 &0 &0\\
		\end{array}
	\right]
	\quad \text{and} \quad
	\mathds{E}_2^{\textrm{iso}} \equalhat \frac{1}{6\sqrt{5}}\left[
		\begin{array}{rrrrrr}
			4 &-2 &-2 &0 &0 &0\\
			-2 &4 &-2 &0 &0 &0\\
			-2 &-2 &4 &0 &0 &0\\
			0 &0 &0 &3 &0 &0\\
			0 &0 &0 &0 &3 &0\\
			0 &0 &0 &0 &0 &3\\
		\end{array}
	\right].
\end{equation}
For the cubic symmetry \eqref{eq:theory_unitCells_symmetryGroupCell_cubic}, the projector takes the form~\cite{Gazis1963}
\begin{equation}\label{eq:SymmetryInformedStrats_projectors_Qtensor_projection_cubic}
	\mathcal{P}^{\cub}::\mathds{Q} = \sum_{i=1}^3 \left\langle \mathds{Q}, \mathds{E}_i^{\cub} \right\rangle \, \mathds{E}_i^{\cub}
\end{equation}
with the two isotropic tensors \eqref{eq:SymmetryInformedStrats_projectors_Qtensor_projection_isotropic2}, $\mathds{E}_i^{\cub} = \mathds{E}_i^{\textrm{iso}}$ ($i=1,2$), and the third tensor
\begin{equation}\label{eq:SymmetryInformedStrats_projectors_Qtensor_projection_cubic2}
	\mathds{E}_3^{\cub} \equalhat \frac{1}{\sqrt{30}}\left[
		\begin{array}{rrrrrr}
			-2 &1 &1 &0 &0 &0\\
			1 &-2 &1 &0 &0 &0\\
			1 &1 &-2 &0 &0 &0\\
			0 &0 &0 &1 &0 &0\\
			0 &0 &0 &0 &1 &0\\
			0 &0 &0 &0 &0 &1\\
		\end{array}
	\right].
\end{equation}
In case of transverse isotropy w.r.t. the $\fe_3$-axis, in addition to the two isotropic tensors \eqref{eq:SymmetryInformedStrats_projectors_Qtensor_projection_isotropic2}, $\mathds{E}_i^{\textrm{ti}} = \mathds{E}_i^{\textrm{iso}}$ ($i=1,2$),
the three tensors
\begin{equation}\label{eq:SymmetryInformedStrats_projectors_Qtensor_projection_ti2}
	\mathds{E}^{\textrm{ti}}_3 \equalhat \frac{1}{6}\left[
		\begin{array}{rrrrrr}
			-2 &-2 &1 &0 &0 &0\\
			-2 &-2 &1 &0 &0 &0\\
			1 &1 &4 &0 &0 &0\\
			0 &0 &0 &0 &0 &0\\
			0 &0 &0 &0 &0 &0\\
			0 &0 &0 &0 &0 &0\\
		\end{array}
	\right], \quad
	\mathds{E}^{\textrm{ti}}_4 \equalhat \frac{6}{\sqrt{30}}\left[
		\begin{array}{rrrrrr}
			-8 &4 &4 &0 &0 &0\\
			4 &-8 &4 &0 &0 &0\\
			4 &4 &-8 &0 &0 &0\\
			0 &0 &0 &9 &0 &0\\
			0 &0 &0 &0 &9 &0\\
			0 &0 &0 &0 &0 &-6\\
		\end{array}
	\right]
\end{equation}
and
\begin{equation}\label{eq:SymmetryInformedStrats_projectors_Qtensor_projection_ti3}
	\mathds{E}^{\textrm{ti}}_5 \equalhat \frac{6}{\sqrt{6}}\left[
		\begin{array}{rrrrrr}
			1 &-5 &4 &0 &0 &0\\
			-5 &1 &4 &0 &0 &0\\
			4 &4 &-8 &0 &0 &0\\
			0 &0 &0 &0 &0 &0\\
			0 &0 &0 &0 &0 &0\\
			0 &0 &0 &0 &0 &3\\
		\end{array}
	\right]
\end{equation}
need to be considered to build the projector~\cite{Gazis1963}
\begin{equation}\label{eq:SymmetryInformedStrats_projectors_Qtensor_projection_ti}
	\mathcal{P}^{\textrm{ti}}::\mathds{Q} = \sum_{i=1}^5 \left\langle \mathds{Q}, \mathds{E}_i^{\textrm{ti}} \right\rangle \, \mathds{E}_i^{\textrm{ti}}.
\end{equation}
Last but not least, the projection onto the space of orthotropic tensors \eqref{eq:special_ortho_isotropic_Qtensor} takes the form
\begin{equation}\label{eq:SymmetryInformedStrats_projectors_Qtensor_projection_orth}
	\mathcal{P}^{\textrm{orth}}::\mathds{Q} \equalhat \left[
		\begin{array}{rrrrrr}
			Q_{11} & Q_{12} & Q_{13} & 0 & 0 & 0\\
			Q_{12} & Q_{22} & Q_{23} & 0 & 0 & 0\\
			Q_{13} & Q_{23} & Q_{33} & 0 & 0 & 0\\
			0 & 0 & 0 & Q_{44} & 0 & 0\\
			0 & 0 & 0 & 0 & Q_{55} & 0\\
			0 & 0 & 0 & 0 & 0 & Q_{66}\\
		\end{array}
	\right].
\end{equation}

\subsection{A-posteriori imposition of symmetries}
\label{sec:SymmetryInformedStrats_postprocessing}

In section \ref{sec:theory_symmetryProperties}, we learned that, for a symmetry group $\mathcal{G}$ of the ensemble, which is a closed subgroup of the orthogonal group $O(d)$, the effective conductivity $\fA^{\eff}$ of a $\mathcal{G}$-invariant ensemble is also $\mathcal{G}$-invariant, i.e., it resides in the space \eqref{eq:SymmetryInformedStrats_projectors_invariant_set}
\begin{equation}\label{eq:SymmetryInformedStrats_postprocessing_Aeff_invariant}
	\fA^{\eff} \in \Sym{d}^{\mathcal{G}}
\end{equation}
of $\mathcal{G}$-invariant symmetric tensors. However, this symmetry is only valid in the large-scale limit \eqref{eq:theory_unitCells_Aapp_convergence}. More precisely, neither the computed apparent property $\fA^{\app}_Y$ nor its expectation $\mean{\fA^{\app}_Y}$ are $\mathcal{G}$-invariant, in general. In the section at hand, we study whether we may infuse this knowledge of invariance into our computational results without doing harm, i.e., introducing bias. More precisely, we consider the projection
\begin{equation}\label{eq:SymmetryInformedStrats_postprocessing_Aapp_projected}
	\mathds{P}^{\mathcal{G}}: \fA^{\app}_Y = \int_{\mathcal{G}} \mathds{L}_{\fQ}: \fA^{\app}_Y \, d\mu_{\mathcal{G}}(\fQ)
\end{equation}
onto the set of $\mathcal{G}$-invariant tensors. This procedure is illustrated in Tab.~\ref{tab:ramification-empirical}, which considers the same general scenario as in Tab.~\ref{tab:ramification-ortho}, i.e., orthotropic fibers in a cubic cell. We consider three \emph{different} samples of the type of microstructure on the same cell.
\begin{longtable}[t]{ccc}
	\caption{Illustration of the projection \eqref{eq:SymmetryInformedStrats_postprocessing_Aapp_projected}}
	\label{tab:ramification-empirical}\\
	\toprule
	\endfirsthead
	\caption*{Demonstration of the projection of the empirical average (cont.)}\\
	\toprule
	\endhead 
	\endfoot 
	\bottomrule
	\endlastfoot
	
	  $\fA$   &  $\fA^{\app}_{Y} $  & $\mathds{P}^{\mathcal{G}}:\fA^{\app}_Y$\\ \noalign{\medskip}
	\midrule
	  \begin{minipage}{.15\textwidth}
		\includegraphics[width=1\linewidth, trim = 1in  0in  10.0in  2in ,clip]{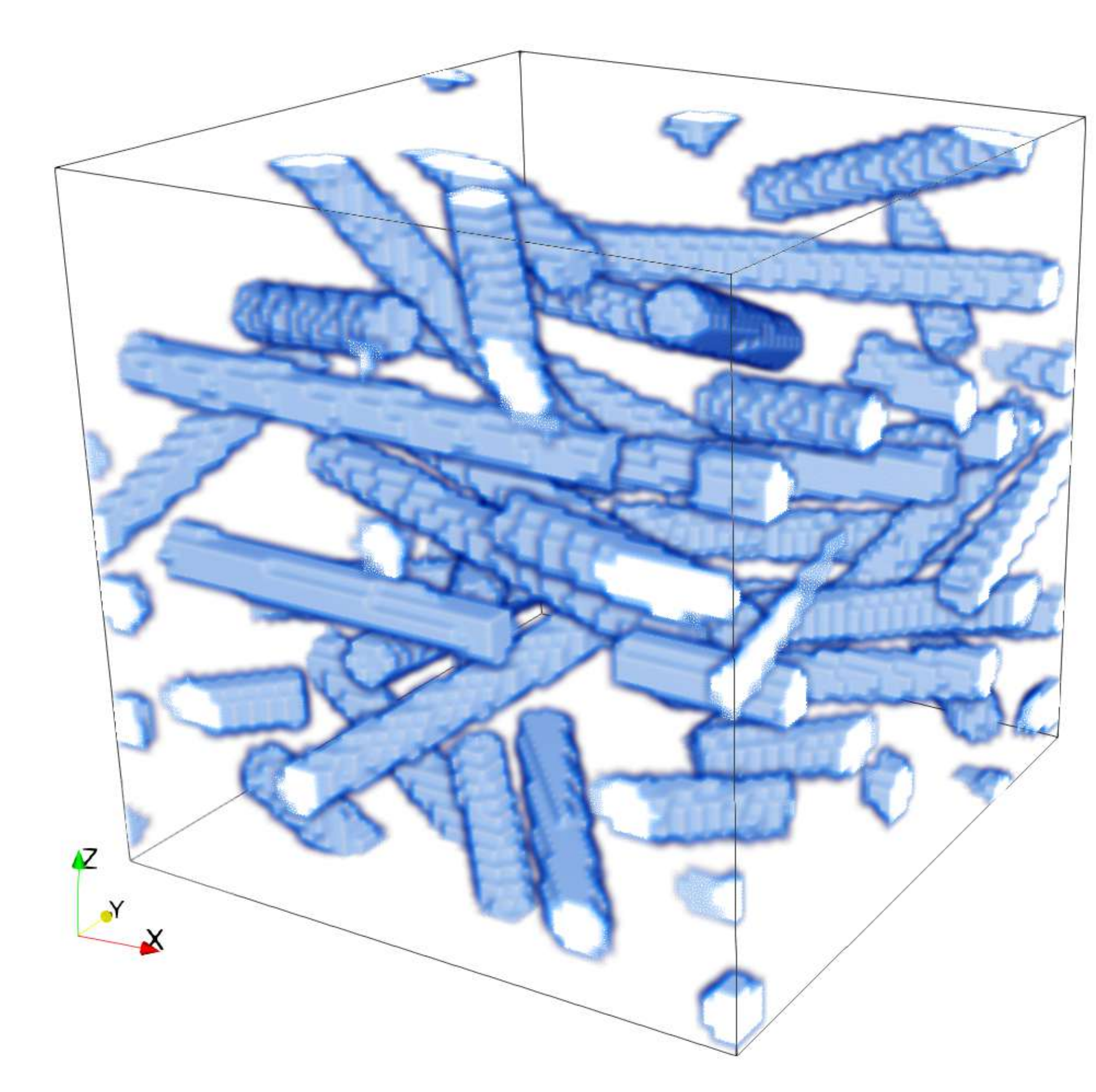}
	\end{minipage} & \resizebox{.305\textwidth}{!}{$\left[\begin{array}{rrr}0.3032 & 0.0003 & -0.0003 \\
		0.0003 & 0.2810  & 0.0003 \\
		-0.0003 & 0.0003 & 0.2668\end{array}\right]$}  & \resizebox{.25\textwidth}{!}{$\mqty[0.3032 & 0.0000 & 0.0000 \\
		0.0000 & 0.2810  & 0.0000 \\
		0.0000 & 0.0000 & 0.2668]$} \\ \noalign{\medskip}
	  \begin{minipage}{.15\textwidth}
		\includegraphics[width=1\linewidth, trim = 1in  0in  10.0in  2in ,clip]{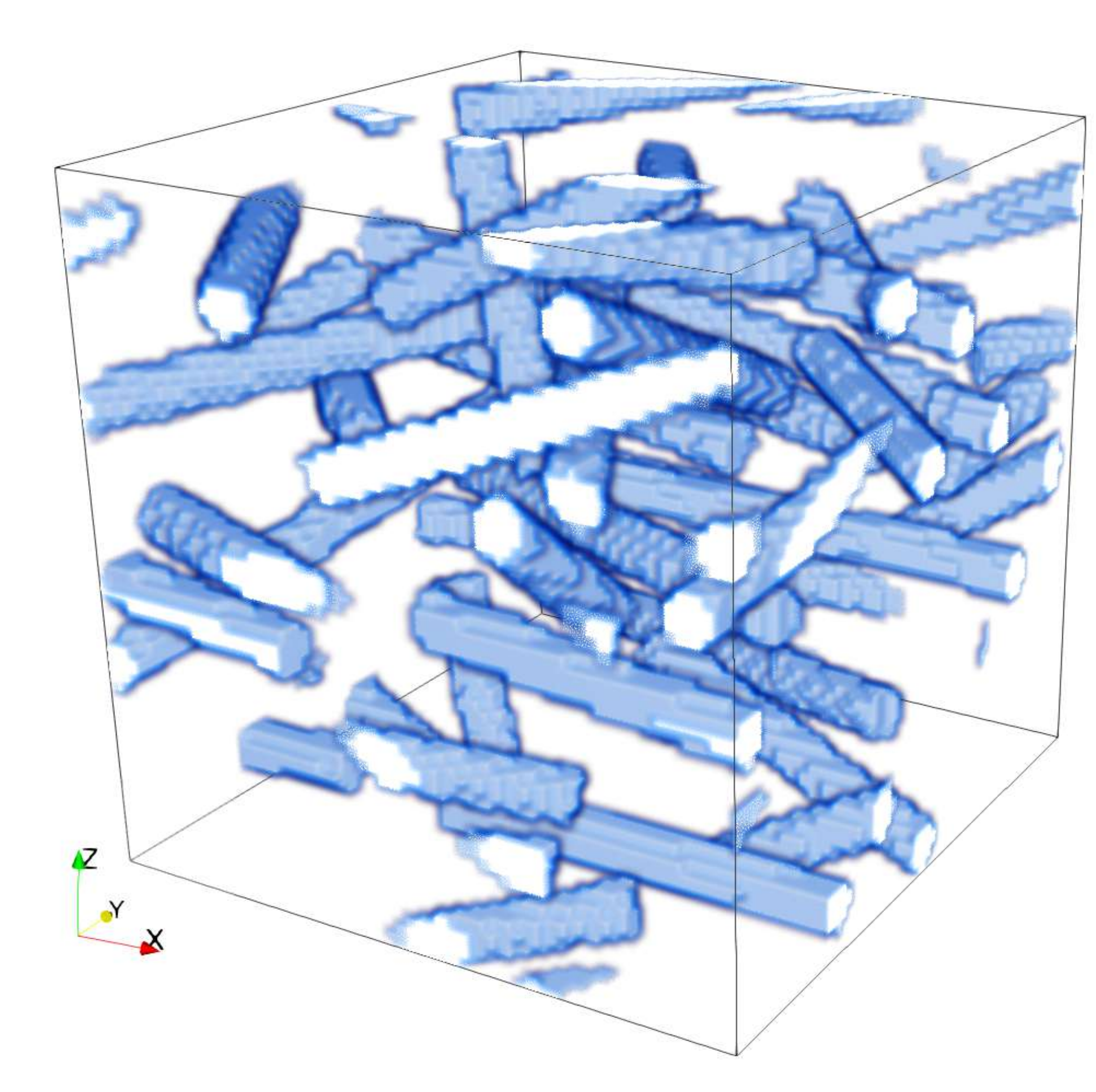}
	\end{minipage} & \resizebox{.305\textwidth}{!}{$\left[\begin{array}{rrr}0.3034 & -0.0001 & 0.0001 \\
		-0.0001 & 0.2809  & 0.0003 \\
		0.0001 & 0.0003 & 0.2665\end{array}\right]$} & \resizebox{.25\textwidth}{!}{$\mqty[0.3034 & 0.0000 & 0.0000 \\
		0.0000 & 0.2809  & 0.0000 \\
		0.0000 & 0.0000 & 0.2665]$} \\ \noalign{\medskip}
	  \begin{minipage}{.15\textwidth}
		\includegraphics[width=1\linewidth, trim =1in  0in  10.0in  2in ,clip]{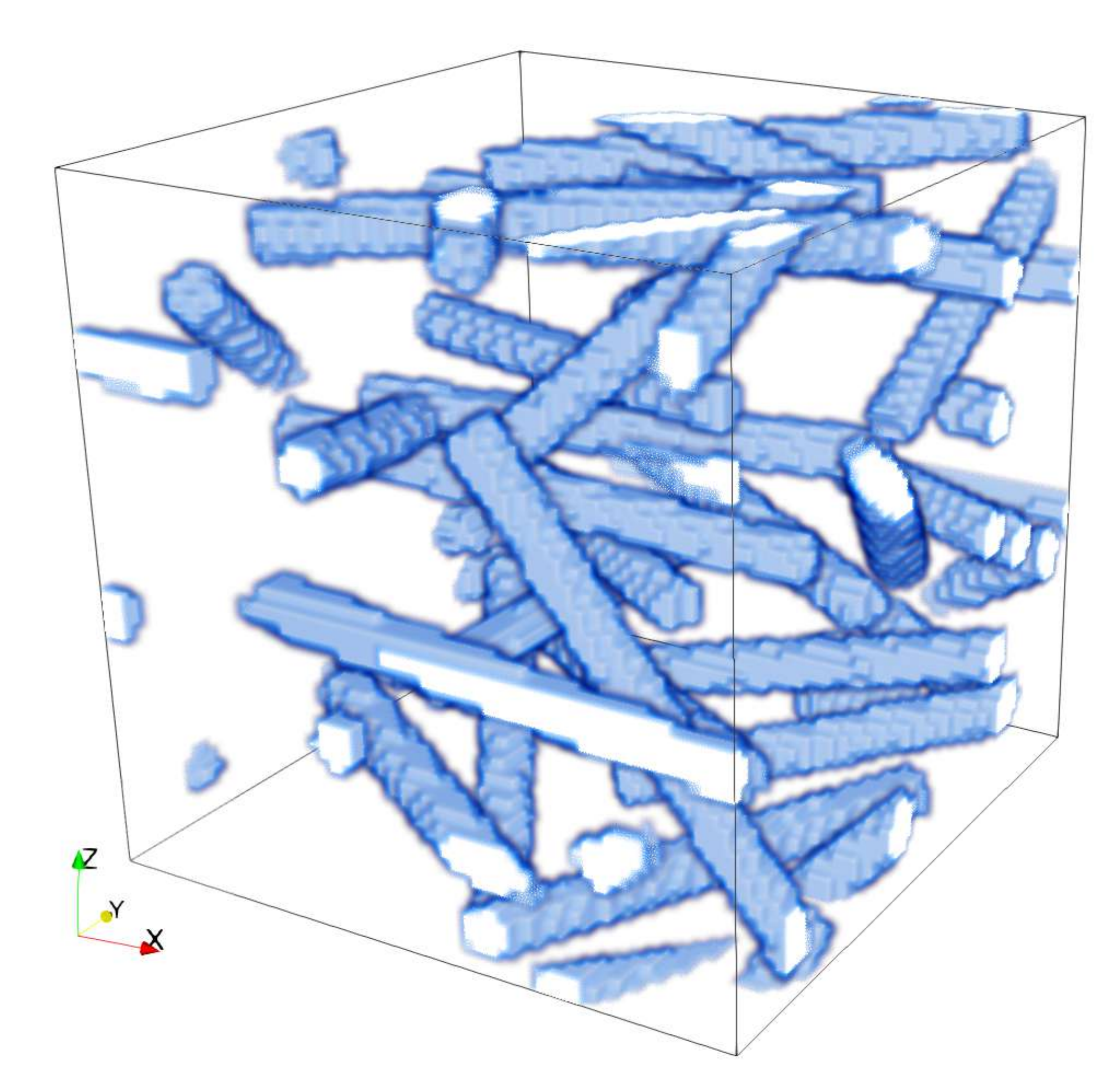}
	\end{minipage} & \resizebox{.305\textwidth}{!}{$\left[\begin{array}{rrr}0.3030 & -0.0001 & -0.0003 \\
		-0.0001 & 0.2809  & 0.0002 \\
		-0.0003 & 0.0002 & 0.2672\end{array}\right]$} & \resizebox{.25\textwidth}{!}{$\mqty[0.3030 & 0.0000 & 0.0000 \\
		0.0000 & 0.2809  & 0.0000 \\
		0.0000 & 0.0000 & 0.2672]$} \\ \noalign{\medskip}
	\midrule
 	Averaging  & \resizebox{.305\textwidth}{!}{$\left[\begin{array}{rrr}0.3032 & 0.0000 & -0.0002 \\
		0.0000 & 0.2809  & 0.0003\\
		-0.0002 & 0.0003 & 0.2668\end{array}\right]$} & \resizebox{.25\textwidth}{!}{$\mqty[0.3032 & 0.0000 & 0.0000 \\
		0.0000 & 0.2809  & 0.0000\\
		0.0000 & 0.0000 & 0.2668]$}\\ \noalign{\medskip}
%		\midrule
%	Projection  & \resizebox{.25\textwidth}{!}{$\mqty[0.3032 & 0.0000 & 0.0000 \\
%		0.0000 & 0.2809  & 0.0000\\
%		0.0000 & 0.0000 & 0.2668]$} & \resizebox{.25\textwidth}{!}{$\mqty[0.3032 & 0.0000 & 0.0000 \\
%		0.0000 & 0.2809  & 0.0000\\
%		0.0000 & 0.0000 & 0.2668]$}\\ 
		\noalign{\medskip}
\end{longtable}
For the projected apparent tensor \eqref{eq:SymmetryInformedStrats_postprocessing_Aapp_projected}, the following properties hold:
\begin{enumerate}
	\item If the ensemble is concentrated on the set \eqref{eq:theory_setup_space} of tensors with eigenvalues in the interval $[\alpha_-,\alpha_+]$, the projected tensor satisfies the bounds
	\begin{equation}\label{eq:SymmetryInformedStrats_postprocessing_bounds}
		\alpha_- \, \|\fxi\|^2 \leq \fxi \cdot \left( \mathds{P}^{\mathcal{G}}:\fA^{\app}_Y\right) \fxi \leq \alpha_+ \, \|\fxi\|^2, \quad \fxi \in \R^d.
	\end{equation}
	In particular, the projection \eqref{eq:SymmetryInformedStrats_postprocessing_Aapp_projected} produces thermodynamically reasonable conductivities.
	\item The projection \eqref{eq:SymmetryInformedStrats_postprocessing_Aapp_projected} does not increase the total error
	\begin{equation}\label{eq:SymmetryInformedStrats_postprocessing_totalError}
		\left\| \mathds{P}^{\mathcal{G}}: \fA^{\app}_Y - \fA^{\eff} \right\| \leq \left\| \fA^{\app}_Y - \fA^{\eff} \right\|
	\end{equation}
	for almost every realization $\fA \in \mathcal{A}$.
	\item The projection \eqref{eq:SymmetryInformedStrats_postprocessing_Aapp_projected} does not increase the dispersion
	\begin{equation}\label{eq:SymmetryInformedStrats_postprocessing_dispersion}
		\left\| \mathds{P}^{\mathcal{G}}: \fA^{\app}_Y - \mean{\mathds{P}^{\mathcal{G}}: \fA^{\app}_Y} \right\| \leq \left\| \fA^{\app}_Y - \mean{\fA^{\app}_Y} \right\|
	\end{equation}
	for almost every realization $\fA \in \mathcal{A}$. In particular, the random error \eqref{eq:theory_unitCells_random_error} does not increase.
	\item The projection \eqref{eq:SymmetryInformedStrats_postprocessing_Aapp_projected} does not increase the systematic error \eqref{eq:theory_unitCells_systematic_error}
	\begin{equation}\label{eq:SymmetryInformedStrats_postprocessing_systematicError}
		\left\| \mean{\mathds{P}^{\mathcal{G}}: \fA^{\app}_Y} - \fA^{\eff} \right\| \leq \left\| \mean{\fA^{\app}_Y} - \fA^{\eff} \right\|.
	\end{equation}
\end{enumerate}
The validity of these properties is easily confirmed by using the non-expansivity \eqref{eq:SymmetryInformedStrats_projectors_nonExpansive} of an orthogonal projector and the invariance \eqref{eq:SymmetryInformedStrats_postprocessing_Aeff_invariant} of the effective conductivity. To expose the line of arguments, we take a look at property \eqref{eq:SymmetryInformedStrats_postprocessing_totalError}. By consequence \eqref{eq:SymmetryInformedStrats_projectors_nonExpansive} of the Pythagorean Theorem \eqref{eq:SymmetryInformedStrats_projectors_Pythagoras} for orthogonal projectors, we observe
\begin{equation}\label{eq:SymmetryInformedStrats_postprocessing_sample_argument}
	\left\| \mathds{P}^{\mathcal{G}}: \left[\fA^{\app}_Y - \fA^{\eff}\right] \right\| \leq \left\| \fA^{\app}_Y - \fA^{\eff} \right\|.
\end{equation}
With little effort, it may be shown that the symmetry error
\begin{equation}\label{eq:SymmetryInformedStrats_postprocessing_symmetry_error_estimate}
	\texttt{err}^{\mathcal{G}}_{\fA} := \left\| \mathds{Q}^{\mathcal{G}}: \fA^{\app}_Y  \right\|
\end{equation}
provides a lower bound to the total error \eqref{eq:theory_unitCells_total_error} without knowing the reference conductivity \eqref{eq:theory_setup_Aeff}. Indeed, using the Pythagorean Theorem \eqref{eq:SymmetryInformedStrats_projectors_Pythagoras}, we may write
\begin{equation}\label{eq:SymmetryInformedStrats_postprocessing_symmetry_error_estimate_argument}
	\left\| \fA^{\app}_Y - \fA^{\eff} \right\|^2 = \left\| \mathds{P}^{\mathcal{G}}: \left[\fA^{\app}_Y - \fA^{\eff}\right] \right\|^2 + \left\| \mathds{Q}^{\mathcal{G}}: \left[\fA^{\app}_Y - \fA^{\eff}\right] \right\|^2,
\end{equation}
where $\mathds{Q}^{\mathcal{G}} = \mathds{I} - \mathds{P}^{\mathcal{G}}$ stands for the complementary projector. As the effective tensor $\fA^{\eff}$ is $\mathcal{G}$-invariant, it is an element of the kernel of the complementary projector. In particular, we observe the identity
\begin{equation}\label{eq:SymmetryInformedStrats_postprocessing_symmetry_error_estimate_argument2}
	\left\| \fA^{\app}_Y - \fA^{\eff} \right\|^2 = \left\| \mathds{P}^{\mathcal{G}} : \fA^{\app}_Y - \fA^{\eff} \right\|^2 + \left\| \mathds{Q}^{\mathcal{G}}: \fA^{\app}_Y  \right\|^2,
\end{equation}
and the estimate \eqref{eq:SymmetryInformedStrats_postprocessing_symmetry_error_estimate} emerges. Similarly, the symmetry error
\begin{equation}\label{eq:SymmetryInformedStrats_postprocessing_symmetry_error_estimate_sys}
	\texttt{err}^{\mathcal{G}}_{\textrm{sys}} := \left\| \mathds{Q}^{\mathcal{G}}: \mean{\fA^{\app}_Y}  \right\|
\end{equation}
provides a lower bound to the systematic error without knowing the reference.\\
Similar conclusions apply to the \Qtensor{} \eqref{eq:theory_unitCells_muQ}. For a start, the projection \eqref{eq:SymmetryInformedStrats_projectors_projection_via_averaging} preserves the eigenvalue bounds \eqref{eq:theory_unitCells_muQ_positive}
\begin{equation}\label{eq:SymmetryInformedStrats_postprocessing_Qtensor_bounds}
	0 \leq \fB : \left[ \mathcal{P}^{\mathcal{G}} :: \Q_{Y} \right] : \fB \leq \sup_{\|\fC\| \leq \|\fB\|} \fC : \Q_{Y} : \fC, \quad \fB \in \Sym{d},
\end{equation}
valid for the \Qtensor{} \eqref{eq:theory_unitCells_muQ}. Moreover, splitting the error \eqref{eq:theory_unitCells_muQ_convergence_quantitative}
\begin{equation}\label{eq:SymmetryInformedStrats_postprocessing_Qtensor_Pythagoras}
	\begin{split}
		\left\| \Q_{Y} - \Q \right\|^2 &= \left\| \mathcal{P}^{\mathcal{G}} :: \left[\Q_{Y} - \Q\right] \right\|^2 + \left\| \mathcal{Q}^{\mathcal{G}} :: \left[\Q_{Y} - \Q\right] \right\|^2\\
		&= \left\| \mathcal{P}^{\mathcal{G}} :: \Q_{Y} - \Q \right\|^2 + \left\| \mathcal{Q}^{\mathcal{G}} :: \Q_{Y}\right\|^2\\
	\end{split}
\end{equation}
via the Pythagorean Theorem \eqref{eq:SymmetryInformedStrats_projectors_Pythagoras} and using the invariance \eqref{eq:theory_setup_muQ_invariance} of the \Qtensor{}, we are led to the conclusion
\begin{equation}\label{eq:SymmetryInformedStrats_postprocessing_Qtensor_error_projected}
	\left\| \mathcal{P}^{\mathcal{G}} :: \left[\Q_{Y} - \Q\right] \right\| \leq \left\| \Q_{Y} - \Q \right\| 
\end{equation}
that the orthogonal projection does not increase the error \eqref{eq:theory_unitCells_muQ_convergence_quantitative}. Moreover, the $\mathcal{G}$-error provides a lower bound to the error \eqref{eq:theory_unitCells_muQ_convergence_quantitative}.
\begin{equation}\label{eq:SymmetryInformedStrats_postprocessing_Qtensor_error_symmetry}
	\left\| \mathcal{Q}^{\mathcal{G}} :: \Q_{Y}\right\| \leq \left\| \Q_{Y} - \Q \right\|.
\end{equation}

%\subsection{A-priori enforcement of symmetries}
%\label{sec:SymmetryInformedStrats_projectors_preprocessing}
%
%In the previous section, we introduced and studies a number of techniques for reducing the error of the RVE method applied to conducting composites with random microstructure. However, the approach did not directly reduce the cost of the computational homogenization procedure.\\
%The section at hand aims to reduce the effort of \todo{\ldots}
%\todo{
%\begin{itemize}
%	\item here, only a limited number of load cases is run, and the tensors are constructed based on these considerations
%	\item should retain both the random and the systematic error compared to a naive technique, or? -> think so
%	\item should be inferior to a projection technique, or? -> depends on the symmetry
%\end{itemize}
%}

%%%%%%%%%%%%%%%%%%%%%%%%%%%%%%%%%%%%%%%%%%%%%%%%%%%%%%%%%%%%%%%%%%%%%%%%%%%%%%%%%%%%%%%%%%%%%%%%%%%%%%%%%%%%%%%%%%%%%%%%%%%%%%%%%%%%%%%%%%%%%%%%%%%%%%%%%%%%%%%%%%%%%%%%%%%%%%%%%%%%%%%%%%%%

%\newpage

\section{Computational investigations}
\label{sec:computations}

\subsection{Setup}
\label{sec:computations_setup}
\begin{figure}[h!]
	\begin{subfigure}{.08\textwidth}
		\includegraphics[trim = 50 20 1950 920, clip, width=\textwidth]{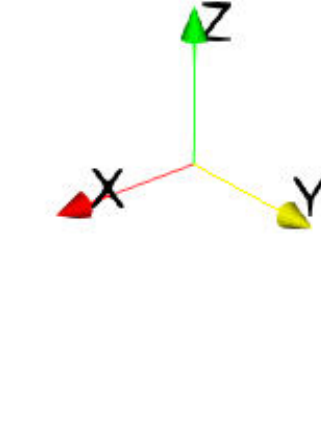}
	\end{subfigure}
	\begin{subfigure}{.4\textwidth}
		\includegraphics[trim = 580 20 580 140, clip, width=\textwidth]{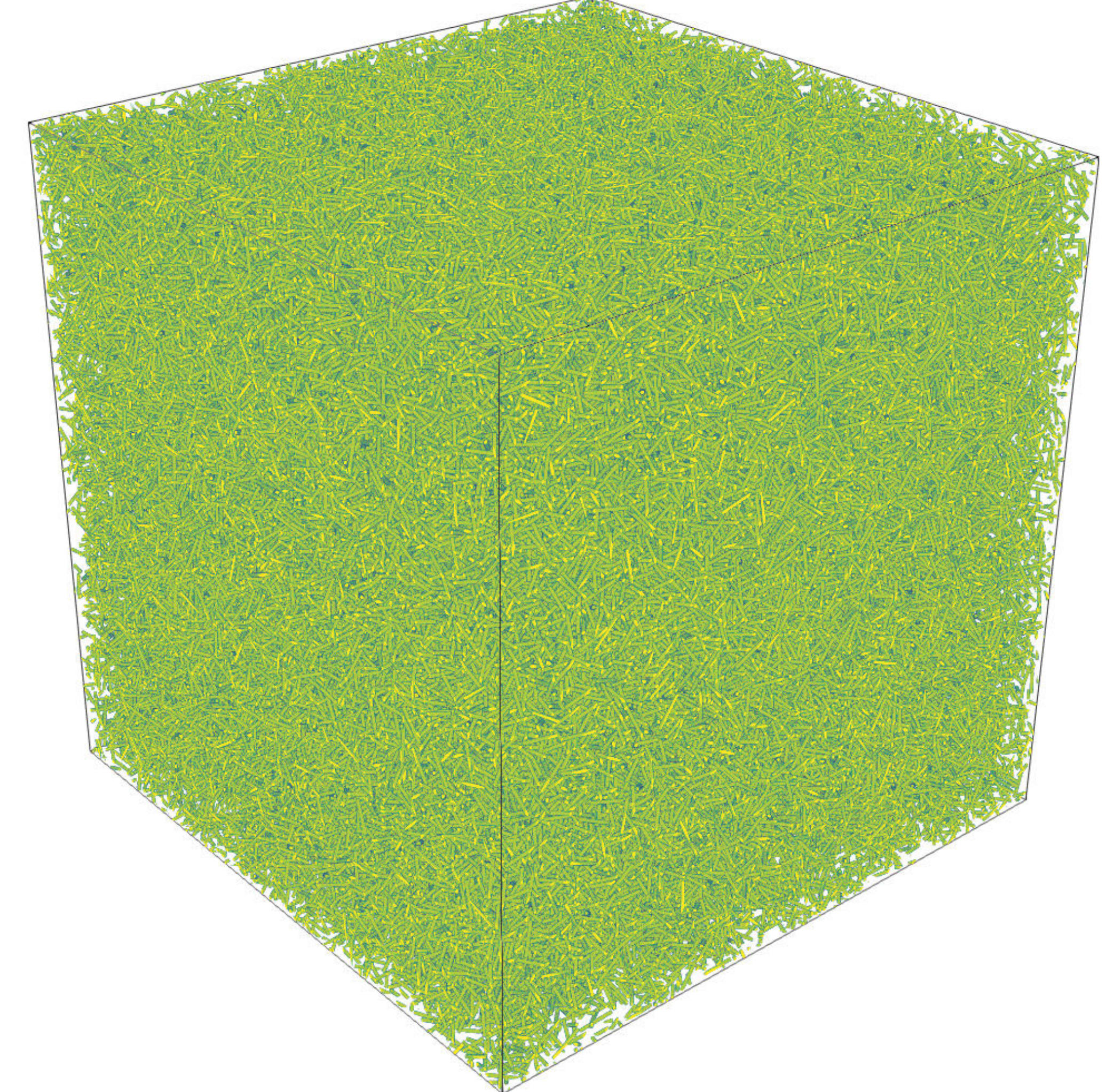}
		\caption{Isotropic ensemble}
		\label{fig:iso_1024}
	\end{subfigure}
	\begin{subfigure}{.4\textwidth}
		\includegraphics[trim = 580 20 580 140, clip, width=\textwidth]{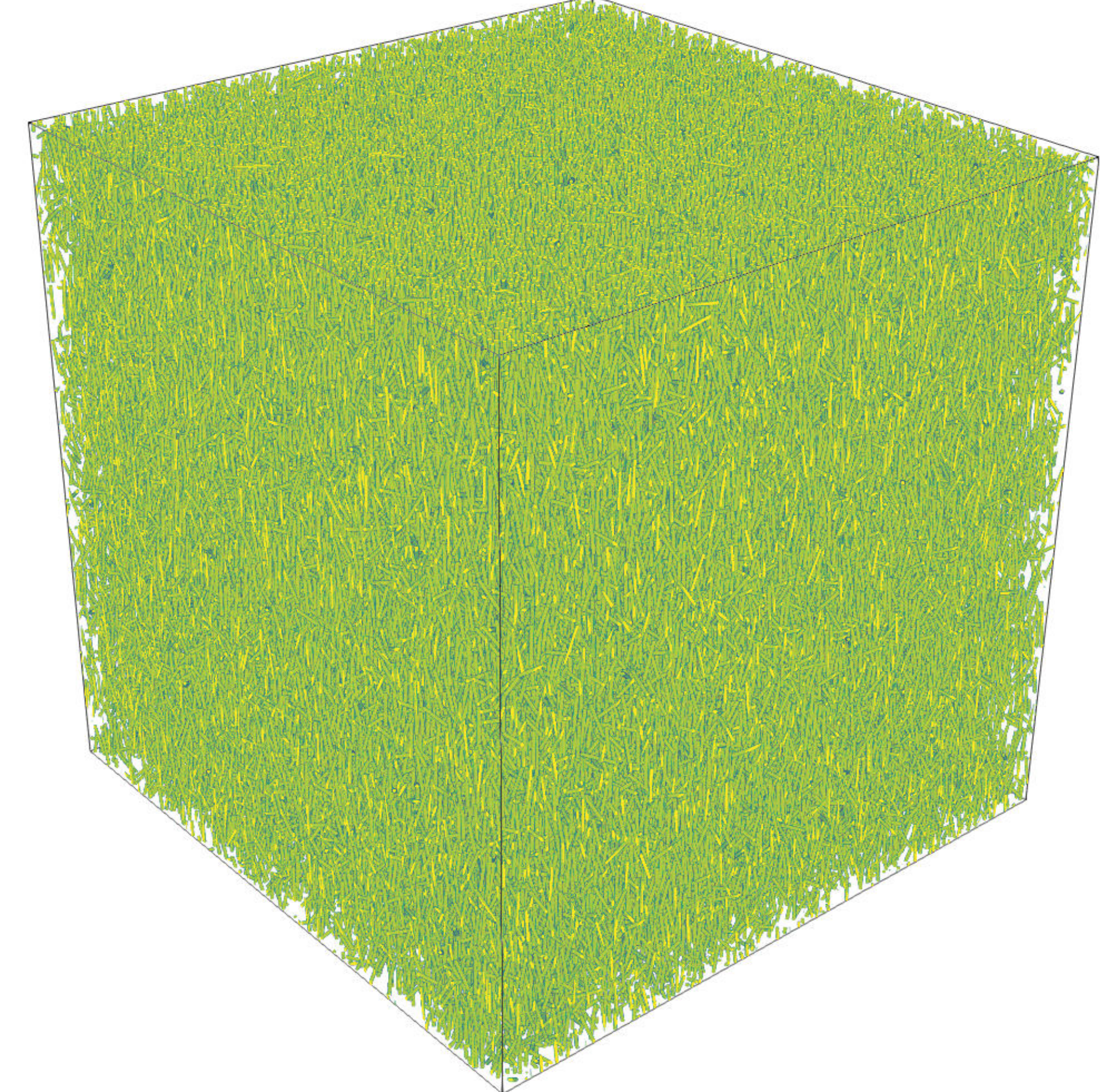}
		\caption{Transversely isotropic ensemble}
		\label{fig:tran_1024}
	\end{subfigure}	
	\centering
	\begin{subfigure}{.4\textwidth}
		\includegraphics[trim = 580 20 580 140, clip, width=\textwidth]{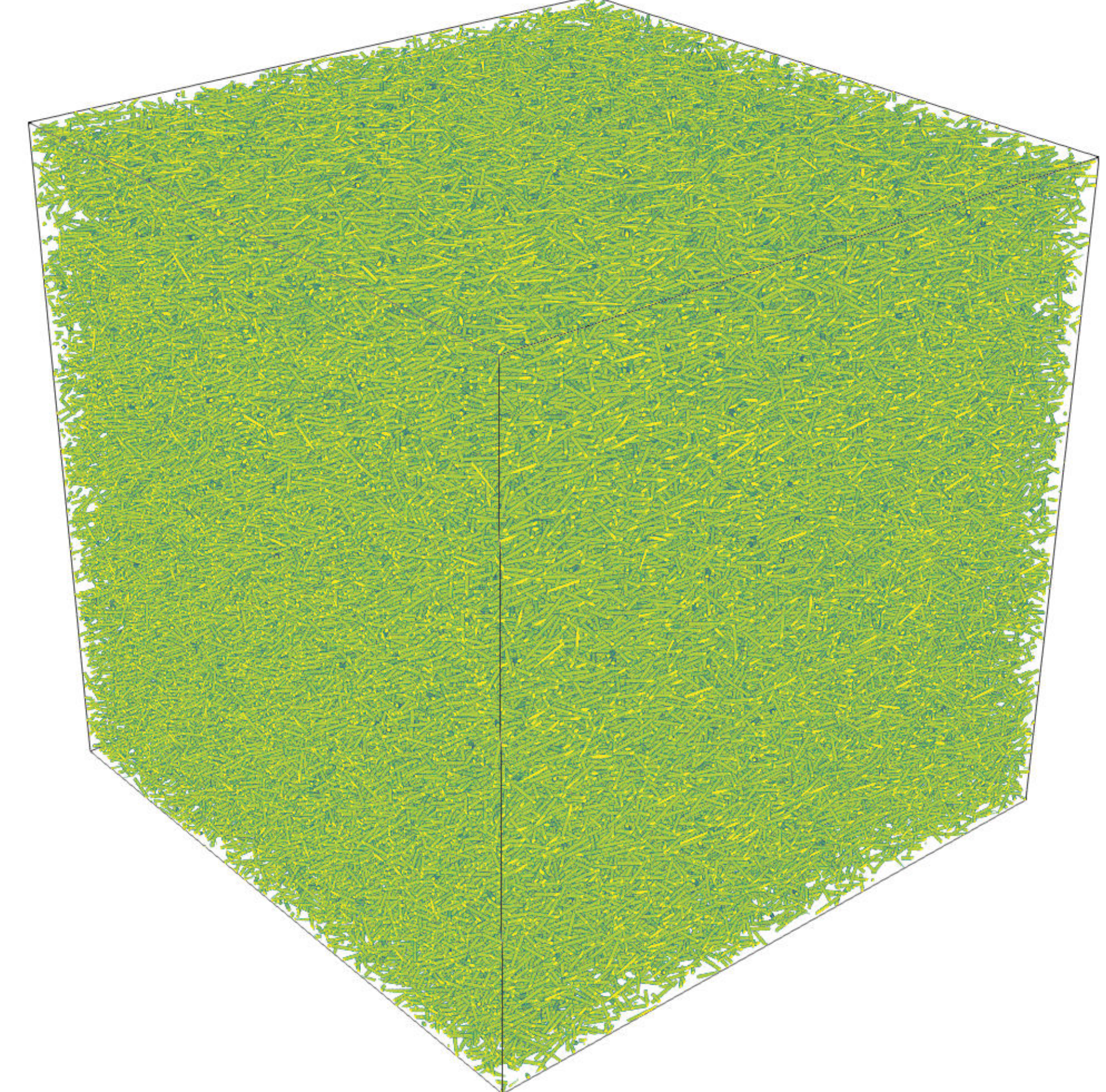}
		\caption{Orthotropic ensemble}
		\label{fig:ortho_1024}
	\end{subfigure}	
	\caption{Realizations of reference cells}
	\label{fig:periodic_cells_reference}
\end{figure}

In this section, we investigate the apparent conductivity \eqref{eq:theory_unitCells_Aapp} and the \Qtensor{} \eqref{eq:theory_unitCells_muQ} for various microstructure symmetries and volume element sizes via computational experiments. We employ a Monte-Carlo procedure to estimate the mean of both the apparent conductivity 
\begin{equation}
	\label{eq:setup-empirical-mean}
	\bar{\fA}^{\app,N}_Y = \frac{1}{N}\sum_{i=1}^{N} \fA^{\app,i}_Y
\end{equation}
and the \Qtensor{}
\begin{equation}
	\label{eq:setup-empirical-muQ-tensor}
	\Q^{N}_Y =  \frac{\text{vol}(Y)}{N-1}\sum_{i=1}^{N}\left( \fA^{\app,i}_Y - \bar{\fA}^{\app,N}_Y \right)^{\otimes 2}
\end{equation}
of fiber-reinforced composites via FFT-based computational homogenization~\cite{schneider2022representative}. Here, $N$ denotes the number of realizations, and $\fA^{\app,i}_Y$ refers to the conductivity of the $i^{th}$ (identically but independently distributed) realization. \review{The computational cells are sampled from periodized ensembles, and periodic boundary conditions are imposed for the temperature fluctuations, similar to the consideration in Schneider et al.~\cite{schneider2022representative}.} To estimate the accuracy of the empirical mean \eqref{eq:setup-empirical-mean}, we consider the unbiased estimation of the standard deviation
\begin{equation}
	\label{eq:setup-empirical-deviation}
	\vb{s}^{N}_Y = \qty(\frac{1}{N-1}\sum_{i=1}^{N} \qty(\fA^{\app,i}_Y - \bar{\fA}^{\app,N}_Y)\odot\qty(\fA^{\app,i}_Y - \bar{\fA}^{\app,N}_Y) )^{\odot 1/2},
\end{equation}
which indicates the component-wise confidence interval 
\begin{equation}
	\label{eq:set-confidence-interval}
	\qty[\bar{\fA}^{\app,N}_Y - \textbf{\texttt{e}}^N_Y, \bar{\fA}^{\app,N}_Y + \textbf{\texttt{e}}^N_Y ],
\end{equation}
with the component-wise error
\begin{equation}
	\label{eq:setup-confidence-interval-error}
	\textbf{\texttt{e}}^N_Y = t_{1-\frac{\alpha}{2}}^{N-1}\frac{\vb{s}^N_Y}{\sqrt{N}}.
\end{equation}
For the paper at hand, we consider the quantile $t_{1-\frac{\alpha}{2}}^{N-1}$ of student's t-distribution with $N-1$ degrees of freedom for the confidence level $\alpha=1\,\%$. Additionally, to evaluate the systematic error \eqref{eq:theory_unitCells_systematic_error}, we monitor the normalized error
\begin{equation}
	\label{eq:setup-monitor-systematic-error}
	\vb*{\epsilon}^{\sys} = \frac{ \qty( \qty(\bar{\fA}^{\app,N}_Y - \fA^{\textrm{ref}})\odot\qty(\bar{\fA}^{\app,N}_Y - \fA^{\textrm{ref}}) )^{\odot1/2} }{\fA^{\textrm{ref}}_{11}},
\end{equation}
where $\fA^{\textrm{ref}}_{11}$ is the 11-component of the reference conductivity matrix $\fA^{\textrm{ref}}$, computed on a rather large volume element of edge length $L=2048\,\mu m$ for $N^*=200$ realizations. For demonstration, Fig.~\ref{fig:periodic_cells_reference} shows three reference cells of the three ensembles that are investigated in the next sections.
Similarly, the random error \eqref{eq:theory_unitCells_random_error} is assessed by monitoring the normalized error
\begin{equation}
	\label{eq:setup-monitor-random-error}
	\vb*{\epsilon}^{\ran} = \frac{ \vb{s}^N_Y }{\fA^{\textrm{ref}}_{11}}.
\end{equation}
Moreover, the normalized confidence interval error \eqref{eq:setup-confidence-interval-error}
\begin{equation}
	\label{eq:setup-confidence-interval-error-normalized}
	\vb*{\epsilon}^{\textrm{ci}} = \frac{\textbf{\texttt{e}}^N_Y }{\fA^{\textrm{ref}}_{11}}
\end{equation}
is employed to justify the number of necessary realizations. On the other hand, the a posteriori symmetry informed projected apparent tensor \eqref{eq:SymmetryInformedStrats_postprocessing_Aapp_projected} results in the projected empirical mean 
\begin{equation}
	\label{eq:setup-empirical-mean-projected}
	\bar{\fA}^{\app,N,\mathds{P}}_Y = \frac{1}{N}\sum_{i=1}^{N} \qty(\mathds{P}^{\mathcal{G}}:\fA^{\app,i}_Y),
\end{equation}
and the corresponding normalized systematic, random and confidence interval errors
\begin{subequations}
	\begin{align}
		\label{eq:setup-monitor-systematic-error-projected}
%		\vb*{\epsilon}^{\sys,\mathds{P}} &= \frac{\abs {\bar{\fA}^{\app,N,\mathds{P}}_Y - \fA^{\textrm{ref}}} }{\fA^{\textrm{ref}}_{11}}, \\
		\vb*{\epsilon}^{\sys,\mathds{P}} &=\frac{ \qty( \qty(\bar{\fA}^{\app,N,\mathds{P}}_Y - \fA^{\textrm{ref}})\odot\qty(\bar{\fA}^{\app,N,\mathds{P}}_Y - \fA^{\textrm{ref}}) )^{\odot1/2} }{\fA^{\textrm{ref}}_{11}},\\
		\label{eq:setup-monitor-random-error-projected}
		\vb*{\epsilon}^{\ran,\mathds{P}} &= \frac{ \vb{s}^{N,\mathds{P}}_Y }{\fA^{\textrm{ref}}_{11}},
%		\label{eq:setup-confidence-interval-error-normalized-projected}
%		\vb*{\epsilon}^{\textrm{ci},\mathds{P}} &= \frac{\textbf{\texttt{e}}^{N,\mathds{P}}_Y }{\fA^{\textrm{ref}}_{11}},
	\end{align}
\end{subequations}
where the projected unbiased standard deviation is computed as
%\begin{subequations}
%	\begin{align}
%		\label{eq:setup-empirical-deviation-projected}
%		\vb{s}^{N,\mathds{P}}_Y &= \sqrt{\frac{1}{N-1}\sum_{i=1}^{N} \qty( \mathds{P}:\fA^{\app,i}_Y - \bar{\fA}^{\app,N,\mathds{P}}_Y)^2 },\\
%		\label{eq:setup-confidence-interval-error-projected}
%		\textbf{\texttt{e}}^{N,\mathds{P}}_Y &= t_{1-\frac{\alpha}{2}}^{N-1}\frac{\vb{s}^{N,\mathds{P}}_Y}{\sqrt{N}}
%	\end{align}
%\end{subequations}
\begin{equation}
	\label{eq:setup-empirical-deviation-projected}
	\vb{s}^{N,\mathds{P}}_Y = \qty(\frac{1}{N-1}\sum_{i=1}^{N} \qty( \mathds{P}:\fA^{\app,i}_Y - \bar{\fA}^{\app,N,\mathds{P}}_Y) \odot \qty( \mathds{P}:\fA^{\app,i}_Y - \bar{\fA}^{\app,N,\mathds{P}}_Y) )^{\odot 1/2}.
\end{equation}

The composites are assumed to be made of E-glass short fibers in a polypropylene matrix, whose isotropic thermal conductivity are $1.2$ W/(mK) and $0.2$ W/(mK) \review{\cite{weidenfeller2004thermal}}, respectively. The short fibers are of 100 $\mu m$ length, 10 $\mu m$ diameter, with the smallest separation distance of 2 $\mu m$ and volume fraction of 10$\%$. To realize the isotropic, transversely isotropic and orthotropic ensemble, we prescribe the second-order fiber orientation tensor $\textbf{\texttt{A}}$ and employ the angular central Gaussian closure of the fourth order fiber orientation tensor approximation~\cite{montgomery2011fast} as described in the publication~\cite{schneider2017sequential}. Following the similar framework of Schneider et al.~\cite{schneider2022representative}, we generate multiple microstructures with increasing edge lengths of $L=128\,\mu m$, $L=256\,\mu m$, $L=512\,\mu m$ by the sequential addition and migration algorithm~\cite{schneider2017sequential}. For each type of microstructure ensemble, we generate $N=15000$ realizations for the smallest volume element edge length $L=128\,\mu m$ and $N=10000$ realizations for the case $L=256\,\mu m$ and $L=512\,\mu m$. The resulting microstructures are resolved by voxels with an edge length of 2 $\mu m$, equivalent to the voxel sizes of $64^3$, $128^3$ and $256^3$, followed by the discretization on a rotated staggered grid~\cite{willot2015fourier} and solved by the linear conjugate gradient method~\cite{zeman2010accelerating}. In total, we performed 75600 simulations on a computing workstation with two AMD EPYC 9354 48 core processors, 1.12 TB of RAM and Ubuntu version 22.04.4. 
\subsection{An isotropic ensemble}
\label{sec:computation_isotropic_ensemble}
%\begin{figure}[h!]
%	\centering
%	\includegraphics[width=1\textwidth, trim = 0 0 0 0, clip]{figures/Spheres/Combined_2.eps}
%	\caption{Volume elements for spherical inclusions with diameter of $10\,\mu m$ and  $10\%$ volume fraction. Volume-element sizes from left to right: L=32\,$\mu m$, L=64\,$\mu m$, L=128\,$\mu m$, , L=256\,$\mu m$}
%	\label{fig:iso-volume-elements}
%\end{figure}
\begin{figure}[h!]
	\begin{subfigure}{.098\textwidth}
		\includegraphics[trim = 50 20 1850 920, clip, width=\textwidth]{data/Iso/iso_64.pdf}
	\end{subfigure}
	\begin{subfigure}{.151\textwidth}
		\includegraphics[trim = 600 20 480 125, clip, width=\textwidth]{data/Iso/iso_64.pdf}
		\caption{$64^3$}
		\label{fig:iso_64}
	\end{subfigure}
	\begin{subfigure}{.285\textwidth}
		\includegraphics[trim = 600 20 480 125, clip, width=\textwidth]{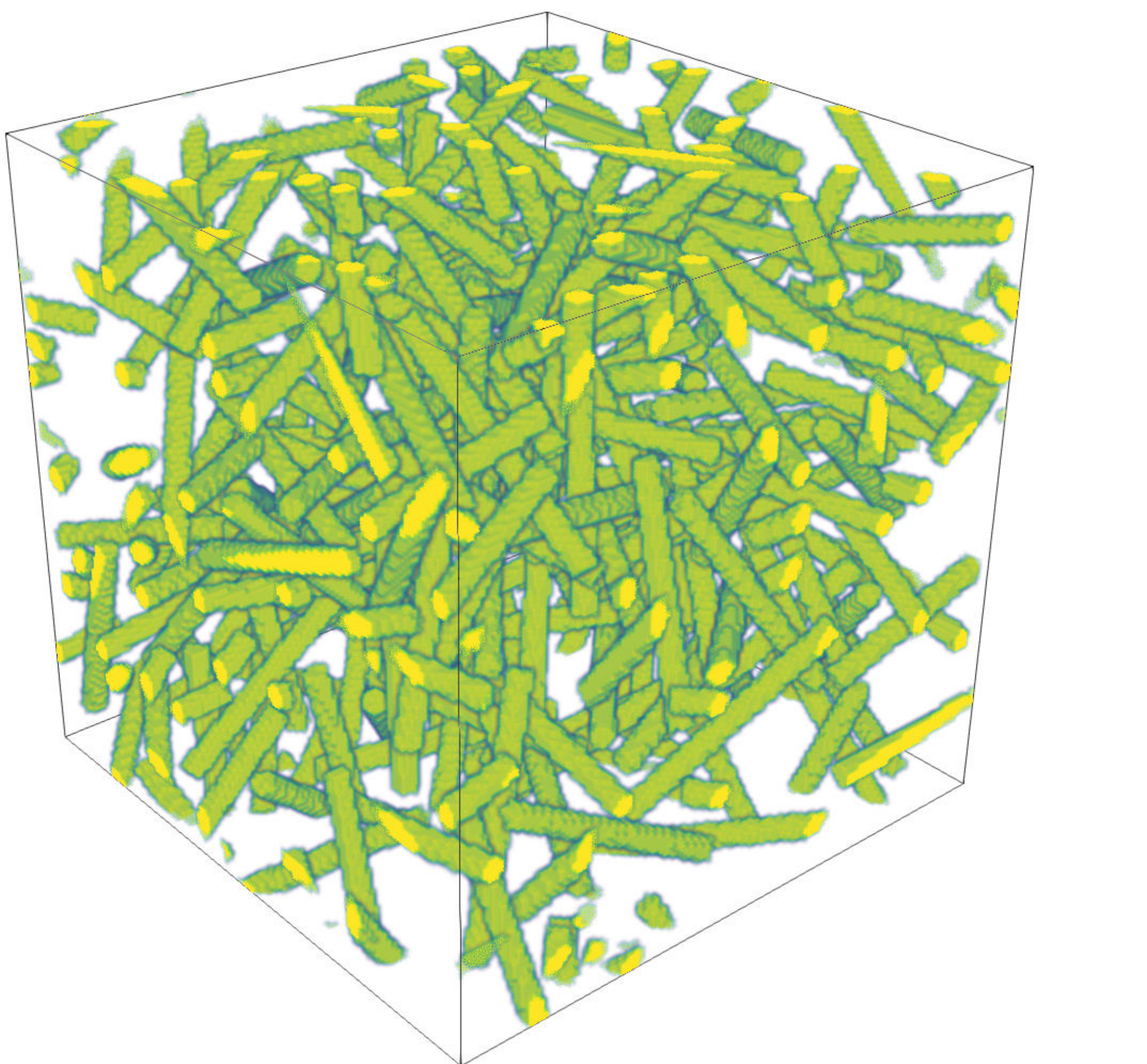}
		\caption{$128^3$}
		\label{fig:iso_128}
	\end{subfigure}
	\begin{subfigure}{.48\textwidth}
		\includegraphics[trim = 600 20 580 125, clip, width=\textwidth]{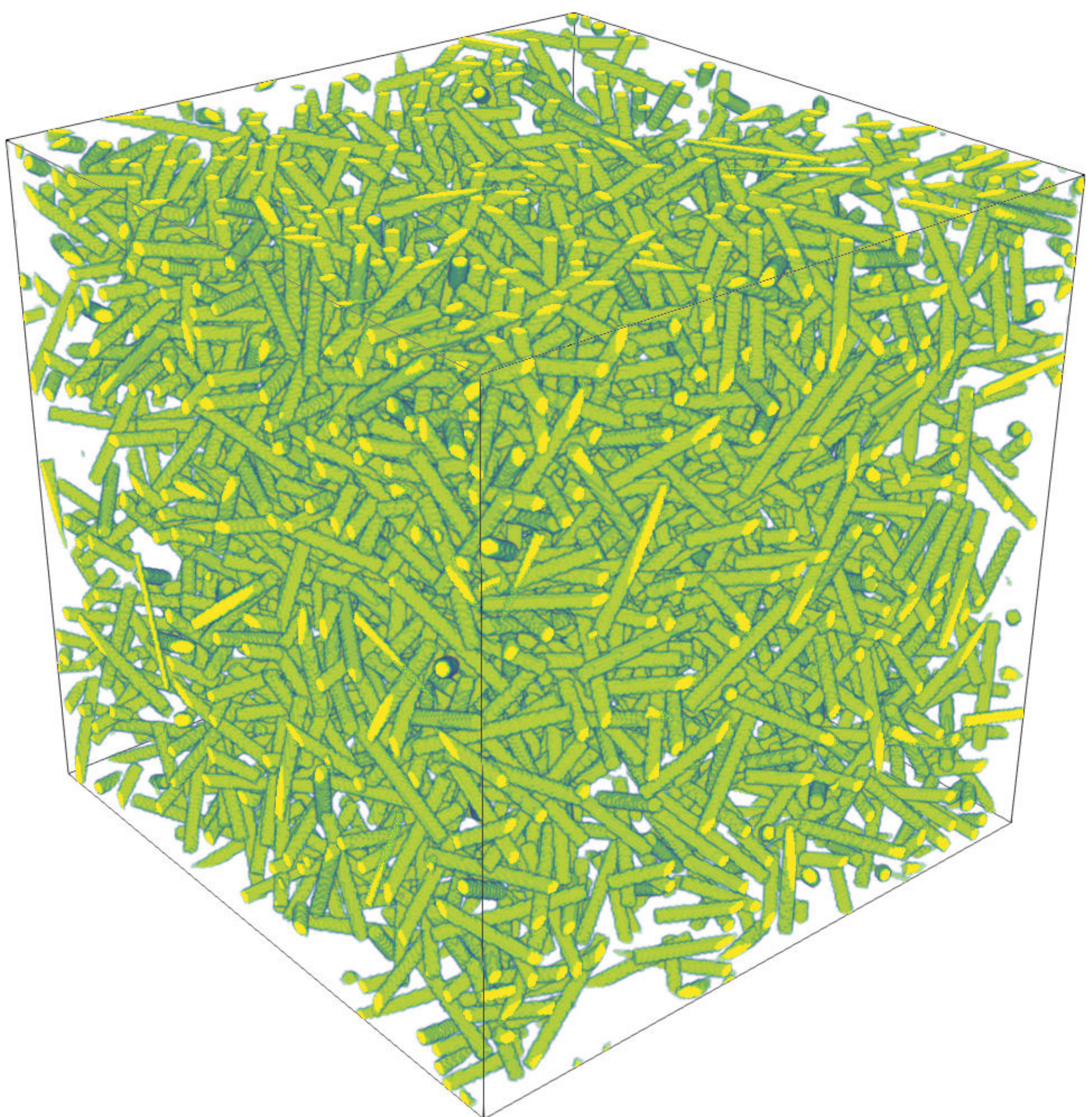}
		\caption{$256^3$}
		\label{fig:iso_256}
	\end{subfigure}
	\caption{Volume elements of the isotropically distributed fibers.}
	\label{fig:iso-volume-elements}
\end{figure}

In this section, we consider an isotropic ensemble by prescribing the second-order fiber orientation tensor 
\begin{equation}
	\label{eq:computation-isotropic-orientation-tensor}
	\textbf{\texttt{A}}^{\texttt{iso}} = \mqty[1/3 & 0 & 0 \\ 0 & 1/3 & 0 \\ 0 & 0 &1/3].
\end{equation}
The resulting
\begin{figure}[!h]
	\centering	
	\begin{minipage}[b]{0.3\textwidth}
		\centering
		\includegraphics[width=1\textwidth]{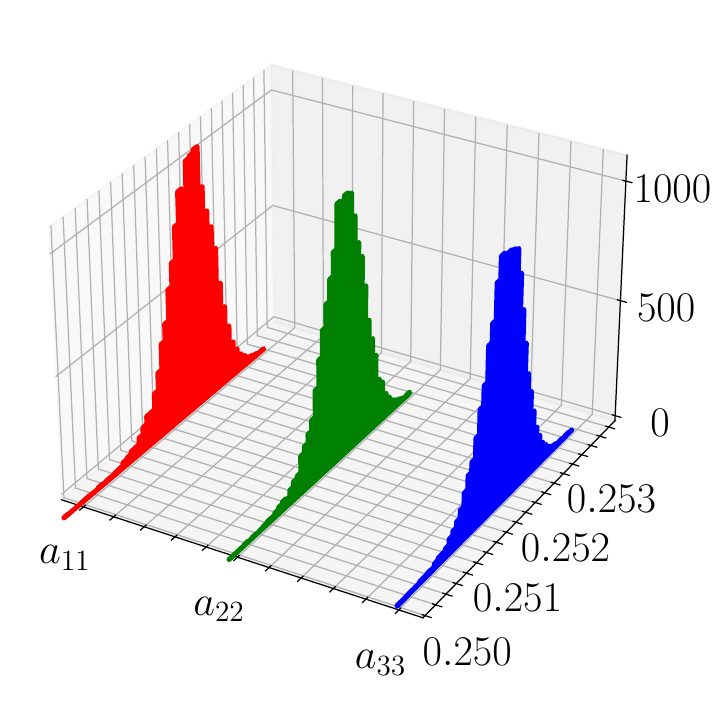}
		\subcaption{L=128\,$\mu m$}
		\label{fig:plot_histogram_diag_Iso_64}
	\end{minipage}	
	\hfill
	\begin{minipage}[b]{0.3\textwidth}
		\centering
		\includegraphics[width=1\textwidth]{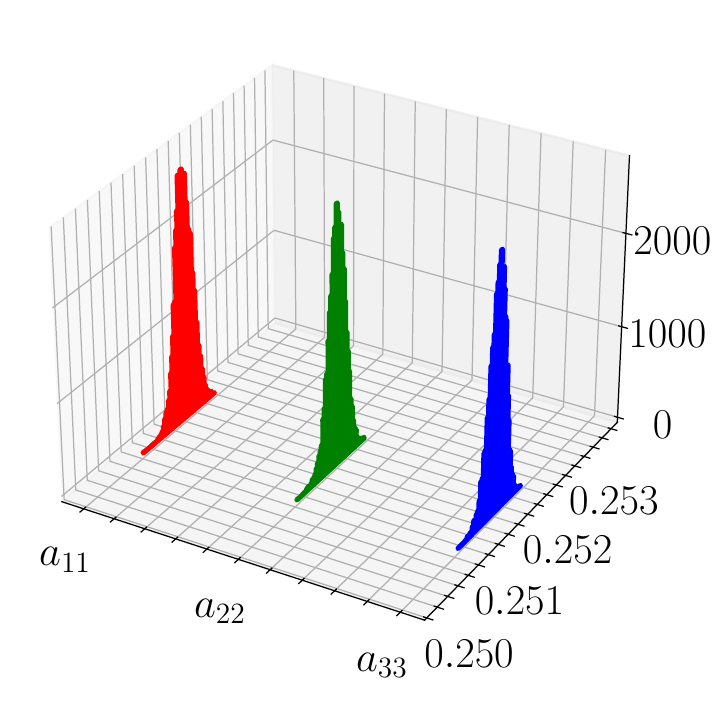}
		\subcaption{L=256\,$\mu m$}
		\label{fig:plot_histogram_diag_Iso_128}
	\end{minipage}		
	\hfill
	\begin{minipage}[b]{0.3\textwidth}
		\centering
		\includegraphics[width=1\textwidth]{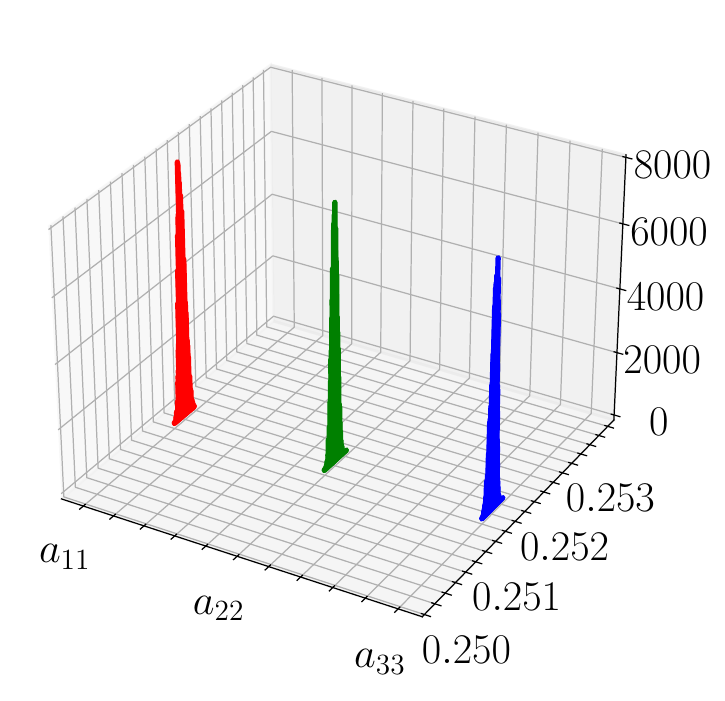}
		\subcaption{L=512\,$\mu m$}
		\label{fig:plot_histogram_diag_Iso_256}
	\end{minipage}		
	
	\begin{minipage}[b]{0.3\textwidth}
		\centering
		\includegraphics[width=1\textwidth]{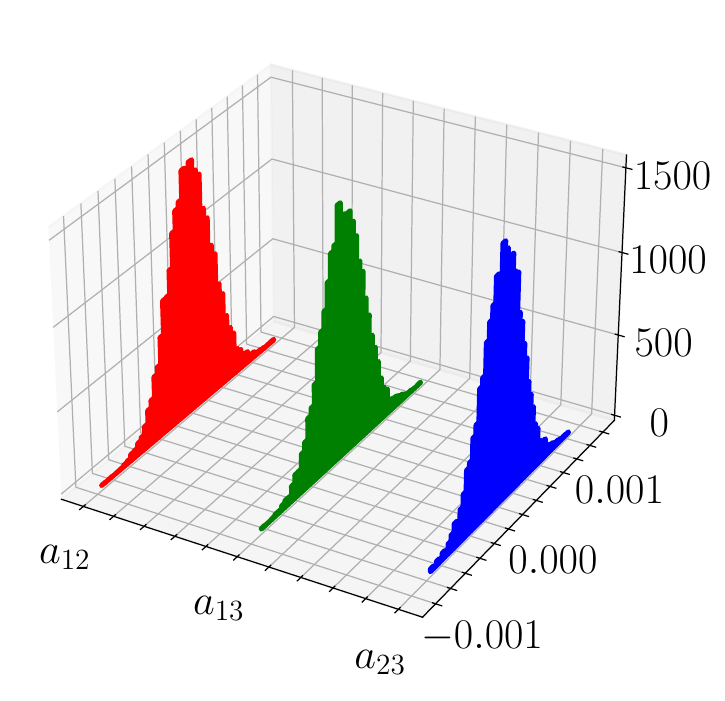}
		\subcaption{L=128\,$\mu m$}
		\label{fig:plot_histogram_offdiag_Iso_64}
	\end{minipage}	
	\hfill
	\begin{minipage}[b]{0.3\textwidth}
		\centering
		\includegraphics[width=1\textwidth]{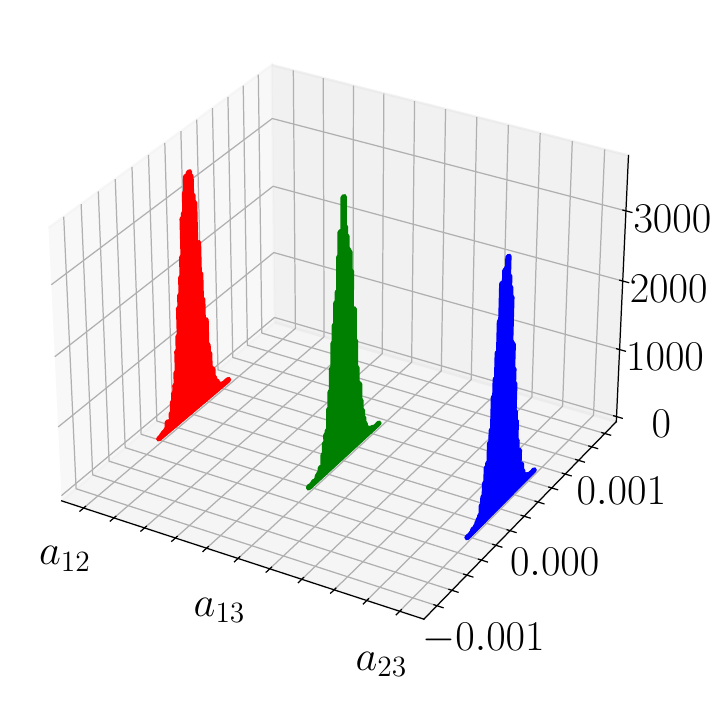}
		\subcaption{L=256\,$\mu m$}
		\label{fig:plot_histogram_offdiag_Iso_128}
	\end{minipage}	
	\hfill	
	\begin{minipage}[b]{0.3\textwidth}
		\centering
		\includegraphics[width=1\textwidth]{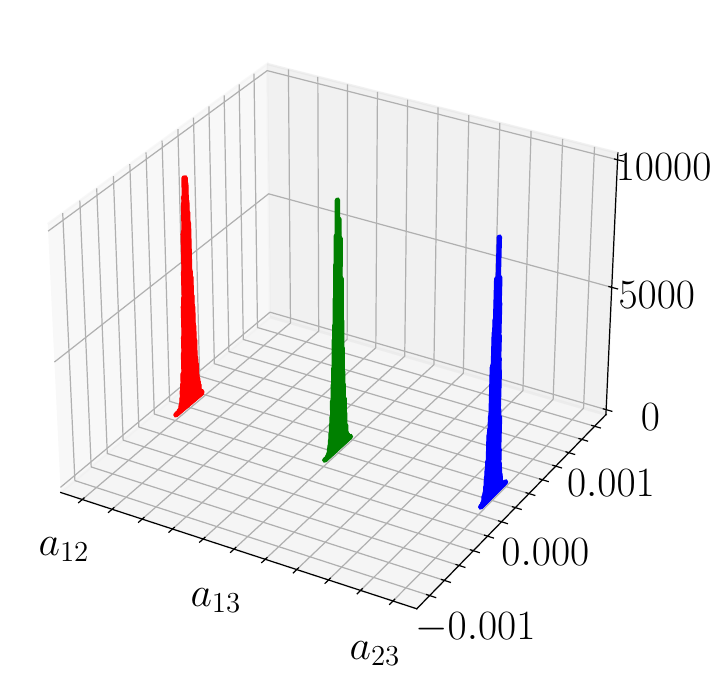}
		\subcaption{L=512\,$\mu m$}
		\label{fig:plot_histogram_offdiag_Iso_256}
	\end{minipage}			
	\caption{Frequency density of isotropic apparent thermal conductivity coefficients for increasing volume element sizes. (a-b-c) Diagonal coefficients $a_{11}$, $a_{22}$ and $a_{33}$; (d-e-f) Off-diagonal coefficients $a_{23}$, $a_{13}$ and $a_{12}$. The statistics of these coefficients are presented in Table \ref{tab:isotropic-ensemble}.}
	\label{fig:iso-histogram-apparent}
\end{figure}
microstructure images are shown in Fig.~\ref{fig:iso-volume-elements} for the sizes of $64^3$, $128^3$ and $256^3$, while their corresponding apparent conductivity coefficients are plotted in Fig.~\ref{fig:iso-histogram-apparent}. For the considered microstructures, the computed apparent conductivity tensor appears to be rather close to an isotropic tensor as demonstrated by the histogram of their diagonal and off-diagonal coefficients in Fig.~\ref{fig:iso-histogram-apparent}. All three diagonal coefficients exhibit similar distributions of the mean value around 0.252 $W/(mK)$ and variances of $4\times10^{-4} \,W^2/(mK)^2$, $1.5\times10^{-4}\,W^2/(mK)^2$ and $0.5\times10^{-4}\,W^2/(mK)^2$ for edge length $L=128\mu m$, $L=256\mu m$ and $L=512\mu m$, respectively, whereas the off-diagonal coefficients all distribute around zero. The detailed results are given in Table \ref{tab:isotropic-ensemble}. We make the following observations.
\begin{table}[h!]
	\caption{Asymptotic behavior of isotropic apparent conductivity. The units of $L$, mean values and standard deviation are $\mu m$, $W/(mK)$ and $W/(mK)$, respectively.}
	\label{tab:isotropic-ensemble}
	\centering
	\begin{tabular}{ll|rl|rrrrl}
		L & A & Mean & Std. & Sys. error & Proj.  & Rand. error & Proj. & CI error \\
		& &  &  &  & sys. error &  &  rand. error & \\
		\hline\hline
		\multirow{6}{*}{128}& $a_{11}$ & $0.25175$ & $0.00040$ & 0.06458\% & 0.06421\% & 0.15819\% & 0.04972\% & 0.00333\%  \\
		& $a_{22}$ & $0.25175$ & $0.00040$ & 0.06372\% & 0.06421\% & 0.15782\% & 0.04972\% & 0.00332\%  \\
		& $a_{33}$ & $0.25175$ & $0.00040$ & 0.06435\% & 0.06421\% & 0.15979\% & 0.04972\% & 0.00336\%  \\
		& $a_{12}$ & $0.00000$ & $0.00028$ & 0.00142\% & -- & 0.10934\% & -- & 0.00230\%  \\
		& $a_{13}$ & $0.00001$ & $0.00028$ & 0.00242\% & -- & 0.10986\% & -- & 0.00231\%  \\
		& $a_{23}$ & $0.00000$ & $0.00027$ & 0.00035\% & -- & 0.10857\% & -- & 0.00228\%  \\
		\hline
		\multirow{6}{*}{256}& $a_{11}$ & $0.25161$ & $0.00015$ & 0.00866\% & 0.00765\% & 0.06156\% & 0.02125\% & 0.00159\%  \\
		& $a_{22}$ & $0.25160$ & $0.00016$ & 0.00663\% & 0.00765\% & 0.06194\% & 0.02125\% & 0.00160\%  \\
		& $a_{33}$ & $0.25161$ & $0.00016$ & 0.00767\% & 0.00765\% & 0.06199\% & 0.02125\% & 0.00160\%  \\
		& $a_{12}$ & $0.00000$ & $0.00012$ & 0.00095\% & -- & 0.04711\% & -- & 0.00121\%  \\
		& $a_{13}$ & $0.00000$ & $0.00012$ & 0.00014\% & -- & 0.04769\% & -- & 0.00123\%  \\
		& $a_{23}$ & $0.00000$ & $0.00012$ & 0.00006\% & -- & 0.04792\% & -- & 0.00123\%  \\
		\hline
		\multirow{6}{*}{512}& $a_{11}$ & $0.25159$ & $0.00005$ & 0.00116\% & 0.00128\% & 0.02163\% & 0.00769\% & 0.00056\%  \\
		& $a_{22}$ & $0.25159$ & $0.00005$ & 0.00148\% & 0.00128\% & 0.02135\% & 0.00769\% & 0.00055\%  \\
		& $a_{33}$ & $0.25159$ & $0.00005$ & 0.00119\% & 0.00128\% & 0.02182\% & 0.00769\% & 0.00056\%  \\
		& $a_{12}$ & $0.00000$ & $0.00004$ & 0.00005\% & -- & 0.01763\% & -- & 0.00045\%  \\
		& $a_{13}$ & $0.00000$ & $0.00004$ & 0.00004\% & -- & 0.01743\% & -- & 0.00045\%  \\
		& $a_{23}$ & $0.00000$ & $0.00004$ & 0.00008\% & -- & 0.01760\% & -- & 0.00045\%  \\
		\hline
		\multirow{6}{*}{2048}& $a_{11}$ & $0.25159$ & $0.00001$ & -- & -- & 0.00254\% & 0.00100\% & 0.00047\%  \\
		& $a_{22}$ & $0.25159$ & $0.00001$ & -- & -- & 0.00235\% & 0.00100\% & 0.00043\%  \\
		& $a_{33}$ & $0.25159$ & $0.00001$ & -- & -- & 0.00271\% & 0.00100\% & 0.00050\%  \\
		& $a_{12}$ & $0.00000$ & $0.00001$ & -- & -- & 0.00232\% & -- & 0.00043\%  \\
		& $a_{13}$ & $0.00000$ & $0.00001$ & -- & -- & 0.00208\% & -- & 0.00038\%  \\
		& $a_{23}$ & $0.00000$ & $0.00001$ & -- & -- & 0.00217\% & -- & 0.00040\%  \\
		\hline
	\end{tabular}
\end{table}

\begin{figure}[!h]
	\begin{center}
		\begin{subfigure}{\textwidth}
			\centering
			\includegraphics[height=.022\textheight]{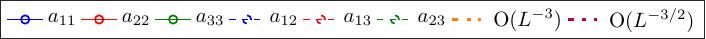}
		\end{subfigure}
	\end{center}
	\centering
	\begin{minipage}[b]{0.3\textwidth}
		\includegraphics[width=1\textwidth]{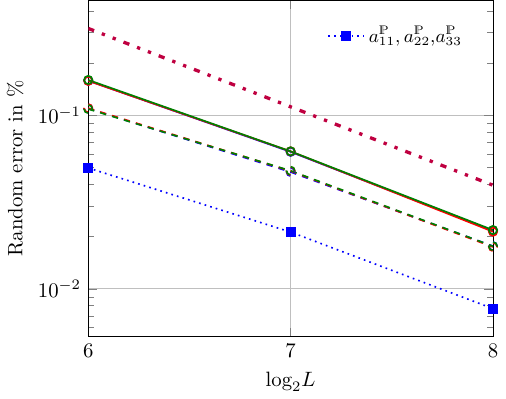}
		\subcaption{Random errors}
		\label{fig:Iso_Fiber_plot_RandError_convergence}
	\end{minipage}
	\hfill
	\begin{minipage}[b]{0.3\textwidth}
		\includegraphics[width=1\textwidth]{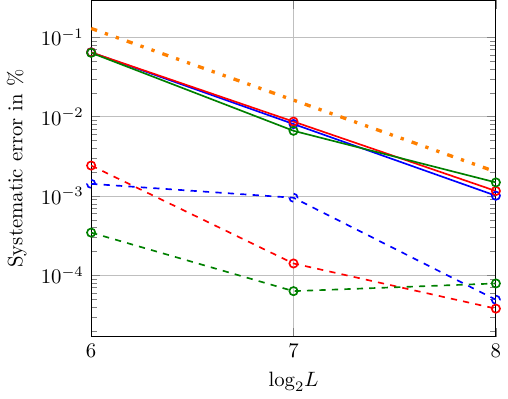}
		\subcaption{Systematic errors}
		\label{fig:Iso_Fiber_plot_SysError_convergence}
	\end{minipage}
	\hfill
	\begin{minipage}[b]{0.3\textwidth}
		\includegraphics[width=1\textwidth]{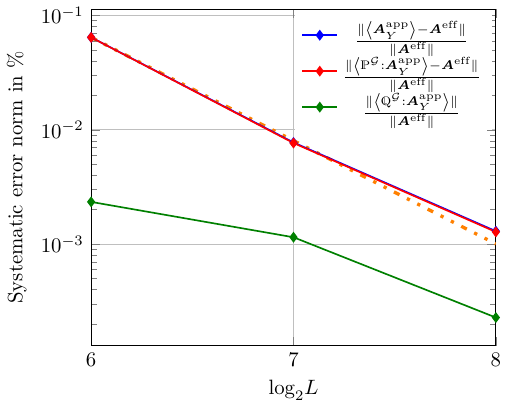}
		\subcaption{Norm of systematic errors}
		\label{fig:Iso_Fiber_plot_normSysError_convergence}
	\end{minipage}	
	\caption{Convergence of: (a) Random errors; (b) Systematic errors; (c) Frobenius norm of systematic errors. $a^{\mathds{P}}_{11},a^{\mathds{P}}_{22},a^{\mathds{P}}_{33}$ denote the diagonal components of the projected empirical mean \eqref{eq:setup-empirical-mean-projected}.}
	\label{fig:Iso_Fiber_sys_rand_errors}		
\end{figure}

We notice the confidence interval error is on the order of $10^{-3}\,\%$, indicating the number of realizations used in the computation is exceedingly sufficient. The systematic errors are rather small. For the smallest volume element size of $64^3$, the systematic error of the diagonal and off-diagonal coefficients are about $0.6\%$ and $0.001\%$, respectively, which are consistently smaller than their random error counterparts by a factor of 2 and 10, respectively. In terms of scaling, we observe the expected decay rate of $L^{-3/2}$ for the random errors as shown in Fig.~\ref{fig:Iso_Fiber_plot_RandError_convergence}. Meanwhile, in contrast to the related study~\cite{schneider2022representative}, we observe the optimal scaling at the rate $L^{-3}$ for the diagonal conductivity coefficients as shown in Fig.~\ref{fig:Iso_Fiber_plot_SysError_convergence}. This convergence rate is, however, less clear for the off-diagonal coefficients, especially the $a_{23}$ coefficient, presumably due to the smallness of these coefficients as they are one to two orders of magnitude smaller than the systematic errors of the diagonal coefficients.
For the isotropic ensemble, we employ the projection \eqref{eq:SymmetryInformedStrats_projectors_SymD_explicitProjectors_iso} on the computed apparent conductivity tensors. As a result, the random errors of the projected tensors are consistently reduced by a factor of 3 with the same convergence rate of $L^{-3/2}$, see the blue dotted curve with square marker in Fig.~\ref{fig:Iso_Fiber_plot_RandError_convergence}. On the other hand, the projection does not have a significant effect on the systematic errors, the improvement is about $1\%$, although the same convergence of $L^{-3}$ is achieved as shown in Fig.~\ref{fig:Iso_Fiber_plot_normSysError_convergence} in terms of Frobenius norm. We also notice that the norms of  systematic error and its projected counterpart are bounded by the normalized lower bound, namely the empirical average of the symmetry error, shown by the green curve in Fig.~\ref{fig:Iso_Fiber_plot_normSysError_convergence}. It should be noted that the projected systematic error could slightly increase component-wise, for example the coefficient $a_{22}$. This observation does not contradict with Eq.~\eqref{eq:SymmetryInformedStrats_postprocessing_systematicError}, as the latter concerns the \emph{whole tensor}, not just a component.

\begin{figure}[!h]
	\begin{center}
		\begin{subfigure}{\textwidth}
			\centering
			\includegraphics[height=.033\textheight]{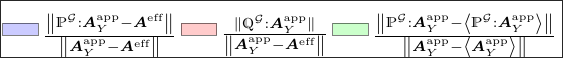}
		\end{subfigure}
	\end{center}
	\centering
	\begin{minipage}[b]{0.3\textwidth}
		\includegraphics[width=1\textwidth]{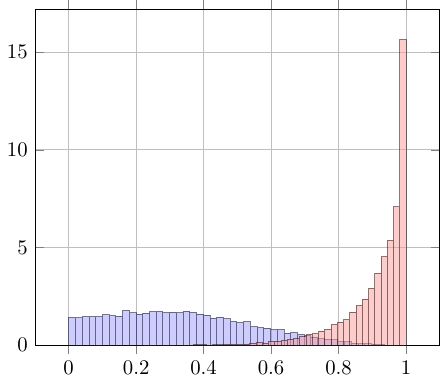}
		\subcaption{$L=128\mu m$}
		\label{fig:plot_histogram_lowerBound_ratio_Iso_64}
	\end{minipage}
	\hfill
	\begin{minipage}[b]{0.3\textwidth}
		\includegraphics[width=1\textwidth]{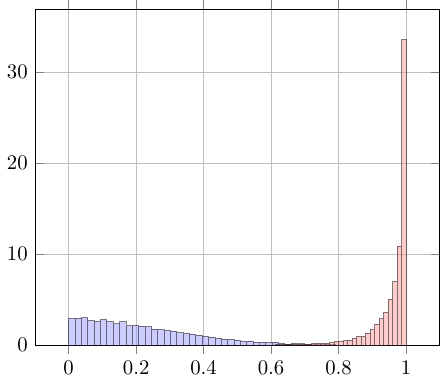}
		\subcaption{$L=256\mu m$}
		\label{fig:plot_histogram_lowerBound_ratio_Iso_128}
	\end{minipage}
	\hfill
	\begin{minipage}[b]{0.3\textwidth}
		\includegraphics[width=1\textwidth]{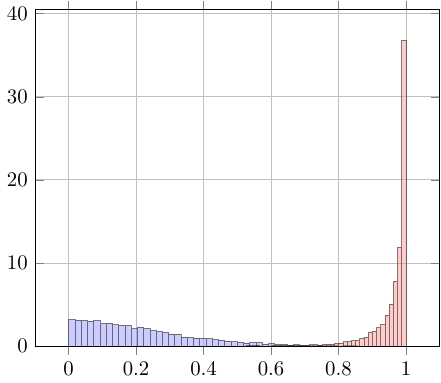}
		\subcaption{$L=512\mu m$}
		\label{fig:plot_histogram_lowerBound_ratio_Iso_256}
	\end{minipage}
	
	\begin{minipage}[b]{0.3\textwidth}
		\includegraphics[width=1\textwidth]{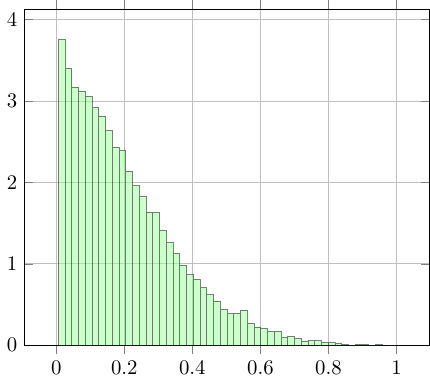}
		\subcaption{$L=128\mu m$}
		\label{fig:plot_ProjectedDispersionError_ratio_Iso_64}
	\end{minipage}
	\hfill
	\begin{minipage}[b]{0.3\textwidth}
		\includegraphics[width=1\textwidth]{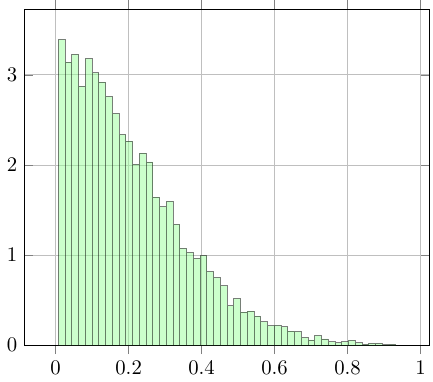}
		\subcaption{$L=256\mu m$}
		\label{fig:plot_ProjectedDispersionError_ratio_Iso_128}
	\end{minipage}
	\hfill
	\begin{minipage}[b]{0.3\textwidth}
		\includegraphics[width=1\textwidth]{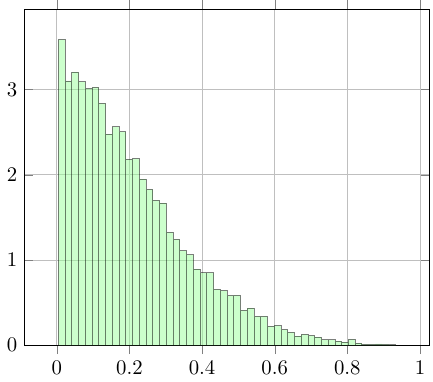}
		\subcaption{$L=512\mu m$}
		\label{fig:plot_ProjectedDispersionError_ratio_Iso_256}
	\end{minipage}	
	\caption{Frequency density of the projection-reduced total error $\left\| \mean{\mathds{P}^{\mathcal{G}}: \fA^{\app}_Y} - \fA^{\eff} \right\|/\left\| \mean{\fA^{\app}_Y} - \fA^{\eff} \right\|$, symmetry error $\left\| \mathds{Q}^{\mathcal{G}}: \fA^{\app}_Y  \right\|/\left\| \mean{\fA^{\app}_Y} - \fA^{\eff} \right\|$ and dispersion error $\left\| \mathds{P}^{\mathcal{G}}: \fA^{\app}_Y - \mean{\mathds{P}^{\mathcal{G}}: \fA^{\app}_Y} \right\|/\left\| \fA^{\app}_Y - \mean{\fA^{\app}_Y} \right\|$, in blue, red and green, respectively.}
	\label{fig:Iso_Fiber_total_dispersion_errors}		
\end{figure}

\review{To evaluate the effectiveness of the orthogonal projection strategy for a change in microstructure characteristics, such as the volume fraction, we performed a similar analysis as for the isotropic ensemble but with an increased volume fraction of 25$\%$. Specifically, 5000 volume elements with edge lengths of $L=128\mu m$ and $L=256\mu m$ were generated and used for devising suitable statistics. The reference values were obtained from 25 volume elements with an edge length $L=2048\mu m$. As reported in Table \ref{tab:isotropic-ensemble-vf-25}, a higher mean value of the apparent conductivity coefficients, about 0.339 $W/(mK)$, emerges due to the higher proportion of conductive fibers in the composite. Nevertheless, the random and systematic errors as well as their corresponding projected errors remain in the same range as those for a volume fraction of $10\%$. Moreover, the orthogonal projection strategy \eqref{eq:SymmetryInformedStrats_projectors_SymD_explicitProjectors_iso} turns out to be effective in this scenario with a higher volume fraction, as well.
	\begin{table}
	\caption{\review{Asymptotic behavior of isotropic apparent conductivity with $25\%$ volume fraction.}}
	\label{tab:isotropic-ensemble-vf-25}
	\centering
	\review{
		\begin{tabular}{ll|rl|rrrrl}
			L & A & Mean & Std. & Sys. error & Proj.  & Rand. error & Proj. & CI error \\
			& &  &  &  & sys. error &  &  rand. error & \\
			\hline\hline
			\multirow{6}{*}{256}& $a_{11}$ & $0.33964$ & $0.00050$ & 0.04986\% & 0.04923\% & 0.14841\% & 0.05612\% & 0.00541\%  \\
			& $a_{22}$ & $0.33963$ & $0.00050$ & 0.04883\% & 0.04923\% & 0.14640\% & 0.05612\% & 0.00534\%  \\
			& $a_{33}$ & $0.33964$ & $0.00050$ & 0.04901\% & 0.04923\% & 0.14776\% & 0.05612\% & 0.00538\%  \\
			& $a_{12}$ & $-0.00001$ & $0.00041$ & 0.00287\% & -- & 0.12095\% & -- & 0.00441\%  \\
			& $a_{13}$ & $0.00000$ & $0.00040$ & 0.00060\% & -- & 0.11654\% & -- & 0.00425\%  \\
			& $a_{23}$ & $0.00000$ & $0.00040$ & 0.00069\% & -- & 0.11808\% & -- & 0.00430\%  \\
			\hline
			\multirow{6}{*}{512}& $a_{11}$ & $0.33949$ & $0.00018$ & 0.00758\% & 0.00631\% & 0.05448\% & 0.02135\% & 0.00281\%  \\
			& $a_{22}$ & $0.33949$ & $0.00019$ & 0.00611\% & 0.00631\% & 0.05495\% & 0.02135\% & 0.00283\%  \\
			& $a_{33}$ & $0.33949$ & $0.00019$ & 0.00525\% & 0.00631\% & 0.05528\% & 0.02135\% & 0.00285\%  \\
			& $a_{12}$ & $0.00000$ & $0.00015$ & 0.00043\% & -- & 0.04517\% & -- & 0.00233\%  \\
			& $a_{13}$ & $0.00000$ & $0.00015$ & 0.00036\% & -- & 0.04543\% & -- & 0.00234\%  \\
			& $a_{23}$ & $0.00000$ & $0.00015$ & 0.00028\% & -- & 0.04452\% & -- & 0.00230\%  \\
			\hline
			\multirow{6}{*}{2048}& $a_{11}$ & $0.33947$ & $0.00001$ & -- & -- & 0.00173\% & 0.00077\% & 0.00097\%  \\
			& $a_{22}$ & $0.33947$ & $0.00001$ & -- & -- & 0.00167\% & 0.00077\% & 0.00094\%  \\
			& $a_{33}$ & $0.33947$ & $0.00001$ & -- & -- & 0.00250\% & 0.00077\% & 0.00140\%  \\
			& $a_{12}$ & $0.00000$ & $0.00001$ & -- & -- & 0.00196\% & -- & 0.00110\%  \\
			& $a_{13}$ & $0.00000$ & $0.00001$ & -- & -- & 0.00153\% & -- & 0.00086\%  \\
			& $a_{23}$ & $0.00000$ & $0.00001$ & -- & -- & 0.00225\% & -- & 0.00126\%  \\
			\hline
			\hline
		\end{tabular}
	}
\end{table}
}

Concerning the influence of the projected apparent conductivity tensors on the total error \eqref{eq:SymmetryInformedStrats_postprocessing_totalError} 
and the dispersion error \eqref{eq:SymmetryInformedStrats_postprocessing_dispersion}, we plot in Fig.~\ref{fig:Iso_Fiber_total_dispersion_errors} the frequency density of the normalized error ratios with respect to their corresponding denominators. We observe that these ratios are bounded between 0 and 1 for all considered isotropic realizations, validating the non-expansiveness of the projection $\mathds{P}^{\textrm{iso}}$ \eqref{eq:SymmetryInformedStrats_projectors_SymD_explicitProjectors_iso}. For the smallest edge length $L=128\mu m$, Fig.~\ref{fig:plot_histogram_lowerBound_ratio_Iso_64} shows that the projected total error ratio, in blue, is broadly distributed from 0 to 0.9 then gradually decrease towards 1. Meanwhile, for the increasing volume sizes as in Figs.~\ref{fig:plot_histogram_lowerBound_ratio_Iso_128} and \ref{fig:plot_histogram_lowerBound_ratio_Iso_256}, this ratio skews towards 0 with the cut-off around 0.6, indicating the improvement by the projection on the apparent conductivity tensors. In these figures, we also observe that the normalized symmetry errors, in red, are highly distributed close to 1, particularly for larger volume sizes, demonstrating the inevitable symmetry error of the volume element. We also recognize that the discrepancy due to non-invariant part of the apparent conductivity tensors is the dominant source of error. Additionally, Fig.~\ref{fig:plot_ProjectedDispersionError_ratio_Iso_64}-Fig.~\ref{fig:plot_ProjectedDispersionError_ratio_Iso_256} present the normalized distribution of the projected dispersion error ratios, that peak close to 0 and gradually decrease towards 1 for all volume element sizes, showing the benefits of the projection $\mathds{P}^{\textrm{iso}}$ in reducing dispersion errors greatly, as also shown earlier with the random errors. 

\begin{figure}[!h]
	\centering
	\begin{minipage}[b]{0.3\textwidth}
		\includegraphics[width=1\textwidth]{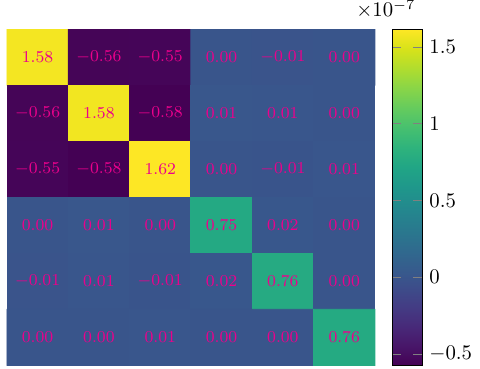}		
		\subcaption{$L=128\mu m$}
		\label{fig:plot_Qtensor_Iso_64}
	\end{minipage}
	\hfill
	\begin{minipage}[b]{0.3\textwidth}
		\includegraphics[width=1\textwidth]{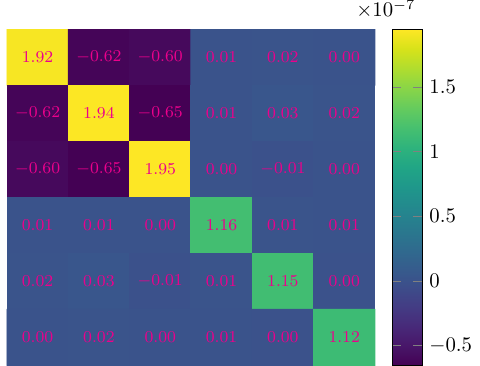}
		\subcaption{$L=256\mu m$}
		\label{fig:plot_Qtensor_Iso_128}
	\end{minipage}	
	\hfill
	\begin{minipage}[b]{0.3\textwidth}
		\includegraphics[width=1\textwidth]{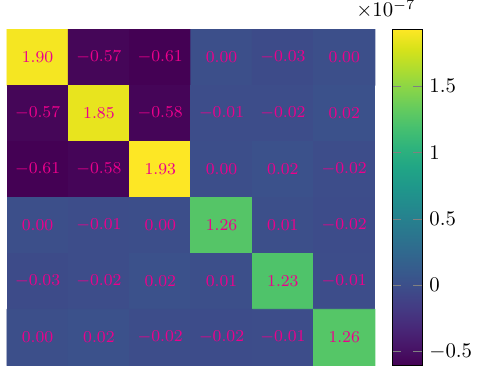}
		\subcaption{$L=512\mu m$}
		\label{fig:plot_Qtensor_Iso_256}
	\end{minipage}	
	\caption{\Qtensor{} of the isotropic ensemble in matrix form. Unit of the \Qtensor{} components is $(W^2/(m^3K^2))$.}% \todo{The unit $(W^2/(km^3K^2))$ seems a bit awkward :(}}
	\label{fig:Iso-muQ-tensor-matrix}		
\end{figure}

\begin{figure}[!h]
	\begin{center}
		\begin{subfigure}{\textwidth}
			\centering
			\includegraphics[height=.022\textheight]{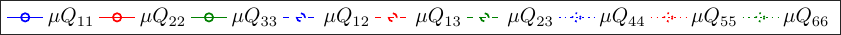}
		\end{subfigure}
	\end{center}
	\centering
	\begin{minipage}[b]{0.45\textwidth}
		\includegraphics[width=1\textwidth]{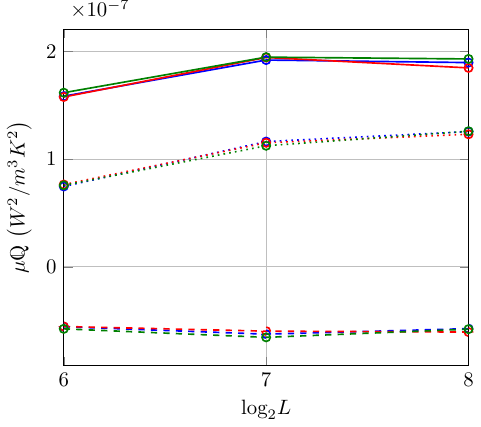}		
		\subcaption{\Qtensor{} components}
		\label{fig:iso_plot_convergence_Qtensor_components}
	\end{minipage}
	\hfill
	\begin{minipage}[b]{0.45\textwidth}
		\includegraphics[width=1\textwidth]{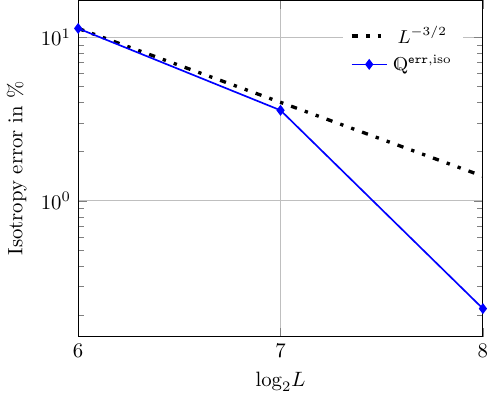}
		\subcaption{Isotropy residual}
		\label{fig:iso_plot_t_isotropy_residue}
	\end{minipage}	
	\caption{(a) Leading-order components of the \Qtensor{}. (b) Convergence of the isotropy error \eqref{eq:computation-isotropic-isotropy-error-Qtensor}. }
	\label{fig:iso-muQ-tensor-components}		
\end{figure}

We turn our attention to the \Qtensor{}s and present their Voigt matrix forms in Fig.~\ref{fig:Iso-muQ-tensor-matrix}. The diagonal coefficients of the \Qtensor{} express the auto-correlations of the conductivity coefficients, which are stronger for the diagonal apparent conductivities than for the off-diagonal terms. The components $\mu Q_{12}$, $\mu Q_{13}$ and $\mu Q_{23}$ are negative and represent negative correlations among the diagonal apparent conductivities. The remaining components of the \Qtensor{} are associated with the off-diagonal apparent conductivities and thus are also rather small. As a result, the \Qtensor{} has nine leading-order coefficients, that are plotted in Fig.~\ref{fig:iso_plot_convergence_Qtensor_components}. Furthermore, to assess the asymptotic behavior of the \Qtensor{}, we compute the isotropy error defined as
\begin{equation}
	\label{eq:computation-isotropic-isotropy-error-Qtensor}
	\mathds{Q}^{\texttt{err},\textrm{iso}} = \frac{\|\mathds{Q} - \mathcal{P}^{\textrm{iso}}::\mathds{Q} \|}{\|\mathcal{P}^{\textrm{iso}}::\mathds{Q}\|},
\end{equation}
and plot the result in Fig.~\ref{fig:iso_plot_t_isotropy_residue}. For the smallest volume element of edge length $L=128\mu m$, the isotropy error is rather large at $11.3\%$ and decreases at the rate $L^{-3/2}$, as expected~\cite{Duerinckx2020muQ}, as the edge length increases to $L=256\mu m$ and reaches $1.41\%$ at $L=512\mu m$. We wish to end this section with one remark. The obtained \Qtensor{} of the considered isotropic ensemble exhibits a good approximation for the isotropic tensor \eqref{eq:special_isotropic_Qtensor} which has \textit{two} independent coefficients, that differs from the related study by Khoromskaia et al.~\cite{khoromskaia2020numerical} where three coefficients of the \Qtensor{} emerge from a cubic ensemble of square inclusions embedded in square volume elements, although their dependencies are not specified.

%\begin{figure}[!h]
%	\centering
%	\begin{minipage}[b]{0.3\textwidth}
%		\includegraphics[width=1\textwidth]{data/Iso/plot_Qtensor_visualization_Y_64.pdf}
%		\subcaption{L=32\,$\mu m$}
%		\label{fig:plot_Young_Sphere_32}
%	\end{minipage}
%	\begin{minipage}[b]{0.3\textwidth}
%		\includegraphics[width=1\textwidth]{data/Iso/plot_Qtensor_visualization_Y_128.pdf}
%		\subcaption{L=64\,$\mu m$}
%		\label{fig:plot_Young_Sphere_64}
%	\end{minipage}	
%	\begin{minipage}[b]{0.3\textwidth}
%		\includegraphics[width=1\textwidth]{data/Iso/plot_Qtensor_visualization_Y_256.pdf}
%		\subcaption{L=128\,$\mu m$}
%		\label{fig:plot_Young_Sphere_128}
%	\end{minipage}	
%	\begin{minipage}[b]{0.3\textwidth}
%		\includegraphics[width=1\textwidth]{data/Trans/plot_Qtensor_visualization_nu_ave_64.pdf}
%		\subcaption{L=32\,$\mu m$}
%		\label{fig:plot_Poisson_average_Sphere_32}
%	\end{minipage}
%	\begin{minipage}[b]{0.3\textwidth}
%		\includegraphics[width=1\textwidth]{data/Trans/plot_Qtensor_visualization_nu_ave_128.pdf}
%		\subcaption{L=64\,$\mu m$}
%		\label{fig:plot_Poisson_average_Sphere_64}
%	\end{minipage}	
%	\begin{minipage}[b]{0.3\textwidth}
%%		\resizebox{\textwidth}{!}{\import{figures/Spheres/}{Poisson_average_Sphere_128.pgf}}
%		\includegraphics[width=1\textwidth]{data/Trans/plot_Qtensor_visualization_nu_ave_256.pdf}
%		\subcaption{L=128\,$\mu m$}
%		\label{fig:plot_Poisson_average_Sphere_128}
%	\end{minipage}	
%	\caption{Visual representation of \Qtensor{} for increasing volume element size. (a-b-c) Young's-modulus-like coefficient. (d-e-f) Poisson-ratio-like coefficient.}
%	\label{fig:sphere-muQ-graphical}		
%\end{figure}

\subsection{A transversely isotropic ensemble}
\label{sec:computation_transverse_isotropy}
%\begin{figure}[!h]
%	\centering	
%	\includegraphics[width=1\textwidth, trim = 0 0 0 0, clip]{figures/Trans_Fiber/Combined2.eps}	
%	\caption{Transversely isotropic volumes elements of short fibers of 100$\mu$m length, 10$\mu$m diameter with 10$\%$ volume fraction. Volume element size from left to right: $L=64\,\mu m$, $L=128\,\mu m$, $L=256\,\mu m$ and  $L=512\,\mu m$.}
%	\label{fig:trans-fiber}
%\end{figure}
\begin{figure}[h!]
	\begin{subfigure}{.098\textwidth}
		\includegraphics[trim = 50 20 1850 920, clip, width=\textwidth]{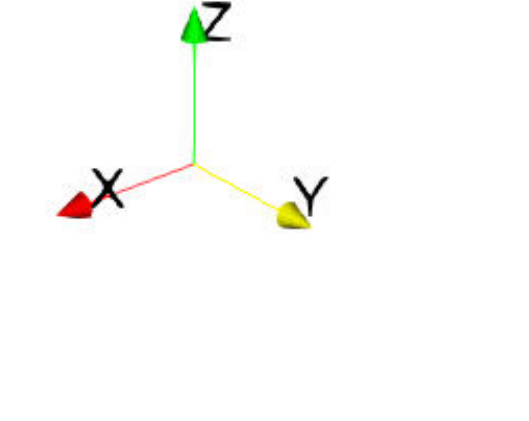}
	\end{subfigure}
	\begin{subfigure}{.151\textwidth}
		\includegraphics[trim = 600 20 480 125, clip, width=\textwidth]{data/Trans/tran_64.pdf}
		\caption{$64^3$}
		\label{fig:tran_64}
	\end{subfigure}
	\begin{subfigure}{.285\textwidth}
		\includegraphics[trim = 600 20 480 125, clip, width=\textwidth]{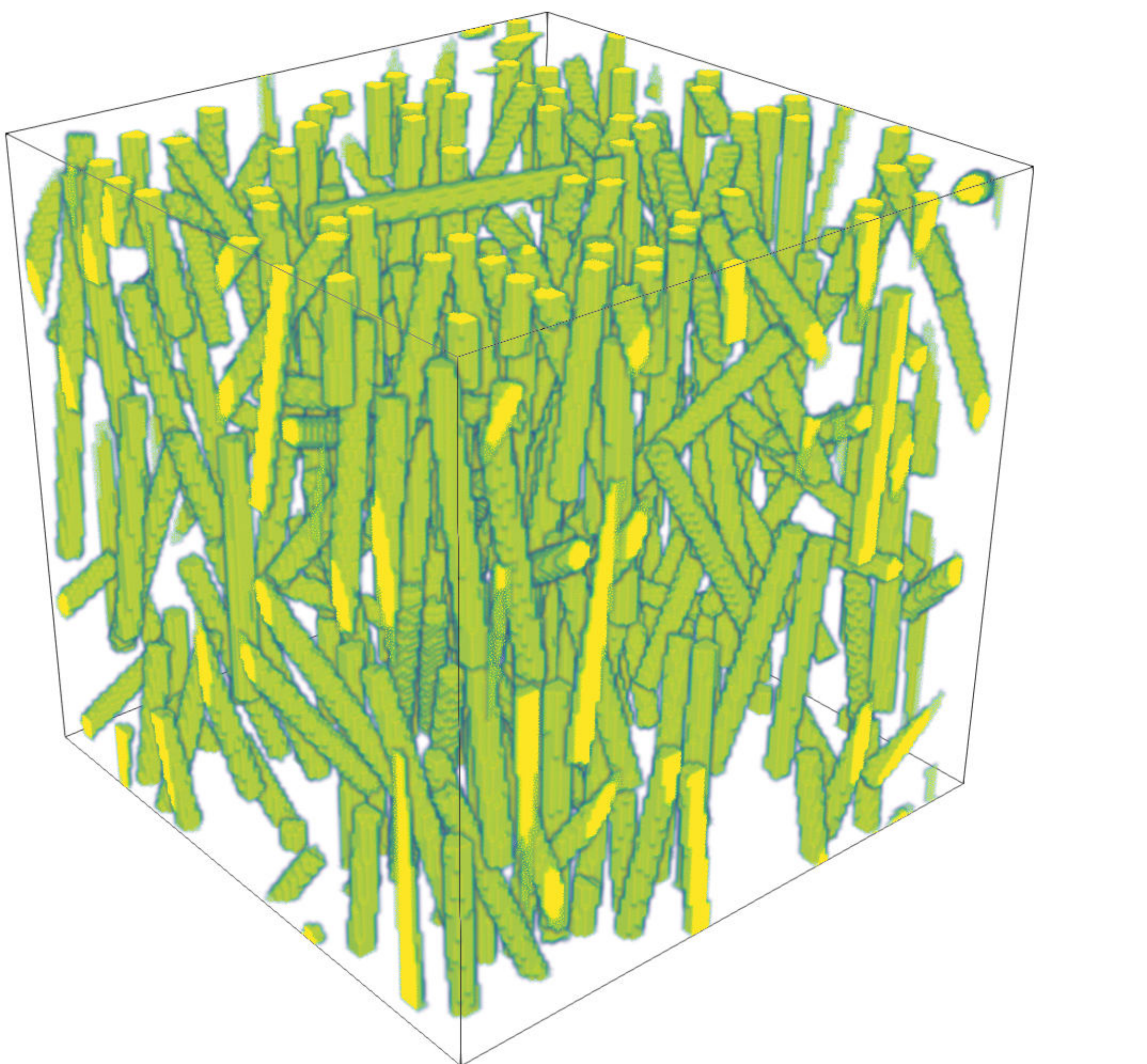}
		\caption{$128^3$}
		\label{fig:tran_128}
	\end{subfigure}
	\begin{subfigure}{.48\textwidth}
		\includegraphics[trim = 600 20 580 125, clip, width=\textwidth]{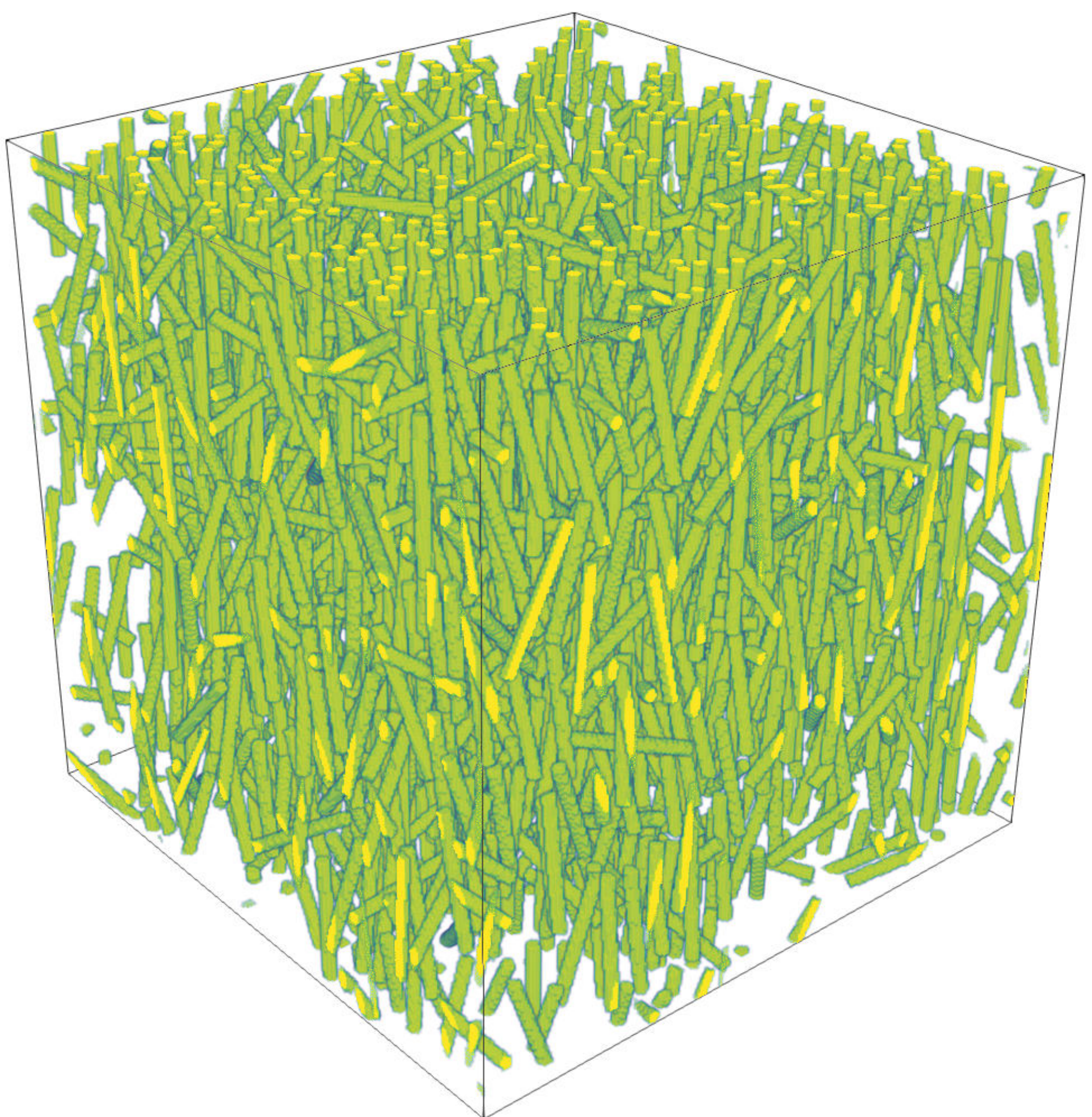}
		\caption{$256^3$}
		\label{fig:tran_256}
	\end{subfigure}
	\caption{Volume elements of the transverse isotropically distributed fibers.}
	\label{fig:trans-fiber}
\end{figure}
\begin{figure}[!h]
	\centering	
	\begin{minipage}[b]{0.3\textwidth}
		\centering
		\includegraphics[width=1\textwidth]{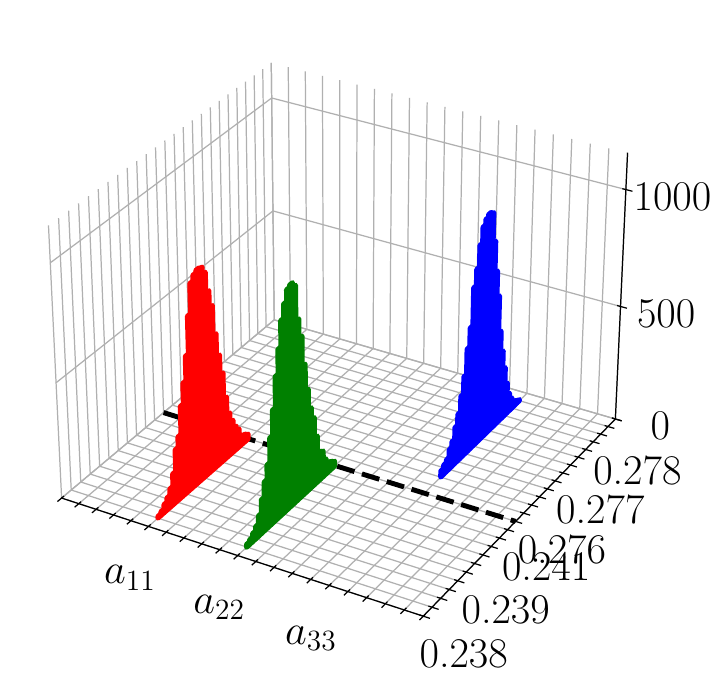}
		\subcaption{L=128\,$\mu m$}
		\label{fig:plot_histogram_diag_zOrient_64}
	\end{minipage}	
	\begin{minipage}[b]{0.3\textwidth}
		\centering
		\includegraphics[width=1\textwidth]{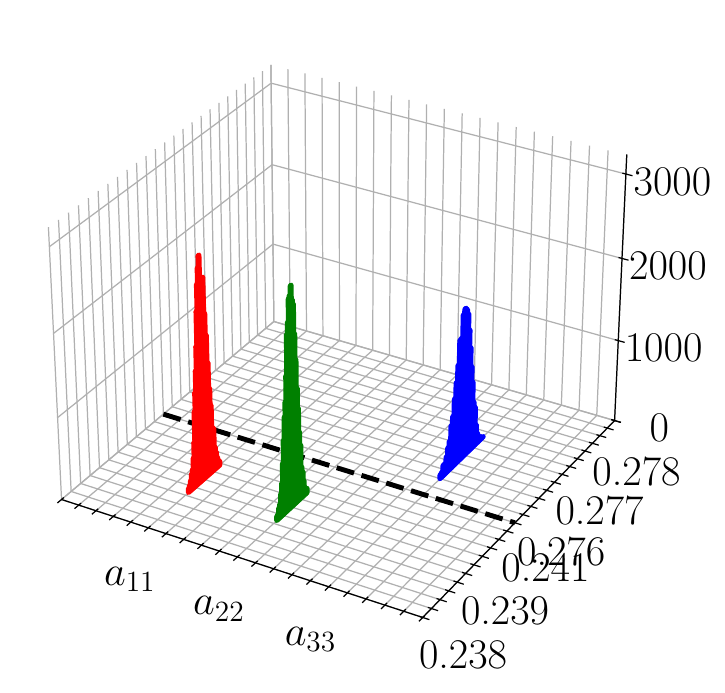}
		\subcaption{L=256\,$\mu m$}
		\label{fig:plot_histogram_diag_zOrient_128}
	\end{minipage}		
	\begin{minipage}[b]{0.3\textwidth}
		\centering
		\includegraphics[width=1\textwidth]{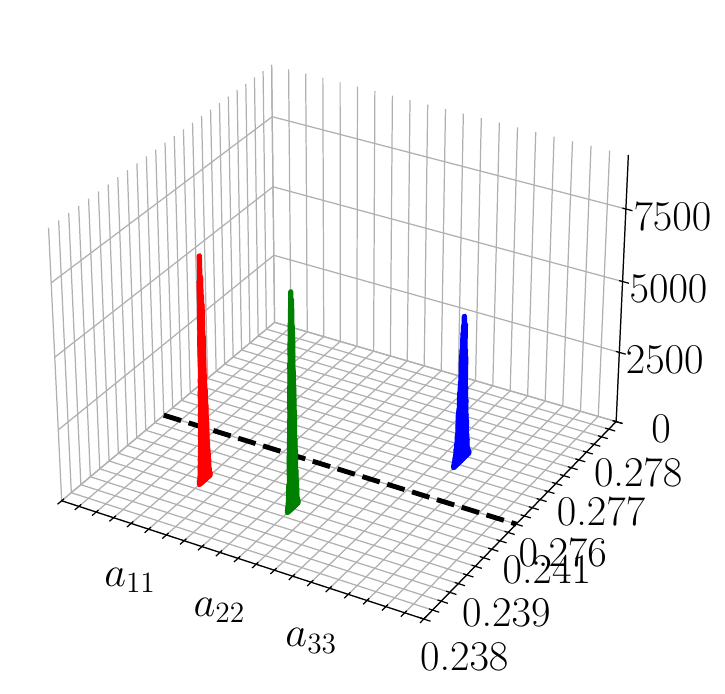}
		\subcaption{L=512\,$\mu m$}
		\label{fig:plot_histogram_diag_zOrient_256}
	\end{minipage}		
	
	\begin{minipage}[b]{0.3\textwidth}
		\centering
		\includegraphics[width=1\textwidth]{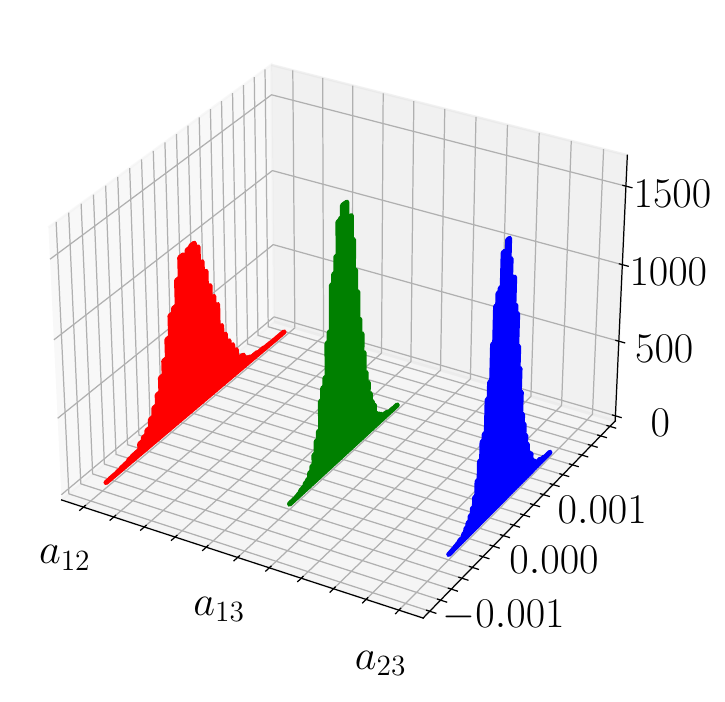}
		\subcaption{L=128\,$\mu m$}
		\label{fig:plot_histogram_offdiag_zOrient_64}
	\end{minipage}	
	\begin{minipage}[b]{0.3\textwidth}
		\centering
		\includegraphics[width=1\textwidth]{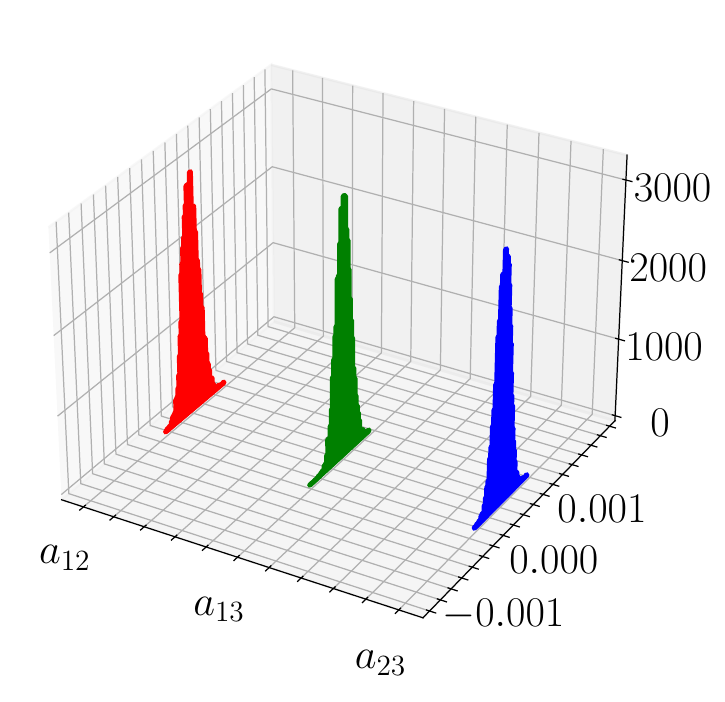}
		\subcaption{L=256\,$\mu m$}
		\label{fig:plot_histogram_offdiag_zOrient_128}
	\end{minipage}		
	\begin{minipage}[b]{0.3\textwidth}
		\centering
		\includegraphics[width=1\textwidth]{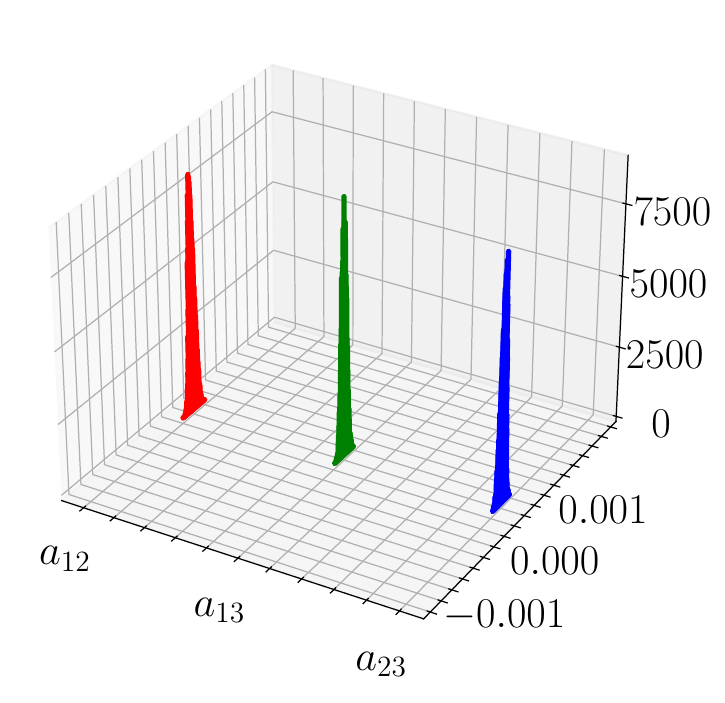}
		\subcaption{L=512\,$\mu m$}
		\label{fig:plot_histogram_offdiag_zOrient_256}
	\end{minipage}		
	\caption{Histogram of transversely isotropic apparent thermal conductivity coefficients for increasing volume element sizes. (a-b-c) Diagonal coefficients $a_{11}$, $a_{22}$ and $a_{33}$; (d-e-f) Off-diagonal coefficients $a_{23}$, $a_{13}$ and $a_{12}$. The statistics of these coefficients are presented in Table \ref{tab:transversely-isotropic-ensemble}.}
	\label{fig:tran-fiber-histogram-apparent}
\end{figure}

In this section, we are concerned with a transversely isotropic ensemble, prescribed by the following second-order fiber orientation tensor
\begin{equation}
	\label{eq:computation-transeverse-isotropy-orientation-tensor}
	\textbf{\texttt{A}}^{\texttt{ti}} = \mqty[0.1 & 0 & 0 \\ 0 & 0.1 & 0 \\ 0 & 0 & 0.8].
\end{equation}
Such orientation tensor results in the microstructures that are invariant with respect to the rotation around the $\fe_3$-direction, as shown in Fig.~\ref{fig:trans-fiber}. The frequency density of the computed apparent conductivity coefficients are presented in Fig.~\ref{fig:tran-fiber-histogram-apparent}. 
We observe that the two conductivity coefficients in the plane of isotropy, $a_{11}$ and $a_{22}$ follow similar distributions with the same mean values around $0.239\,W/(mK)$, whereas the coefficient $a_{33}$ has a distinct mean value of $0.276\,W/(mK)$. The off-diagonal conductivity coefficients are close to zero. The details of the statistical results are reported in Table \ref{tab:transversely-isotropic-ensemble}.
\begin{table}[h!]
	\caption{Asymptotic behavior of transversely isotropic apparent conductivity. The units of $L$, mean values and standard deviation are $\mu m$, $W/(mK)$ and $W/(mK)$, respectively.}
	\label{tab:transversely-isotropic-ensemble}
	\centering
	\begin{tabular}{ll|rl|rrrrl}
		L & A & Mean & Std. & Sys. error & Proj.  & Rand. error & Proj. & CI error \\
		& &  &  &  & sys. error &  &  rand. error & \\
		\hline\hline
		\multirow{6}{*}{128}& $a_{11}$ & $0.23949$ & $0.00045$ & 0.00643\% & 0.00845\% & 0.18627\% & 0.06521\% & 0.00392\%  \\
		& $a_{22}$ & $0.23948$ & $0.00045$ & 0.01048\% & 0.00845\% & 0.18618\% & 0.06521\% & 0.00392\%  \\
		& $a_{33}$ & $0.27687$ & $0.00043$ & 0.23725\% & 0.23725\% & 0.17828\% & 0.17828\% & 0.00375\%  \\
		& $a_{12}$ & $0.00000$ & $0.00038$ & 0.00113\% & -- & 0.16020\% & -- & 0.00337\%  \\
		& $a_{13}$ & $0.00000$ & $0.00025$ & 0.00051\% & -- & 0.10615\% & -- & 0.00223\%  \\
		& $a_{23}$ & $0.00000$ & $0.00025$ & 0.00078\% & -- & 0.10487\% & -- & 0.00221\%  \\
		\hline
		\multirow{6}{*}{256}& $a_{11}$ & $0.23951$ & $0.00016$ & 0.00035\% & 0.00036\% & 0.06506\% & 0.02661\% & 0.00168\%  \\
		& $a_{22}$ & $0.23951$ & $0.00015$ & 0.00037\% & 0.00036\% & 0.06459\% & 0.02661\% & 0.00166\%  \\
		& $a_{33}$ & $0.27628$ & $0.00023$ & 0.02229\% & 0.02229\% & 0.09684\% & 0.09684\% & 0.00249\%  \\
		& $a_{12}$ & $0.00000$ & $0.00014$ & 0.00053\% & -- & 0.05758\% & -- & 0.00148\%  \\
		& $a_{13}$ & $0.00000$ & $0.00013$ & 0.00040\% & -- & 0.05431\% & -- & 0.00140\%  \\
		& $a_{23}$ & $0.00000$ & $0.00013$ & 0.00028\% & -- & 0.05413\% & -- & 0.00139\%  \\
		\hline
		\multirow{6}{*}{512}& $a_{11}$ & $0.23951$ & $0.00005$ & 0.00012\% & 0.00010\% & 0.02262\% & 0.00937\% & 0.00058\%  \\
		& $a_{22}$ & $0.23951$ & $0.00005$ & 0.00008\% & 0.00010\% & 0.02274\% & 0.00937\% & 0.00059\%  \\
		& $a_{33}$ & $0.27622$ & $0.00008$ & 0.00374\% & 0.00374\% & 0.03439\% & 0.03439\% & 0.00089\%  \\
		& $a_{12}$ & $0.00000$ & $0.00005$ & 0.00014\% & -- & 0.02111\% & -- & 0.00054\%  \\
		& $a_{13}$ & $0.00000$ & $0.00005$ & 0.00005\% & -- & 0.02010\% & -- & 0.00052\%  \\
		& $a_{23}$ & $0.00000$ & $0.00005$ & 0.00019\% & -- & 0.02026\% & -- & 0.00052\%  \\
		\hline
		\multirow{6}{*}{2048}& $a_{11}$ & $0.23951$ & $0.00001$ & -- & -- & 0.00316\% & 0.00118\% & 0.00058\%  \\
		& $a_{22}$ & $0.23951$ & $0.00001$ & -- & -- & 0.00301\% & 0.00118\% & 0.00056\%  \\
		& $a_{33}$ & $0.27621$ & $0.00001$ & -- & -- & 0.00421\% & 0.00421\% & 0.00078\%  \\
		& $a_{12}$ & $0.00000$ & $0.00001$ & -- & -- & 0.00260\% & -- & 0.00048\%  \\
		& $a_{13}$ & $0.00000$ & $0.00001$ & -- & -- & 0.00244\% & -- & 0.00045\%  \\
		& $a_{23}$ & $0.00000$ & $0.00001$ & -- & -- & 0.00236\% & -- & 0.00044\%  \\
		\hline
	\end{tabular}
\end{table}
\begin{figure}[!h]
	\begin{center}
		\begin{subfigure}{\textwidth}
			\centering
			\includegraphics[height=.022\textheight]{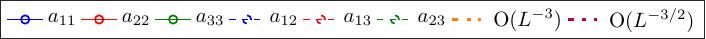}
		\end{subfigure}
	\end{center}
	\centering
	\begin{minipage}[b]{0.3\textwidth}
		\includegraphics[width=1\textwidth]{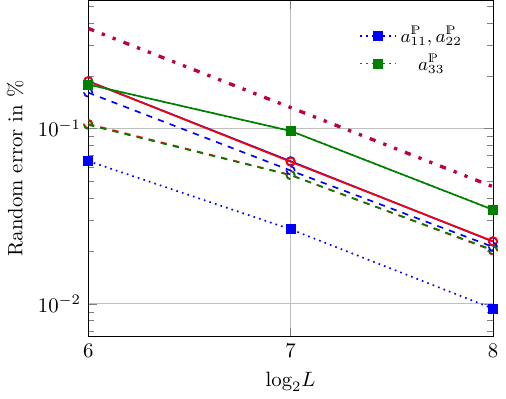}
		\subcaption{Random errors}
		\label{fig:Trans_Fiber_plot_RandError_convergence}
	\end{minipage}
	\hfill
	\begin{minipage}[b]{0.3\textwidth}
		\includegraphics[width=1\textwidth]{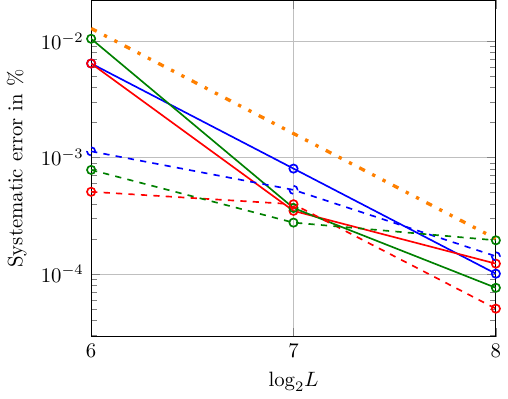}
		\subcaption{Systematic errors}
		\label{fig:Trans_Fiber_plot_SysError_convergence}
	\end{minipage}
	\hfill
	\begin{minipage}[b]{0.3\textwidth}
		\includegraphics[width=1\textwidth]{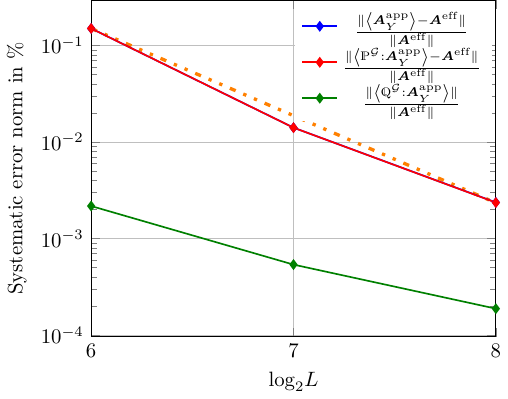}
		\subcaption{Norm of systematic errors}
		\label{fig:Trans_Fiber_plot_normSysError_convergence}
	\end{minipage}
	\caption{Convergence of: (a) Random errors; (b) Systematic errors; (c) Frobenius norm of systematic errors. $a^{\mathds{P}}_{11},a^{\mathds{P}}_{22},a^{\mathds{P}}_{33}$ denote the diagonal components of the projected empirical mean \eqref{eq:setup-empirical-mean-projected}.}
	\label{fig:Trans_Fiber_sys_rand_errors}		
\end{figure}

The confidence interval errors are well below $0.005\%$, reaffirming our choice for the number of realizations. For the smallest volume element, with $L=128\mu m$, we notice that the systematic errors are rather small and consistently smaller than the random errors, similar to the previous section, except for the coefficient $a_{33}$ whose systematic error is 1.4 times higher than its random error counterpart. However, such outlier behavior is no longer observed for larger volume elements. In terms of scaling, the random errors converge at the expected rate of $L^{-3/2}$ for both diagonal and off-diagonal coefficients, as shown in Fig.~\ref{fig:Trans_Fiber_plot_RandError_convergence}. For the systematic errors, we recover the optimal rate of $L^{-3}$ for the diagonal coefficients but not for the off-diagonal ones, shown in Fig.~\ref{fig:Trans_Fiber_plot_SysError_convergence}. This could be attributed to the smallness of these values. Upon invoking the projection \eqref{eq:SymmetryInformedStrats_projectors_SymD_explicitProjectors_trans_iso}, the random errors of the isotropic-plane conductivity coefficients $a_{11}$ and $a_{22}$ are reduced by a factor of 3, whereas that of $a_{33}$ is unaffected, which is also observed by the blue and green dotted line with square markers in Fig.~\ref{fig:Trans_Fiber_plot_RandError_convergence}. The projected random errors are also scaled by $L^{-3/2}$. On the other hand, similar to the previous case, the projection is not effective on the systematic error, only improves it by $0.3\%$ in terms of Frobenius norm, shown by the blue and red curves in Fig.~\ref{fig:Trans_Fiber_plot_normSysError_convergence}, at the rate $L^{-3}$. In this figure, we also plot the empirical average of symmetry error as the lower bound of the systematic error, shown by the green curve. %\todo{I am still missing the justifications for the effectiveness of the projection on the isotropic components.}
\begin{figure}[!h]
	\begin{center}
		\begin{subfigure}{\textwidth}
			\centering
			\includegraphics[height=.033\textheight]{data/Trans/legend_histogram_plot.pdf}
		\end{subfigure}
	\end{center}
	\centering
	\begin{minipage}[b]{0.3\textwidth}
		\includegraphics[width=1\textwidth]{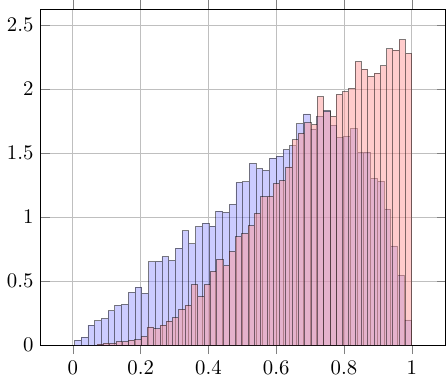}
		\subcaption{$L=128\mu m$}
		\label{fig:plot_histogram_lowerBound_ratio_Trans_64}
	\end{minipage}
	\hfill
	\begin{minipage}[b]{0.3\textwidth}
		\includegraphics[width=1\textwidth]{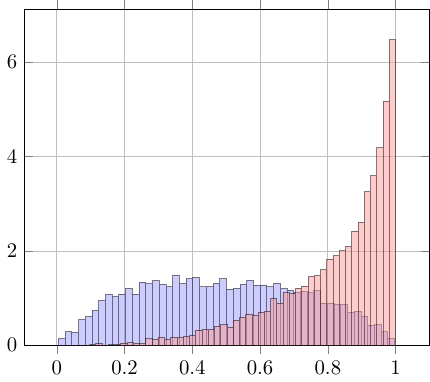}
		\subcaption{$L=256\mu m$}
		\label{fig:plot_histogram_lowerBound_ratio_Trans_128}
	\end{minipage}
	\hfill
	\begin{minipage}[b]{0.3\textwidth}
		\includegraphics[width=1\textwidth]{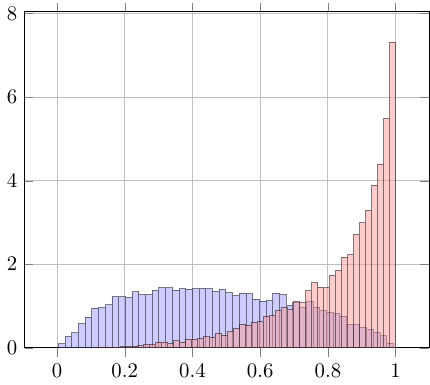}
		\subcaption{$L=512\mu m$}
		\label{fig:plot_histogram_lowerBound_ratio_Trans_256}
	\end{minipage}
	
	\begin{minipage}[b]{0.3\textwidth}
		\includegraphics[width=1\textwidth]{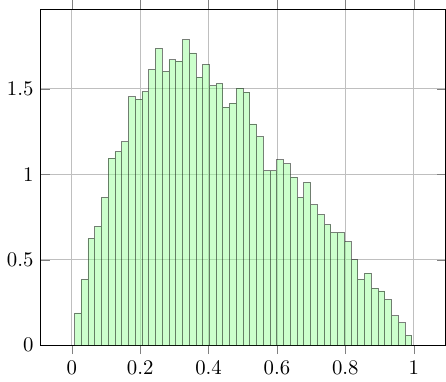}
		\subcaption{$L=128\mu m$}
		\label{fig:plot_ProjectedDispersionError_ratio_Trans_64}
	\end{minipage}
	\hfill
	\begin{minipage}[b]{0.3\textwidth}
		\includegraphics[width=1\textwidth]{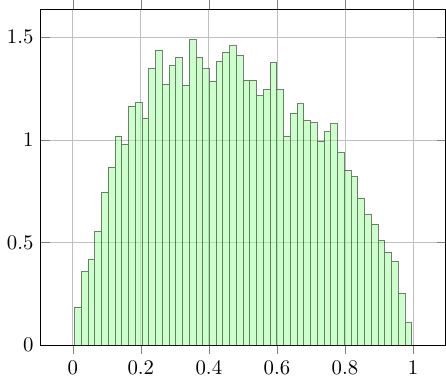}
		\subcaption{$L=256\mu m$}
		\label{fig:plot_ProjectedDispersionError_ratio_Trans_128}
	\end{minipage}
	\hfill
	\begin{minipage}[b]{0.3\textwidth}
		\includegraphics[width=1\textwidth]{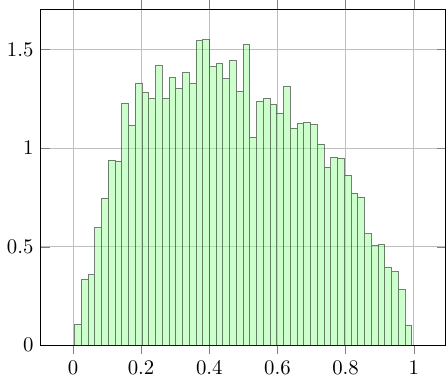}
		\subcaption{$L=512\mu m$}
		\label{fig:plot_ProjectedDispersionError_ratio_Trans_256}
	\end{minipage}
	\caption{Frequency density of the projection-reduced total error $\left\| \mean{\mathds{P}^{\mathcal{G}}: \fA^{\app}_Y} - \fA^{\eff} \right\|/\left\| \mean{\fA^{\app}_Y} - \fA^{\eff} \right\|$, symmetry error $\left\| \mathds{Q}^{\mathcal{G}}: \fA^{\app}_Y  \right\|/\left\| \mean{\fA^{\app}_Y} - \fA^{\eff} \right\|$ and dispersion error $\left\| \mathds{P}^{\mathcal{G}}: \fA^{\app}_Y - \mean{\mathds{P}^{\mathcal{G}}: \fA^{\app}_Y} \right\|/\left\| \fA^{\app}_Y - \mean{\fA^{\app}_Y} \right\|$, in blue, red and green, respectively.}
	\label{fig:Trans_Fiber_total_dispersion_errors}		
\end{figure}

We continue the investigation of the consequences of the projection $\mathds{P}^{\textrm{ti}}$ \eqref{eq:SymmetryInformedStrats_projectors_SymD_explicitProjectors_trans_iso} on the total error \eqref{eq:SymmetryInformedStrats_postprocessing_totalError} 
and the dispersion error \eqref{eq:SymmetryInformedStrats_postprocessing_dispersion} by showing the frequency density of their normalized counterparts in Fig.~\ref{fig:Trans_Fiber_total_dispersion_errors}. These ratios are bounded between 0 and 1 for all considered transversely isotropic realizations, asserting the inequalities \eqref{eq:SymmetryInformedStrats_postprocessing_totalError} and  \eqref{eq:SymmetryInformedStrats_postprocessing_dispersion}. As shown in Fig.~\ref{fig:plot_histogram_lowerBound_ratio_Trans_64} for the smallest volume element of $64^3$, the projected total error ratio, shown in blue, skews to the right and peaks at around 0.7. As the volume element size increases to $128^3$ and $256^3$, see Fig.~\ref{fig:plot_histogram_lowerBound_ratio_Trans_64} and Fig.~\ref{fig:plot_histogram_lowerBound_ratio_Trans_256}, the projected total error ratio spreads broadly with a plateau between 0.2 and 0.6. Hence, the projection $\mathds{P}^{\textrm{ti}}$ is rather ineffective for small volume element size but provides a fair improvement of the total error for larger volume element sizes, albeit the large variability. Also in these figures, the normalized symmetry errors are right-skewed and peak close to 1, but with a larger tail as compared to the previous isotropic ensemble, because the $\fe_3$-direction of the transversely isotropic ensemble actually aligns with one axis of the volume element. Therefore, the non-invariant part of the apparent conductivity tensors is still the dominant source of error. Regarding the dispersion errors, shown in Figs.~\ref{fig:plot_ProjectedDispersionError_ratio_Trans_64}-\ref{fig:plot_ProjectedDispersionError_ratio_Trans_256}, we notice that the projection also brings moderate improvement as their ratios are also broadly distributed and peak around 0.2-0.4, which is in agreement with the observation of random errors. As mentioned earlier, the projected random errors are reduced by a factor of 3 for the conductivity coefficients in the isotropic-plane, but unchanged for the $a_{33}$ component due to the alignment of the invariant axis $\fe_3$. 

\begin{figure}[!h]
	\centering
	\begin{minipage}[b]{0.3\textwidth}
		\includegraphics[width=1\textwidth]{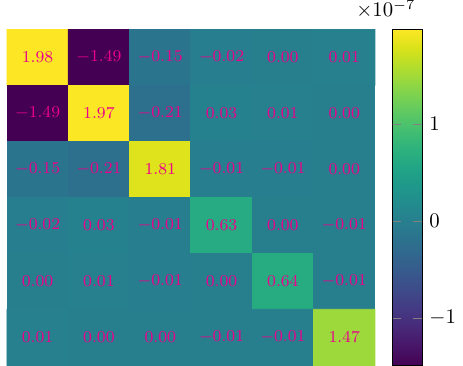}		
		\subcaption{$L=64\mu m$}
		\label{fig:muQ_tensor_zOrient_64}
	\end{minipage}
	\hfill
	\begin{minipage}[b]{0.3\textwidth}
		\includegraphics[width=1\textwidth]{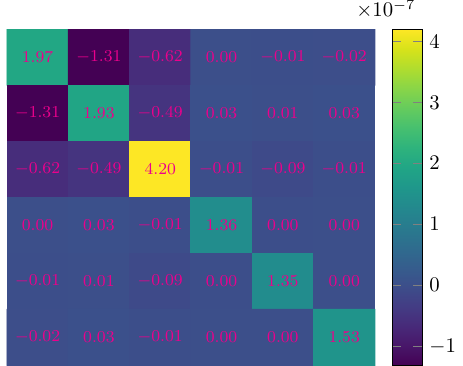}
		\subcaption{$L=128\mu m$}
		\label{fig:muQ_tensor_zOrient_128}
	\end{minipage}	
	\hfill
	\begin{minipage}[b]{0.3\textwidth}
		\includegraphics[width=1\textwidth]{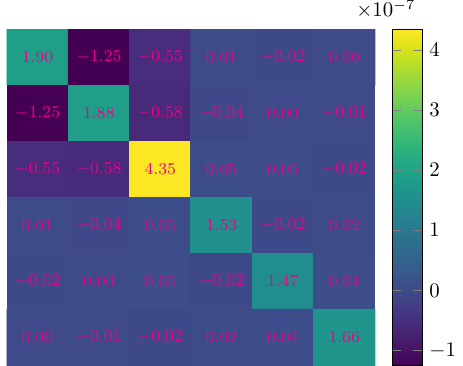}
		\subcaption{$L=256\mu m$}
		\label{fig:muQ_tensor_zOrient_256}
	\end{minipage}	
	\caption{\Qtensor{} of the transversely isotropic ensemble in matrix form. Unit of the \Qtensor{} components is $(W^2/(m^3K^2))$.}
	\label{fig:tran-muQ-tensor-matrix}		
\end{figure}

\begin{figure}[!h]
	\begin{center}
		\begin{subfigure}{\textwidth}
			\centering
			\includegraphics[height=.022\textheight]{data/Iso/legend_Qtensor_components.pdf}
		\end{subfigure}
	\end{center}
	\centering
	\begin{minipage}[b]{0.45\textwidth}
		\includegraphics[width=1\textwidth]{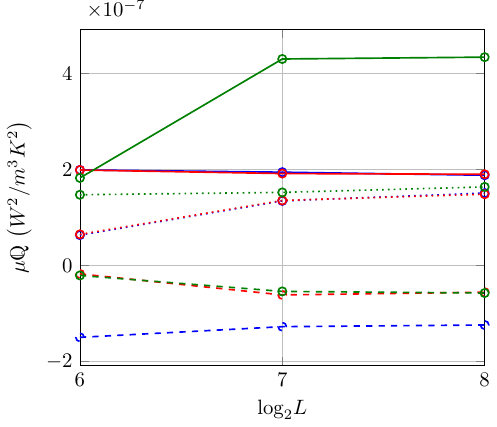}		
		\subcaption{\Qtensor{} components}
		\label{fig:plot_convergence_Qtensor_components}
	\end{minipage}
	\hfill
	\begin{minipage}[b]{0.45\textwidth}
		\includegraphics[width=1\textwidth]{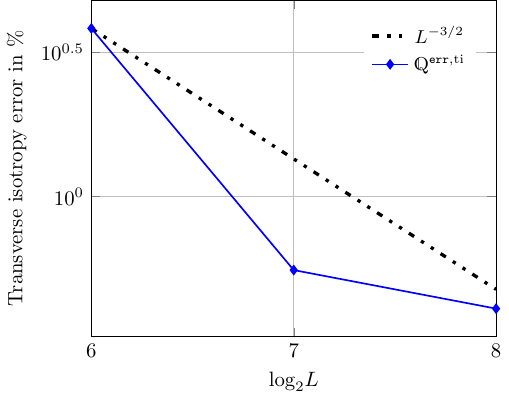}
		\subcaption{Transverse isotropy residual}
		\label{fig:plot_t_isotropy_residue}
	\end{minipage}		
	\caption{(a) Leading-order components of the \Qtensor{}. (b) Convergence of the transverse isotropy error \eqref{eq:computation-trans-isotropic-isotropy-error-Qtensor}. }
	\label{fig:tran-muQ-tensor-components}		
\end{figure}

The apparent \Qtensor{}s of the transversely isotropic ensemble in their matrix forms are presented in Fig.~\ref{fig:tran-muQ-tensor-matrix}. We notice again the nine leading-order coefficients of a \Qtensor{} as in the previous section. The auto-correlation strengths between the diagonal conductivity coefficients are consistently larger than those of the off-diagonal components, analogous to the isotropic ensemble. Besides, the components $\mu Q_{12}$, $\mu Q_{13}$ and $\mu Q_{23}$ continue to show the negative correlations between the diagonal conductivity terms. We present the nine leading-order coefficients of the \Qtensor{} of the transversely isotropic ensemble in Fig.~\ref{fig:plot_convergence_Qtensor_components}. In terms of scaling, we monitor the following error
\begin{equation}
	\label{eq:computation-trans-isotropic-isotropy-error-Qtensor}
	\mathds{Q}^{\texttt{err},\textrm{ti}} = \frac{\|\mathds{Q} - \mathcal{P}^{\textrm{ti}}::\mathds{Q} \|}{\|\mathcal{P}^{\textrm{ti}}::\mathds{Q}\|},
\end{equation}
and present the result in Fig.~\ref{fig:plot_t_isotropy_residue}. For the volume element size of $64^3$, the transverse isotropy error of the \Qtensor{} begins at $3.8\%$, then decreases to $0.56\%$ and $0.41\%$ as the volume element size increases to $128^3$ and $256^3$, respectively. Except the middle data point, the \Qtensor{} converges at the rate $L^{-3/2}$ to a transversely isotropic tensor \eqref{eq:special_trans_isotropic_Qtensor}, characterized by five independent coefficients.

%\begin{figure}[!h]
%	\centering
%	\begin{minipage}[b]{0.3\textwidth}
%		\includegraphics[width=1\textwidth]{data/Trans/plot_Qtensor_visualization_Y_64.pdf}
%		\subcaption{L=64 $\mu m$}
%		\label{fig:plot_Young_zOrient_64}
%	\end{minipage}
%	\begin{minipage}[b]{0.3\textwidth}
%		\includegraphics[width=1\textwidth]{data/Trans/plot_Qtensor_visualization_Y_128.pdf}
%		\subcaption{L=128 $\mu m$}
%		\label{fig:plot_Young_zOrient_128}
%	\end{minipage}	
%	\begin{minipage}[b]{0.3\textwidth}
%		\includegraphics[width=1\textwidth]{data/Trans/plot_Qtensor_visualization_Y_256.pdf}
%		\subcaption{L=256 $\mu m$}
%		\label{fig:plot_Young_zOrient_256}
%	\end{minipage}	
%	
%	\begin{minipage}[b]{0.3\textwidth}
%		\includegraphics[width=1\textwidth]{data/Trans/plot_Qtensor_visualization_nu_ave_64.pdf}
%		\subcaption{L=64 $\mu m$}
%		\label{fig:plot_Poisson_average_zOrient_64}
%	\end{minipage}
%	\begin{minipage}[b]{0.3\textwidth}
%		\includegraphics[width=1\textwidth]{data/Trans/plot_Qtensor_visualization_nu_ave_128.pdf}
%		\subcaption{L=128 $\mu m$}
%		\label{fig:plot_Poisson_average_zOrient_128}
%	\end{minipage}	
%	\begin{minipage}[b]{0.3\textwidth}
%		\includegraphics[width=1\textwidth]{data/Trans/plot_Qtensor_visualization_nu_ave_256.pdf}
%		\subcaption{L=256 $\mu m$}
%		\label{fig:plot_Poisson_average_zOrient_256}
%	\end{minipage}	
%	\caption{Visualization of transversely isotropic \Qtensor{} for increasing volume element size. (a-b-c) Young's-modulus-like coefficient. (d-e-f) Poisson's-ratio-like coefficient.}
%	\label{fig:tran-muQ-graphical}		
%\end{figure}
\subsection{An orthotropic ensemble}
\label{sec:computation_orthotropy}
%\begin{figure}[!h]
%	\centering
%	\includegraphics[width=1\textwidth, trim = 0 0 0 0, clip]{figures/Ortho_Fiber/Combined2.eps}
%	\caption{Orthotropic volume elements of short fibers of 100$\mu$m length, 10$\mu$m diameter with 10$\%$ volume fraction. Volume element sizes from left to right: $L=64\,\mu m$, $L=128\,\mu m$, $L=256\,\mu m$ and $L=512\,\mu m$.}
%	\label{fig:ortho-fiber}
%\end{figure}
\begin{figure}[h!]
	\begin{subfigure}{.098\textwidth}
		\includegraphics[trim = 50 20 1850 920, clip, width=\textwidth]{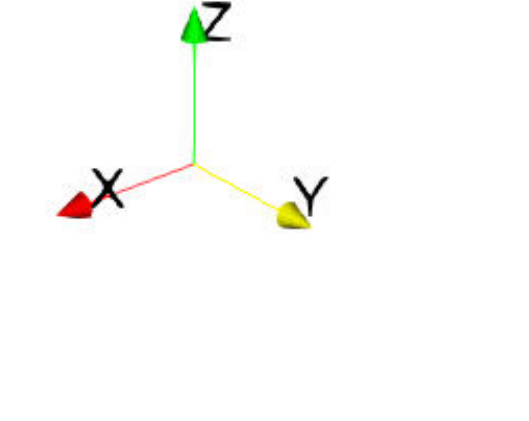}
	\end{subfigure}
	\begin{subfigure}{.151\textwidth}
		\includegraphics[trim = 600 20 480 125, clip, width=\textwidth]{data/Ortho/ortho_64.pdf}
		\caption{$64^3$}
		\label{fig:ortho_64}
	\end{subfigure}
	\begin{subfigure}{.285\textwidth}
		\includegraphics[trim = 600 20 580 125, clip, width=\textwidth]{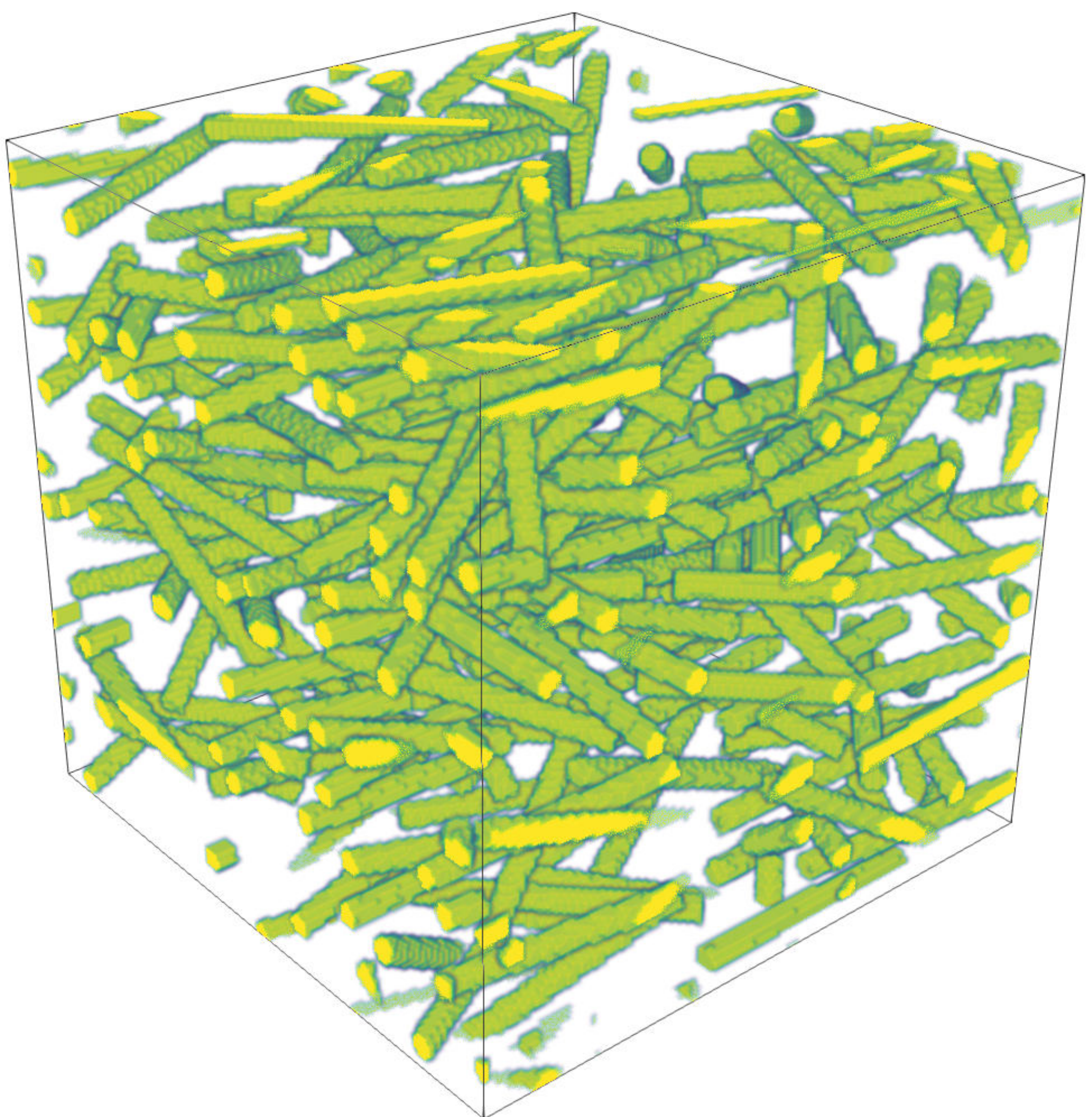}
		\caption{$128^3$}
		\label{fig:ortho_128}
	\end{subfigure}
	\begin{subfigure}{.48\textwidth}
		\includegraphics[trim = 600 20 580 125, clip, width=\textwidth]{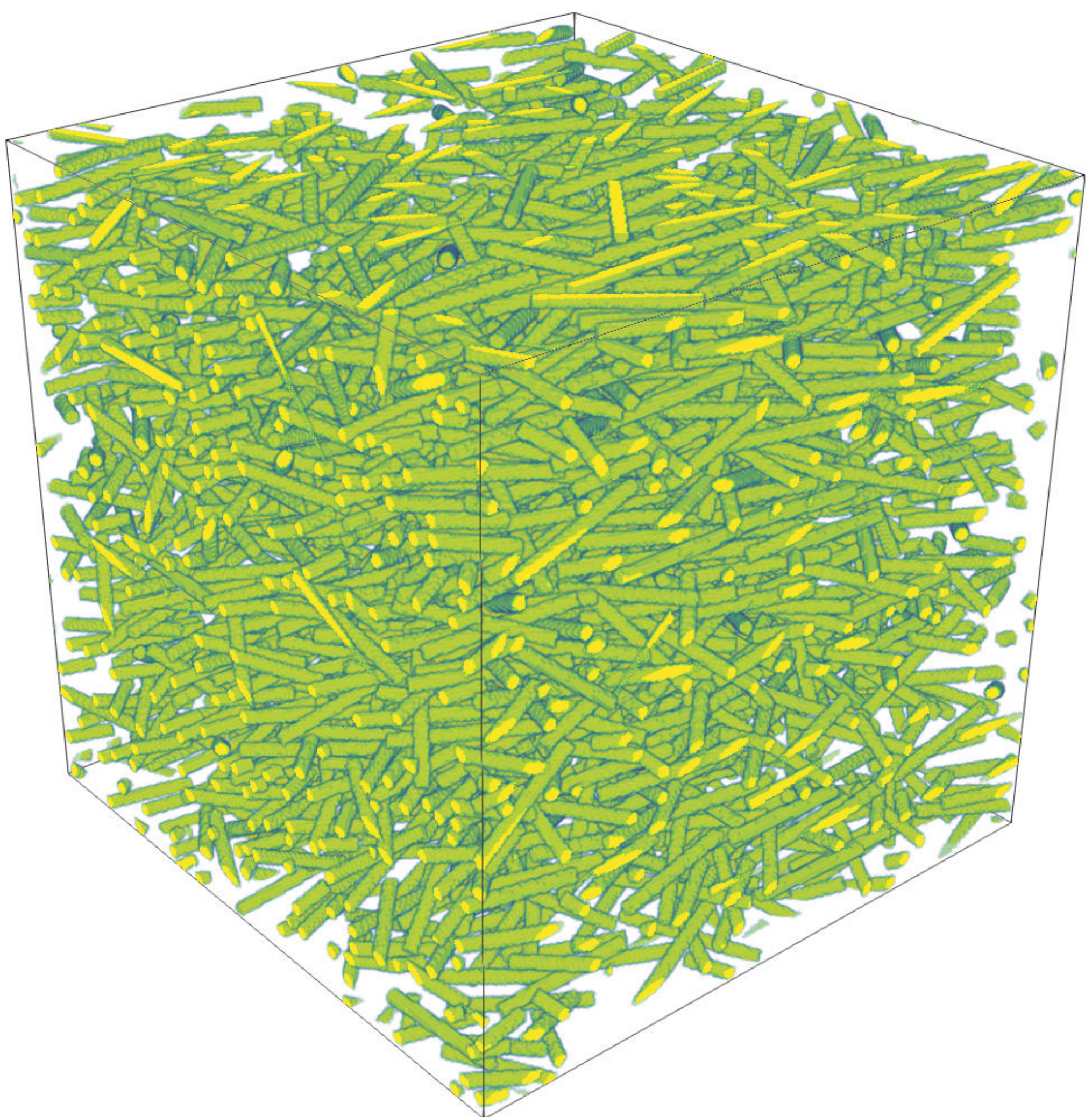}
		\caption{$256^3$}
		\label{fig:ortho_256}
	\end{subfigure}
	\caption{Volume elements of the orthotropically distributed fibers.}
	\label{fig:ortho-fiber}
\end{figure}
\begin{figure}[!h]
	\centering	
	\begin{minipage}[b]{0.3\textwidth}
		\centering
		\includegraphics[width=1\textwidth]{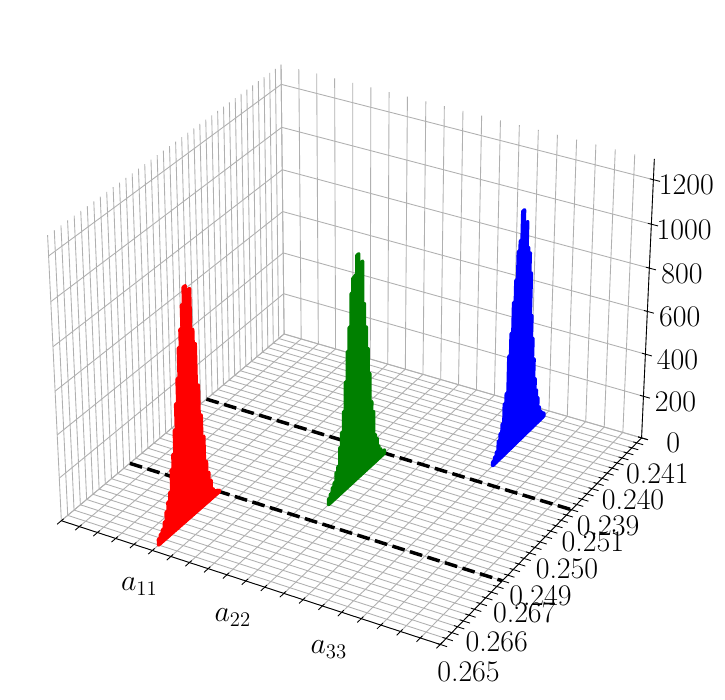}
		\subcaption{L=128\,$\mu m$}
		\label{fig:plot_histogram_diag_Ortho2_64}
	\end{minipage}	
	\begin{minipage}[b]{0.3\textwidth}
		\centering
		\includegraphics[width=1\textwidth]{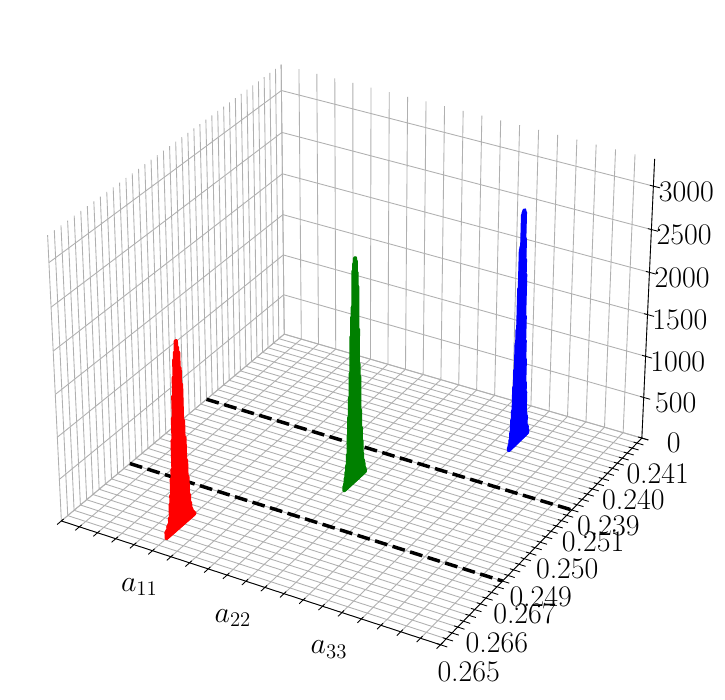}
		\subcaption{L=256\,$\mu m$}
		\label{fig:plot_histogram_diag_Ortho2_128}
	\end{minipage}		
	\begin{minipage}[b]{0.3\textwidth}
		\centering
		\includegraphics[width=1\textwidth]{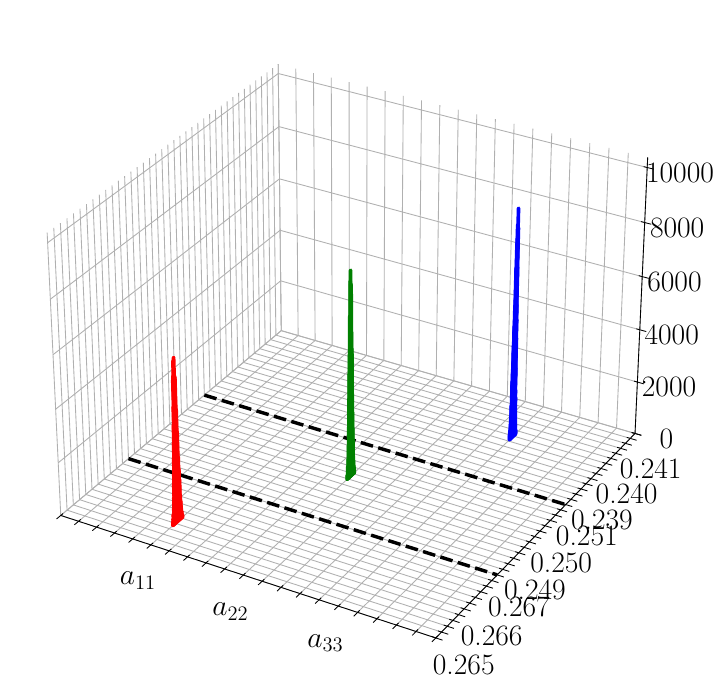}
		\subcaption{L=512\,$\mu m$}
		\label{fig:plot_histogram_diag_Ortho2_256}
	\end{minipage}		
	
	\begin{minipage}[b]{0.3\textwidth}
		\centering
		\includegraphics[width=1\textwidth]{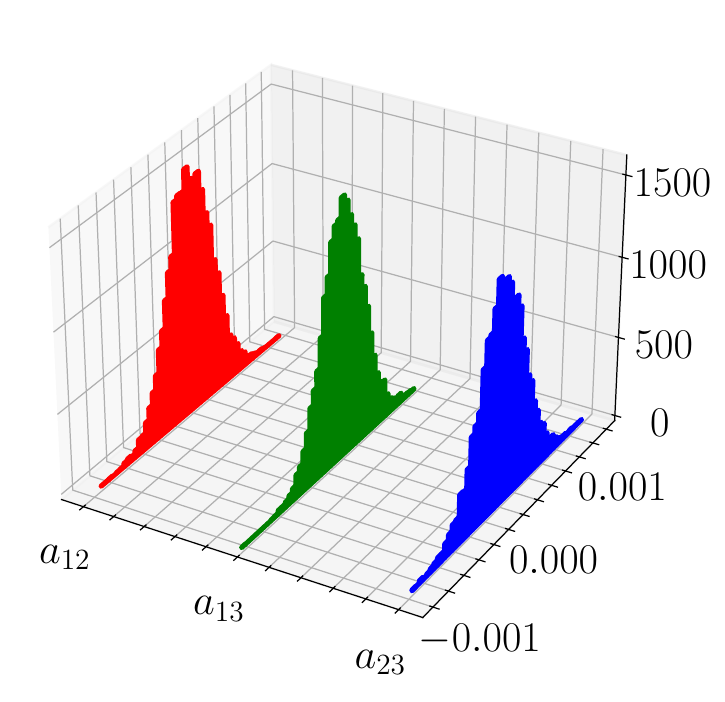}
		\subcaption{L=128\,$\mu m$}
		\label{fig:plot_histogram_offdiag_Ortho2_64}
	\end{minipage}	
	\begin{minipage}[b]{0.3\textwidth}
		\centering
		\includegraphics[width=1\textwidth]{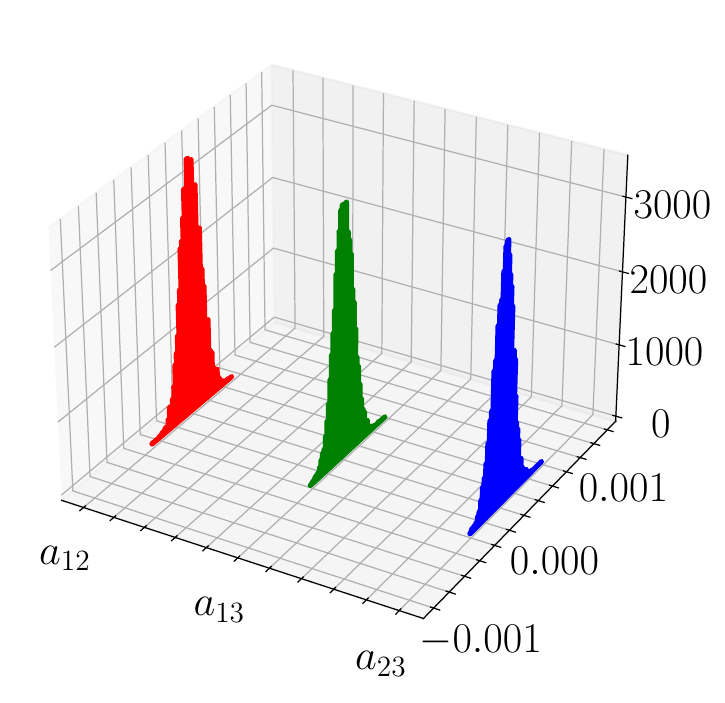}
		\subcaption{L=256\,$\mu m$}
		\label{fig:plot_histogram_offdiag_Ortho2_128}
	\end{minipage}		
	\begin{minipage}[b]{0.3\textwidth}
		\centering
		\includegraphics[width=1\textwidth]{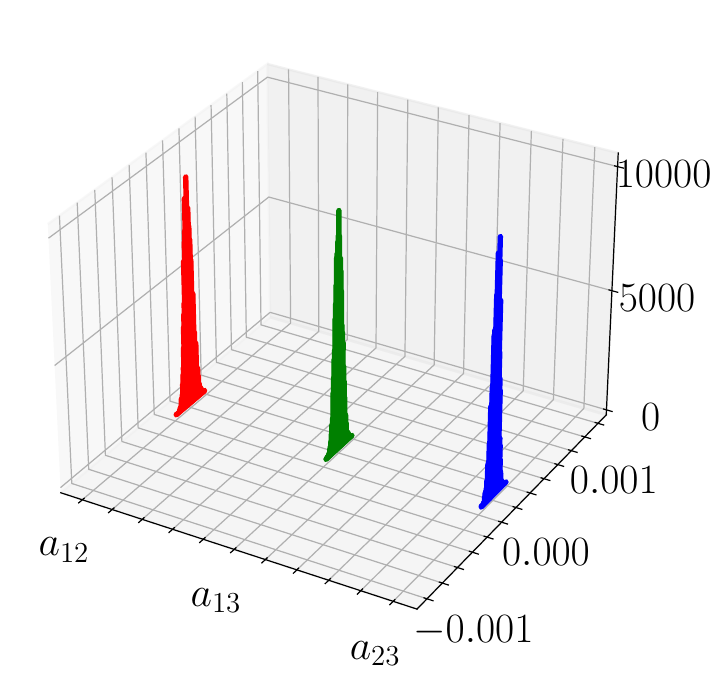}
		\subcaption{L=512\,$\mu m$}
		\label{fig:plot_histogram_offdiag_Ortho2_256}
	\end{minipage}	
	\caption{Histogram of orthotropic apparent thermal conductivity coefficients for increasing volume element sizes. (a-b-c) Diagonal coefficients $a_{11}$, $a_{22}$ and $a_{33}$; (d-e-f) Off-diagonal coefficients $a_{23}$, $a_{13}$ and $a_{12}$. The means and variances of these coefficients are presented in Table \ref{tab:orthotropic-ensemble}. }
	\label{fig:ortho-fiber-histogram-apparent}
\end{figure}
This example considers an ensemble of orthotropically distributed fibers, described via the second-order fiber orientation tensor
\begin{equation}
	\label{eq:computation-orthotropy-orientation-tensor}
	\textbf{\texttt{A}}^{\texttt{orth}} = \mqty[0.6 & 0 & 0 \\ 0 & 0.3 & 0 \\ 0 & 0 & 0.1],
\end{equation}
whose orthotropy axes are aligned with the computational cell. Some representative realizations of this ensemble are shown in Fig.~\ref{fig:ortho-fiber}. The distributions of the computed apparent conductivity tensors are presented in Fig.~\ref{fig:ortho-fiber-histogram-apparent}, where the diagonal conductivity coefficients are distributed with three distinct mean values of $0.265\,W/(mK)$, $0.250\,W/(mK)$ and $0.240\,W/(mK)$ for $a_{11}$, $a_{22}$ and $a_{33}$, respectively. The detailed statistical results are tabulated in Table \ref{tab:orthotropic-ensemble}.
\begin{table}[h!]
	\caption{Asymptotic behavior of orthotropic apparent conductivity. The units of $L$, mean values and standard deviation are $\mu m$, $W/(mK)$ and $W/(mK)$, respectively.}
	\label{tab:orthotropic-ensemble}
	\centering
	\begin{tabular}{ll|rl|rrrrl}
		L & A & Mean & Std. & Sys. error & Proj.  & Rand. error & Proj. & CI error  \\
		& &  &  &  & sys. error &  &  rand. error & \\
		\hline\hline
		\multirow{6}{*}{128}& $a_{11}$ & $0.26578$ & $0.00041$ & 0.16808\% & 0.16808\% & 0.15535\% & 0.15535\% & 0.00400\%  \\
		& $a_{22}$ & $0.24973$ & $0.00041$ & 0.04574\% & 0.04574\% & 0.15324\% & 0.15324\% & 0.00395\%  \\
		& $a_{33}$ & $0.24001$ & $0.00039$ & 0.01529\% & 0.01529\% & 0.14653\% & 0.14653\% & 0.00377\%  \\
		& $a_{12}$ & $0.00000$ & $0.00027$ & 0.00061\% & -- & 0.10006\% & -- & 0.00258\%  \\
		& $a_{13}$ & $0.00000$ & $0.00026$ & 0.00028\% & -- & 0.09833\% & -- & 0.00253\%  \\
		& $a_{23}$ & $0.00000$ & $0.00030$ & 0.00058\% & -- & 0.11415\% & -- & 0.00294\%  \\
		\hline
		\multirow{6}{*}{256}& $a_{11}$ & $0.26542$ & $0.00019$ & 0.01784\% & 0.01784\% & 0.07151\% & 0.07151\% & 0.00184\%  \\
		& $a_{22}$ & $0.24963$ & $0.00015$ & 0.00460\% & 0.00460\% & 0.05793\% & 0.05793\% & 0.00149\%  \\
		& $a_{33}$ & $0.23998$ & $0.00014$ & 0.00185\% & 0.00185\% & 0.05381\% & 0.05381\% & 0.00139\%  \\
		& $a_{12}$ & $0.00000$ & $0.00012$ & 0.00045\% & -- & 0.04650\% & -- & 0.00120\%  \\
		& $a_{13}$ & $0.00000$ & $0.00012$ & 0.00035\% & -- & 0.04533\% & -- & 0.00117\%  \\
		& $a_{23}$ & $0.00000$ & $0.00012$ & 0.00035\% & -- & 0.04629\% & -- & 0.00119\%  \\
		\hline
		\multirow{6}{*}{512}& $a_{11}$ & $0.26539$ & $0.00007$ & 0.00251\% & 0.00251\% & 0.02539\% & 0.02539\% & 0.00073\%  \\
		& $a_{22}$ & $0.24962$ & $0.00005$ & 0.00058\% & 0.00058\% & 0.02044\% & 0.02044\% & 0.00059\%  \\
		& $a_{33}$ & $0.23998$ & $0.00005$ & 0.00022\% & 0.00022\% & 0.01840\% & 0.01840\% & 0.00053\%  \\
		& $a_{12}$ & $0.00000$ & $0.00005$ & 0.00032\% & -- & 0.01748\% & -- & 0.00050\%  \\
		& $a_{13}$ & $0.00000$ & $0.00004$ & 0.00008\% & -- & 0.01689\% & -- & 0.00049\%  \\
		& $a_{23}$ & $0.00000$ & $0.00004$ & 0.00034\% & -- & 0.01667\% & -- & 0.00048\%  \\
		\hline
		\multirow{6}{*}{2048}& $a_{11}$ & $0.26538$ & $0.00001$ & -- & -- & 0.00309\% & 0.00309\% & 0.00057\%  \\
		& $a_{22}$ & $0.24962$ & $0.00001$ & -- & -- & 0.00256\% & 0.00256\% & 0.00047\%  \\
		& $a_{33}$ & $0.23998$ & $0.00001$ & -- & -- & 0.00223\% & 0.00223\% & 0.00041\%  \\
		& $a_{12}$ & $0.00000$ & $0.00001$ & -- & -- & 0.00217\% & -- & 0.00040\%  \\
		& $a_{13}$ & $0.00000$ & $0.00001$ & -- & -- & 0.00200\% & -- & 0.00037\%  \\
		& $a_{23}$ & $0.00000$ & $0.00001$ & -- & -- & 0.00200\% & -- & 0.00037\%  \\
		\hline
	\end{tabular}
\end{table}
\begin{figure}[!h]
	\begin{center}
		\begin{subfigure}{\textwidth}
			\centering
			\includegraphics[height=.022\textheight]{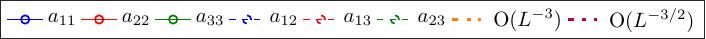}
		\end{subfigure}
	\end{center}
	\centering
	\begin{minipage}[b]{0.3\textwidth}
		\includegraphics[width=1\textwidth]{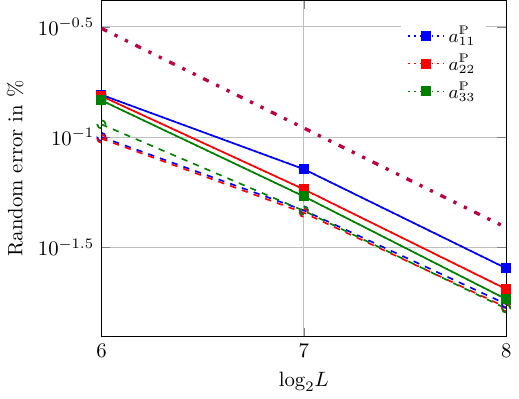}
		\subcaption{Random errors}
		\label{fig:Ortho_Fiber_plot_RandError_convergence}
	\end{minipage}
	\hfill
	\begin{minipage}[b]{0.3\textwidth}
		\includegraphics[width=1\textwidth]{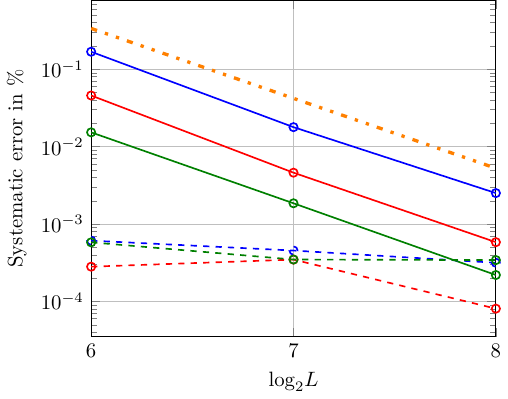}
		\subcaption{Systematic errors}
		\label{fig:Ortho_Fiber_plot_SysError_convergence}
	\end{minipage}
	\hfill
	\begin{minipage}[b]{0.3\textwidth}
		\includegraphics[width=1\textwidth]{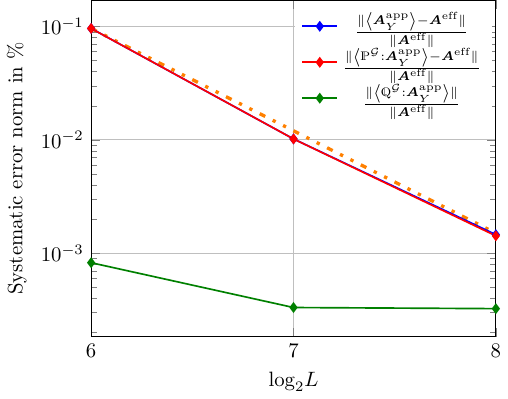}
		\subcaption{Norm of systematic errors}
		\label{fig:Ortho_Fiber_plot_normSysError_convergence}
	\end{minipage}
	\caption{Convergence of: (a) Random errors; (b) Systematic errors; (c) Frobenius norm of systematic errors. $a^{\mathds{P}}_{11},a^{\mathds{P}}_{22},a^{\mathds{P}}_{33}$ denote the diagonal components of the projected empirical mean \eqref{eq:setup-empirical-mean-projected}.}
	\label{fig:Ortho_Fiber_sys_rand_errors}		
\end{figure}

The considered number of realizations is still exceedingly sufficient in this case, with the confidence interval error in the order of $4\times10^{-3}\,\%$ and below. For the smallest volume element size of $64^3$, we again observe that the random errors are consistently larger than their systematic counterparts by one order of magnitude, except for the coefficient $a_{11}$ whose random error is actually $10\%$ smaller. As the volume element size increases, this exception is no longer observed and the random error converges with the rate $L^{-3/2}$, as expected, see Fig.~\ref{fig:Ortho_Fiber_plot_RandError_convergence}. On the other hand, the systematic errors of the diagonal conductivity coefficients $a_{11}$, $a_{22}$ and $a_{33}$ are persistently decreasing in that order by the factors of 4 and 3, presumably related to the description of the fiber orientation $\textbf{\texttt{A}}^{\texttt{orth}}$ \eqref{eq:computation-orthotropy-orientation-tensor}. Moreover, their systematic errors are one to two orders of magnitude larger than those of the off-diagonal coefficients. Hence, in terms of scaling, the systematic errors of the diagonal coefficients converge with the rate $L^{-3}$, as expected, meanwhile the off-diagonal coefficients do not follow this rate, as shown in Fig.~\ref{fig:Ortho_Fiber_plot_normSysError_convergence}. In the next step, we apply the orthotropic projection $\mathds{P}^{\textrm{orth}}$ \eqref{eq:SymmetryInformedStrats_projectors_SymD_explicitProjectors_ortho} to the apparent conductivity tensors. This action does not affect both random and systematic errors as their corresponding projected errors are unchanged for all volume element sizes. As also shown in Fig.~\ref{fig:Ortho_Fiber_plot_RandError_convergence} by the blue, red and green dotted lines with square markers, the random errors of the projected empirical mean of are identical to the original values. Apparently, the projected random errors are scaled by $L^{-3/2}$. In terms of the Frobenius norm, the projected systematic error are not improved for volume element sizes of $64^3$ and $128^3$, while a small improvement of $2.5\%$ is observed for the volume size of $256^3$ as shown in Fig.~\,\ref{fig:Ortho_Fiber_plot_normSysError_convergence}. Additionally, we also provide in this figure the empirical average of symmetry error as the lower bound of the systematic error, shown by the green curve. 
\begin{figure}[!h]
	\begin{center}
		\begin{subfigure}{\textwidth}
			\centering
			\includegraphics[height=.033\textheight]{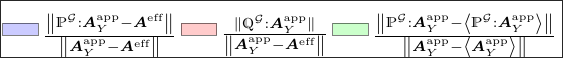}
		\end{subfigure}
	\end{center}
	\centering
	\begin{minipage}[b]{0.3\textwidth}
		\includegraphics[width=1\textwidth]{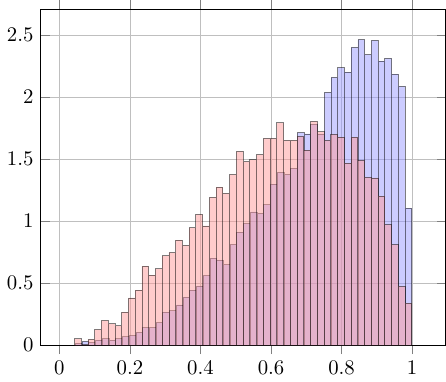}
		\subcaption{$L=128\mu m$}
		\label{fig:plot_histogram_lowerBound_ratio_Ortho_64}
	\end{minipage}
	\hfill
	\begin{minipage}[b]{0.3\textwidth}
		\includegraphics[width=1\textwidth]{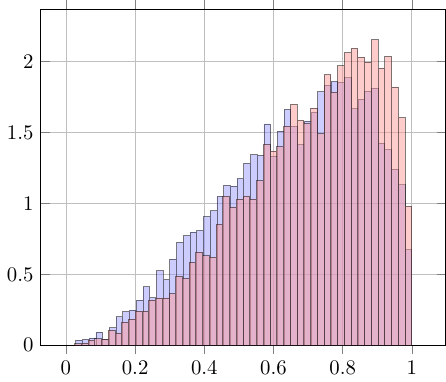}
		\subcaption{$L=256\mu m$}
		\label{fig:plot_histogram_lowerBound_ratio_Ortho_128}
	\end{minipage}
	\hfill
	\begin{minipage}[b]{0.3\textwidth}
		\includegraphics[width=1\textwidth]{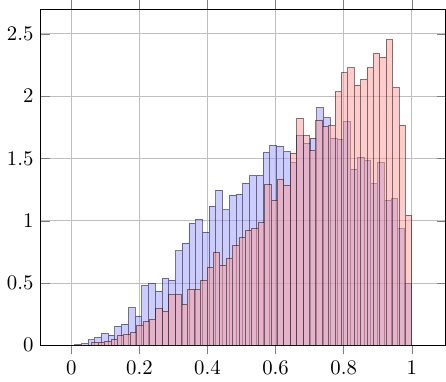}
		\subcaption{$L=512\mu m$}
		\label{fig:plot_histogram_lowerBound_ratio_Ortho_256}
	\end{minipage}
	
	\begin{minipage}[b]{0.3\textwidth}
		\includegraphics[width=1\textwidth]{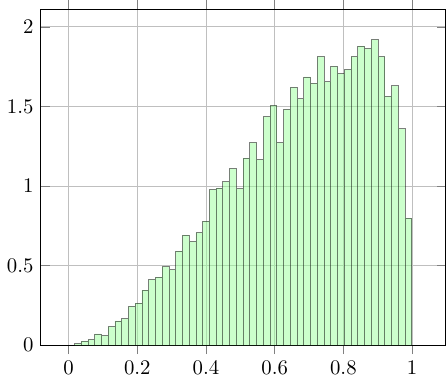}
		\subcaption{$L=128\mu m$}
		\label{fig:plot_ProjectedDispersionError_ratio_Ortho_64}
	\end{minipage}
	\hfill
	\begin{minipage}[b]{0.3\textwidth}<
		\includegraphics[width=1\textwidth]{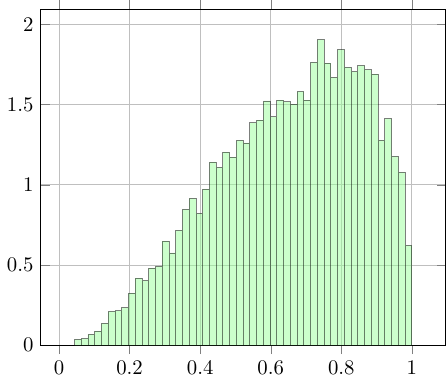}
		\subcaption{$L=256\mu m$}
		\label{fig:plot_ProjectedDispersionError_ratio_Ortho_128}
	\end{minipage}
	\hfill
	\begin{minipage}[b]{0.3\textwidth}
		\includegraphics[width=1\textwidth]{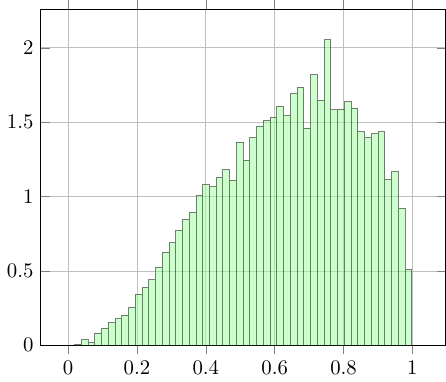}
		\subcaption{$L=512\mu m$}
		\label{fig:plot_ProjectedDispersionError_ratio_Ortho_256}
	\end{minipage}	
	\caption{Frequency density of the projection-reduced total error $\left\| \mean{\mathds{P}^{\mathcal{G}}: \fA^{\app}_Y} - \fA^{\eff} \right\|/\left\| \mean{\fA^{\app}_Y} - \fA^{\eff} \right\|$, symmetry error $\left\| \mathds{Q}^{\mathcal{G}}: \fA^{\app}_Y  \right\|/\left\| \mean{\fA^{\app}_Y} - \fA^{\eff} \right\|$ and dispersion error $\left\| \mathds{P}^{\mathcal{G}}: \fA^{\app}_Y - \mean{\mathds{P}^{\mathcal{G}}: \fA^{\app}_Y} \right\|/\left\| \fA^{\app}_Y - \mean{\fA^{\app}_Y} \right\|$, in blue, red and green, respectively.}
	\label{fig:Ortho_Fiber_total_dispersion_errors}		
\end{figure}

Subsequently, we analyze the effect of the orthotropic projection $\mathds{P}^{\textrm{orth}}$ \eqref{eq:SymmetryInformedStrats_projectors_SymD_explicitProjectors_ortho} on the total error \eqref{eq:SymmetryInformedStrats_postprocessing_totalError} and the dispersion error \eqref{eq:SymmetryInformedStrats_postprocessing_dispersion}. We plot in Fig.~\ref{fig:Ortho_Fiber_total_dispersion_errors} the normalized distribution of their respective ratios, which are bound between 0 and 1 for all computed realizations. The projected total error ratios, shown in blue, are right-skewed and peak around 0.9 for volume element size of $64^3$, slightly shift to 0.8 and 0.75 for the increasing volume element size of $128^3$ and $256^3$, respectively, see Figs.~\ref{fig:plot_histogram_lowerBound_ratio_Ortho_64}-\ref{fig:plot_histogram_lowerBound_ratio_Ortho_256}. Such distributions show that the orthotropic projection $\mathds{P}^{\textrm{orth}}$ provide a modest improvement for the total error. Meanwhile, we also observe in these figures the distributions of the lower bound of the total errors, shown in red. Their distributions are also right-skewed but do not peak at 1 as observed in the two previous ensembles. In addition, the dispersion error ratios, see Figs.~\ref{fig:plot_ProjectedDispersionError_ratio_Ortho_64}-\ref{fig:plot_ProjectedDispersionError_ratio_Ortho_256}, exhibit the similar distribution, namely skewing to the right and peaking around 0.8-0.9. The reason behind such distributions is because of the alignment between the orthotropy axes and the computational cell, reducing the effectiveness of the orthotropic projection.  

\begin{figure}[!h]
	\centering
	\begin{minipage}[b]{0.32\textwidth}
		\includegraphics[width=1\textwidth]{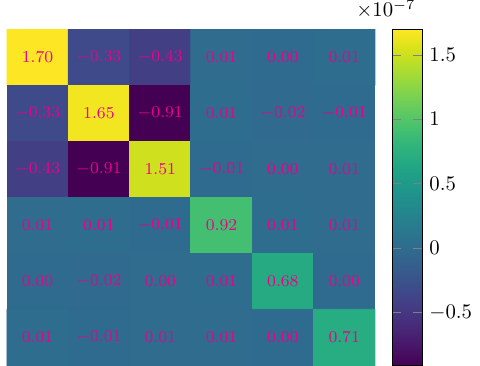}		
		\subcaption{$L=128\mu m$}
		\label{fig:plot_Qtensor_Ortho_64}
	\end{minipage}
	\hfill
	\begin{minipage}[b]{0.3\textwidth}
		\includegraphics[width=1\textwidth]{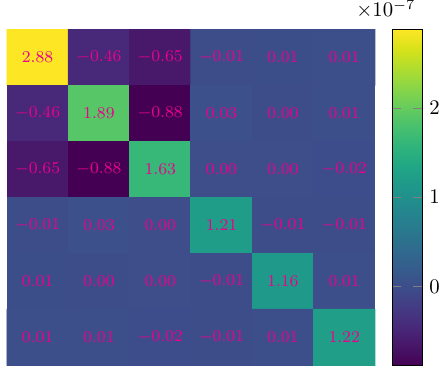}		
		\subcaption{$L=256\mu m$}
		\label{fig:plot_Qtensor_Ortho_128}
	\end{minipage}	
	\hfill
	\begin{minipage}[b]{0.3\textwidth}
		\includegraphics[width=1\textwidth]{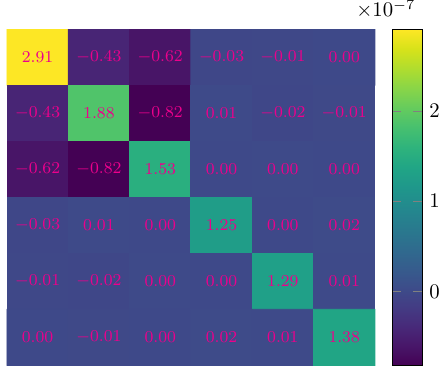}		
		\subcaption{$L=512\mu m$}
		\label{fig:plot_Qtensor_Ortho_256}
	\end{minipage}	
	\caption{\Qtensor{} of the orthotropic ensemble in matrix form. Unit of the \Qtensor{} components is $(W^2/(m^3K^2))$.}
	\label{fig:Ortho-muQ-tensor-matrix}		
\end{figure}
\begin{figure}[!h]
	\begin{subfigure}{\textwidth}
		\centering
		\includegraphics[height=.022\textheight]{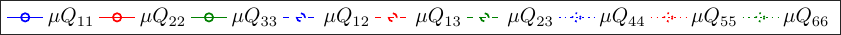}
	\end{subfigure}
	\centering
	\begin{minipage}[b]{0.45\textwidth}
		\includegraphics[width=1\textwidth]{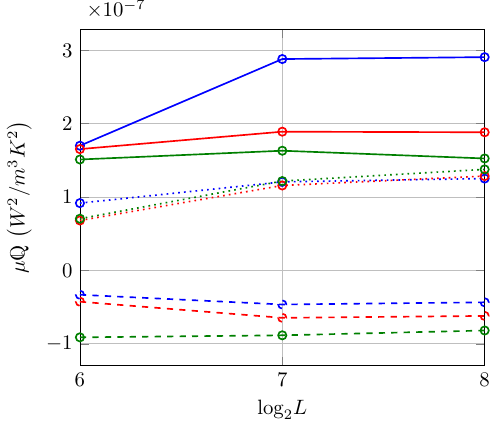}		
		\subcaption{\Qtensor{} components}
		\label{fig:ortho_plot_convergence_Qtensor_components}
	\end{minipage}
	\hfill
	\begin{minipage}[b]{0.45\textwidth}
		\includegraphics[width=1\textwidth]{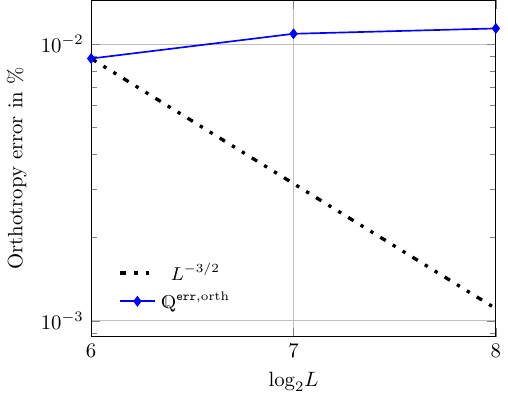}
		\subcaption{Orthotropy residual}
		\label{fig:plot_orthotropy_residue}
	\end{minipage}		
	\caption{(a) Leading-order components of the \Qtensor{}. (b) Convergence of the orthotropy error \eqref{eq:computation-orthotropy-error-Qtensor}. }
	\label{fig:ortho-muQ-tensor-components}		
\end{figure}
Concerning the \Qtensor{} of an orthotropic ensemble, we plot its matrix form in Fig.~\ref{fig:Ortho-muQ-tensor-matrix}. We again focus on the nine leading-order coefficients of the \Qtensor{}, where we observe the dominance of the auto-correlations between diagonal conductivity coefficients as compared to those of the off-diagonal ones, representing by the magnitudes of the diagonal components of the \Qtensor{}. Moreover, the correlations among the diagonal conductivity coefficients are again negative. Fig.~\ref{fig:ortho_plot_convergence_Qtensor_components} presents these leading-order components of the \Qtensor{}, indicating convergence after the volume element size of $128^3$. In case of orthotropic symmetry, the orthotropy error
\begin{equation}
	\label{eq:computation-orthotropy-error-Qtensor}
	\mathds{Q}^{\texttt{err},\textrm{orth}} = \frac{\|\mathds{Q} - \mathcal{P}^{\textrm{orth}}::\mathds{Q} \|}{\|\mathcal{P}^{\textrm{orth}}::\mathds{Q}\|},
\end{equation}
should vanish in expectation. In Fig.~\ref{fig:plot_orthotropy_residue}, we observe that the orthotropy error is indeed very small, two to three orders of magnitude smaller than the isotropy error \eqref{eq:computation-isotropic-isotropy-error-Qtensor} and the transverse isotropy error \eqref{eq:computation-trans-isotropic-isotropy-error-Qtensor}, starting at $8.86\times 10^{-3}\,\%$ for the volume element of size $64^3$ but increasing to $1.10\times10^{-2}\,\%$ and $1.14\times10^{-2}\,\%$ for the volume element of sizes $128^3$ and $256^3$, respectively.

%\begin{figure}
%	\centering
%	\begin{minipage}[b]{0.3\textwidth}
%		\includegraphics[width=1\textwidth]{data/Ortho/plot_Qtensor_visualization_Y_64.pdf}
%		\subcaption{}
%		\label{fig:plot_Young_Ortho2_64}
%	\end{minipage}
%	\begin{minipage}[b]{0.3\textwidth}
%		\includegraphics[width=1\textwidth]{data/Ortho/plot_Qtensor_visualization_Y_128.pdf}
%		\subcaption{}
%		\label{fig:plot_Young_Ortho2_128}
%	\end{minipage}	
%	\begin{minipage}[b]{0.3\textwidth}
%		\includegraphics[width=1\textwidth]{data/Ortho/plot_Qtensor_visualization_Y_256.pdf}
%		\subcaption{}
%		\label{fig:plot_Young_Ortho2_256}
%	\end{minipage}	
%	
%	\begin{minipage}[b]{0.3\textwidth}
%		\includegraphics[width=1\textwidth]{data/Trans/plot_Qtensor_visualization_nu_ave_64.pdf}
%		\subcaption{}
%		\label{fig:plot_Poisson_average_Ortho2_64}
%	\end{minipage}
%	\begin{minipage}[b]{0.3\textwidth}
%		\includegraphics[width=1\textwidth]{data/Trans/plot_Qtensor_visualization_nu_ave_128.pdf}
%		\subcaption{}
%		\label{fig:plot_Poisson_average_Ortho2_128}
%	\end{minipage}	
%	\begin{minipage}[b]{0.3\textwidth}
%		\includegraphics[width=1\textwidth]{data/Trans/plot_Qtensor_visualization_nu_ave_256.pdf}
%		\subcaption{}
%		\label{fig:plot_Poisson_average_Ortho2_256}
%	\end{minipage}	
%	\caption{Orthotropic \Qtensor{} representation}
%	\label{fig:ortho-muQ-graphical}		
%\end{figure}

%%%%%%%%%%%%%%%%%%%%%%%%%%%%%%%%%%%%%%%%%%%%%%%%%%%%%%%%%%%%%%%%%%%%%%%%%%%%%%%%%%%%%%%%%%%%%%%%%%%%%%%%%%%%%%%%%%%%%%%%%%%%%%%%%%%%%%%%%%%%%%%%%%%%%%%%%%%%%%%%%%%%%%%%%%%%%%%%%%%%%%%%%%%%

%\newpage

\section{Conclusion}
\label{sec:conclusion}
This paper explored the ramifications of the ensemble symmetry in stochastic homogenization of conductivity problems and provided a post-processing strategy to improve the RVE method. Upon investigating isotropic, transversely isotropic and orthotropic ensembles of short fiber-reinforced composite, we make the following remarks:
\begin{enumerate}
	\item For a cubic cell, the symmetry-informed projection technique is most effective for an isotropic ensemble, moderately effective for a transversely isotropic ensemble and least effective for an orthotropic ensemble, at least in cases the axes of orthotropy align with the edges of the cell. The latter is not surprising, as no symmetry breaking occurs in this case.
	\item The projection strategy helps to retain the convenience of the cubic cell and avoid the spherical volume elements~\cite{Gluege2012SphericalCell,Gluege2013SphericalCell} in the case isotropic ensemble or cylindrical volume element in the case of transversely isotropic ensemble. In particular, we the devised projection method leads to a highly effective, low-cost alternative to the spherical unit-cell approach.
	\item The random errors are reduced by factors three to five, while the systematic errors are not affected by the projection. Thus, we recommend to apply the post-processing projection strategy on the apparent conductivity tensors for better approximation without additional computational cost. \review{Although our theoretical considerations were perfectly general, our computational considerations were not. In particular, the effects of the introduced symmetry-enforcing technique on microstructures with a higher variety in microstructural characteristics such as volume fraction, fiber length or diameter distribution may be considered.}
	\item The symmetry-informed projection technique may be complementary to other variance reduction techniques such as the antithetic variables or control variates~\cite{costaouec2010variance,legoll2015control,blanc2016some}.
	\item The projected errors attain the same optimal decay rate as their original errors. Please note that we only employ the periodized ensemble for the work at hand, the performance of post-processing projection in snap-shot ensembles~\cite{schneider2022representative} \review{and under Dirichlet or Neumann boundary conditions \cite{risthaus2024imposing,risthaus2024fft}} could be interesting to investigate.
	\item The projection technique also works on the \Qtensor{} and helps to recover its optimal convergence rate for the isotropic and transversely isotropic ensemble.
	\item \review{The paper at hand focuses on thermal conductivity. Extending the symmetry-informed projection framework to linear elasticity is conceptually straightforward, and we refer the the literature~\cite{danescu1997number,ecker2024haar,d2024representation} for the relevant orthogonal projectors enforcing specific symmetries of the apparent elastic tensor. Nevertheless, handling the eighth-order \Qtensor{} associated to linear elasticity requires extra effort.}
\end{enumerate}

\section*{Acknowledgements}

BHN and MS acknowledge support from the European Research Council within the Horizon Europe program - project 101040238. Fruitful discussions with Maximilian Krause (UDE) and Felix Otto (MPI MiS) are acknowledged. \review{We thank the anonymous referees for taking the time to review the manuscript and for providing constructive feedback. We also thank Giulio G. Giusteri for pointing out the correction to the reference~\cite{giusteri2022periodic}.}

\section*{Data availability statement}

The data that support the findings of this study are available from the corresponding author upon reasonable request.

%%%%%%%%%%%%%%%%%%%%%%%%%%%%%%%%%%%%%%%%%%%%%%%%%%%%%%%%%%%%%%%%%%%%%%%%%%%%%%%%%%%%%%%%%%%%%%%%%%%%%%%%%%%%%%%%%%%%%%%%%%%%%%%%%%%%%%%%%%%%%%%%%%%%%%%%%%%%%%%%%%%%%%%%%%%%%%%%%%%%%%%%%%%%

\appendix

\section{Arguments for specific covariance and invariance properties}
\label{apx:theory}

\subsection{Covariance of correctors on symmetric cells}
\label{apx:theory_covariantCorrectors}

The goal of this section is to show the validity of the statement \eqref{eq:theory_unitCells_covariance_corrector}
\begin{equation}\label{eq:apx_theory_covariantCorrectors1}
	\phi_{Y,\fA^{\fQ},\fQ\bar{\fxi}}(\fQ\fx) = \phi_{Y,\fA,\bar{\fxi}}(\fx)
\end{equation}
for the corrector \eqref{eq:theory_unitCells_correctorEqStrong} on the unit cell $Y$, where $\fQ \in O_Y(d)$ stands for a cell-preserving orthogonal transformation \eqref{eq:theory_unitCells_symmetryGroupCell_orthogonal}, $\fA \in \mathcal{A}$ denotes a fixed conductivity tensor, $\bar{\fxi} \in \R^d$ refers to a given vector and the equation \eqref{eq:apx_theory_covariantCorrectors1} should be valid for almost every $\fx \in Y$.\\
For this purpose, we first write down the weak form of the equation \eqref{eq:theory_unitCells_correctorEqStrong} which characterizes the unique solution $\phi_{Y,\fA,\bar{\fxi}}(\fx)$: Find the scalar field $\phi \in H^1_0(Y)$, s.t. the equation
\begin{equation}\label{eq:apx_theory_covariantCorrectors2}
	\int_{Y} \nabla v (\fx) \cdot \fA(\fx) \left[ \bar{\fxi} + \nabla \phi(\fx) \right] \, d\fx = 0
\end{equation}
holds for all test fields $v \in H^1_0(Y)$. Similarly, the function $\psi = \phi_{Y,\fA^{\fQ},\fQ\bar{\fxi}}$ solves the problem
\begin{equation}\label{eq:apx_theory_covariantCorrectors3}
	\int_{Y} \nabla w (\fy) \cdot \fA^{\fQ}(\fy) \left[ \fQ\bar{\fxi} + \nabla \psi(\fy) \right] \, d\fy = 0
\end{equation}
for all fields $w \in H^1_0(Y)$, which we may also rewrite in the form
\begin{equation}\label{eq:apx_theory_covariantCorrectors4}
	0 = \int_{Y} \nabla w (\fy) \cdot \fQ\fA(\fQ^T \fy) \fQ^T \left[ \fQ\bar{\fxi} + \nabla \psi(\fy) \right] \, d\fy
	= \int_{Y} \fQ^T\nabla w (\fy) \cdot \fA(\fQ^T \fy) \left[ \bar{\fxi} + \fQ^T \nabla \psi(\fy) \right] \, d\fy
\end{equation}
in view of eq.~\eqref{eq:theory_setup_group_action_definition}. Next, we change variables via $\fx = \fQ^T \fy$ to observe
\begin{equation}\label{eq:apx_theory_covariantCorrectors5}
	0 = \int_{Y} \nabla w (\fy) \cdot \fQ\fA(\fQ^T \fy) \fQ^T \left[ \fQ\bar{\fxi} + \nabla \psi(\fy) \right] \, d\fy
	= \int_{Y} \fQ^T\nabla w (\fQ \fx) \cdot \fA(\fx) \left[ \bar{\fxi} + \fQ^T \nabla \psi(\fQ \fx) \right] \, d\fx,
\end{equation}
where we used that the transformation $\fQ$ is an element of the orthogonal group and thus preserves the Lebesgue measure. For any function $w \in H^1_0(Y)$, the transformed function
\begin{equation}\label{eq:apx_theory_covariantCorrectors6}
	w^{\fQ}:Y \rightarrow \R, \quad w^{\fQ}(\fx) = w(\fQ \fx),
\end{equation}
defines an element of the Sobolev space $H^1_0(Y)$ and satisfies the transformation formula
\begin{equation}\label{eq:apx_theory_covariantCorrectors7}
	\nabla w^{\fQ}(\fx) = \fQ^T \nabla w(\fQ\fx).
\end{equation}
In fact, for any fixed element $\fQ \in O(d)$, the prescription \eqref{eq:apx_theory_covariantCorrectors6} defines an isometric automorphism of the Hilbert space $H^1_0(Y)$.
Taking into account eq.~\eqref{eq:apx_theory_covariantCorrectors7}, we may re-write the equation \eqref{eq:apx_theory_covariantCorrectors5} in the form
\begin{equation}\label{eq:apx_theory_covariantCorrectors8}
	0 = \int_{Y} \nabla w^{\fQ} (\fx) \cdot \fA(\fx) \left[ \bar{\fxi} + \nabla \psi^{\fQ} \right] \, d\fx.
\end{equation}
Setting $w = v^{\fQ^T}$, i.e., a "{}change of coordinates"{} in the test space, we thus observe that the function $\left(\phi_{Y,\fA^{\fQ},\fQ\bar{\fxi}}\right)^{\fQ}$ solves the equation \eqref{eq:apx_theory_covariantCorrectors2}. Due to the uniqueness of the solution, the desired identity \eqref{eq:apx_theory_covariantCorrectors1} follows.

\subsection{Invariance properties of the effective conductivity}
\label{apx:theory_eff}

The purpose of this section is to validate the statement \eqref{eq:theory_unitCells_Aapp_invariance}
\begin{equation}\label{eq:apx_theory_eff1}
	\fQ \mean{\fA^{\app}_Y} \fQ^T = \mean{\fA^{\app}_Y}
\end{equation}
for any element $\fQ$ of the symmetry group $\mathcal{G}_Y$ \eqref{eq:theory_unitCells_symmetryGroupCell} under the condition \eqref{eq:theory_setup_invariance_assumption}.
To do so, we will show that the identity
\begin{equation}\label{eq:apx_theory_eff2}
	\bar{\fxi}^T\fQ \mean{\fA^{\app}_Y} \fQ^T\bar{\fxi} = \bar{\fxi}^T \mean{\fA^{\app}_Y} \bar{\fxi}
\end{equation}
holds for all vectors $\bar{\fxi} \in \R^d$. As the tensor $\mean{\fA^{\app}_Y}$ is symmetric, the condition \eqref{eq:apx_theory_eff2} implies the statement \eqref{eq:apx_theory_eff1}.\\
We define the integrable function
\begin{equation}\label{eq:apx_theory_eff3}
	F(\fA) = \mean{\bar{\fxi}^T \fA^{\app}_Y \bar{\fxi}}, \quad \fA \in \mathcal{A}.
\end{equation}
By assumption \eqref{eq:theory_setup_invariance_assumption}, the identity
\begin{equation}\label{eq:apx_theory_eff4}
	\mean{\bar{\fxi}^T \fA^{\app}_Y \bar{\fxi}} = F(\fA) = F(\fA^{\fQ}) = \mean{\bar{\fxi}^T \left(\fA^{\fQ} \right)^{\app}_Y \bar{\fxi}}
\end{equation}
follows. By definition \eqref{eq:theory_unitCells_Aapp}, we may write the term inside the expectation on the right-hand side in the form
\begin{equation}\label{eq:apx_theory_eff5}
	\begin{split}
		\bar{\fxi}^T \left(\fA^{\fQ} \right)^{\app}_Y \bar{\fxi} &= \bar{\fxi}^T \dashint_{Y} \fA^{\fQ}(\fx) (\bar{\fxi} + \nabla \phi_{Y,\fA^{\fQ},\bar{\fxi}}(\fx)) \, d\fx \\
		&= \bar{\fxi}^T \dashint_{Y} \fQ\fA(\fQ^T \fx) \fQ^T (\bar{\fxi} + \nabla \phi_{Y,\fA^{\fQ},\bar{\fxi}}(\fx)) \, d\fx\\
		&= \left( \fQ^T \bar{\fxi} \right)^T \dashint_{Y} \fA(\fQ^T \fx) (\fQ^T\bar{\fxi} + \fQ^T\nabla \phi_{Y,\fA^{\fQ},\bar{\fxi}}(\fx)) \, d\fx\\
		&= \left( \fQ^T \bar{\fxi} \right)^T \dashint_{Y} \fA(\fy) (\fQ^T\bar{\fxi} + \fQ^T\nabla \phi_{Y,\fA^{\fQ},\bar{\fxi}}(\fQ \fy)) \, d\fy\\
		&= \left( \fQ^T \bar{\fxi} \right)^T \dashint_{Y} \fA(\fy) (\fQ^T\bar{\fxi} + \nabla \phi_{Y,\fA,\fQ^T\bar{\fxi}}(\fy)) \, d\fy\\
		&= \left( \fQ^T \bar{\fxi} \right)^T  \fA^{\app}_Y \fQ^T \bar{\fxi}\\
		&= \bar{\fxi}^T  \fQ \fA^{\app}_Y \fQ^T \bar{\fxi},\\
	\end{split}
\end{equation}
where we used the definition \eqref{eq:theory_setup_group_action_definition} of the action, changed variables according to the rule $\fy = \fQ^T \fx$ and incorporated the previously established transformation rule \eqref{eq:theory_unitCells_covariance_corrector}.\\
For later use we also record the following consequence of eq.~\eqref{eq:apx_theory_eff5},
\begin{equation}\label{eq:apx_theory_eff6}
	\bar{\fxi}^T \left(\fA^{\fQ} \right)^{\app}_Y \bar{\feta} = \bar{\fxi}^T  \fQ \fA^{\app}_Y \fQ^T \bar{\feta}, \quad \bar{\fxi}, \bar{\feta} \in \R^d,
\end{equation}
that is obtained by using the symmetry of the apparent conductivity tensor. Indeed, suppose that two symmetric $d \times d$-tensors $\fB$ and $\fC$ satisfy the equation
\begin{equation}\label{eq:apx_theory_eff7}
	\bar{\fxi}^T \fB \bar{\fxi} = \bar{\fxi}^T \fC \bar{\fxi} \quad \text{for all} \quad \bar{\fxi} \in \R^d.
\end{equation}
In particular, we may replace the vector $\bar{\fxi}$ by the sum $\bar{\fxi} + \bar{\feta}$ and use the binomial formula to obtain
\begin{equation}\label{eq:apx_theory_eff8}
	\bar{\fxi}^T \fB \bar{\fxi} + 2\bar{\fxi}^T \fB \bar{\feta} + \bar{\feta}^T \fB \bar{\feta}  = \bar{\fxi}^T \fC \bar{\fxi} + 2\bar{\fxi}^T \fC \bar{\feta} + \bar{\feta}^T \fC \bar{\feta}.
\end{equation}
Due to the validity of the identity \eqref{eq:apx_theory_eff7} for both $\fxi$ and $\feta$, we are led to the identity
\begin{equation}\label{eq:apx_theory_eff9}
	\bar{\fxi}^T \fB \bar{\feta} = \bar{\fxi}^T \fC \bar{\feta}, \quad \bar{\fxi}, \bar{\feta} \in \R^d.
\end{equation}
Applying this insight to $\fB = \left(\fA^{\fQ} \right)^{\app}_Y$ and $\fC = \fQ \fA^{\app}_Y \fQ^T$ implies eq.~\eqref{eq:apx_theory_eff6}.\\
Taking expectations on both hands of the equation \eqref{eq:apx_theory_eff5}, we are led to the identity
\begin{equation}\label{eq:apx_theory_eff10}
	\mean{\bar{\fxi}^T \left(\fA^{\fQ} \right)^{\app}_Y \bar{\fxi}} = \mean{\bar{\fxi}^T  \fQ \fA^{\app}_Y \fQ^T \bar{\fxi}}.
\end{equation}
In view of eq.~\eqref{eq:apx_theory_eff4}, we deduce the result
\begin{equation}\label{eq:apx_theory_eff11}
	\mean{\bar{\fxi}^T \fA^{\app}_Y \bar{\fxi}} = \mean{\bar{\fxi}^T \left(\fA^{\fQ} \right)^{\app}_Y \bar{\fxi}} = \mean{\bar{\fxi}^T  \fQ \fA^{\app}_Y \fQ^T \bar{\fxi}},
\end{equation}
which was to be shown, see eq.~\eqref{eq:apx_theory_eff1}, taking into account that both the vector $\bar{\fxi} \in \R^d$ and the group element $\fQ \in \mathcal{G}_Y$ are constant.

\subsection{Invariance properties of the $\mu$Q-tensor}
\label{apx:theory_muQ}

In this section, we wish to show that the invariance property \eqref{eq:theory_unitCells_muQ_invariance}
\begin{equation}\label{eq:apx_theory_muQ1}
	\left(\Q_Y\right)^{\fQ} = \Q_Y
\end{equation}
holds for all elements $\fQ$ of the symmetry group $\mathcal{G}$ for the implicitly defined action
\begin{equation}\label{eq:apx_theory_muQ2}
	\fxi\otimes_s \feta : \Q^{\fQ}_Y : \fxi\otimes_s \feta = \left[(\fQ \fxi) \otimes_s (\fQ \feta)\right] : \Q_Y : \left[(\fQ \fxi) \otimes_s (\fQ \feta)\right], \quad \fxi, \feta \in \R^d,
\end{equation}
and the \Qtensor{} \eqref{eq:theory_unitCells_muQ} on the unit cell $Y$. For our purposes, it is actually convenient to consider the representation \eqref{eq:theory_unitCells_muQ2}
\begin{equation}\label{eq:apx_theory_muQ3}
	\Q_Y = \text{vol}(Y) \left( \mean{\fA^{\app}_Y \otimes \fA^{\app}_Y} - \mean{\fA^{\app}_Y} \otimes \mean{\fA^{\app}_Y} \right).
\end{equation}
We will establish the validity of the equations
\begin{equation}\label{eq:apx_theory_muQ4}
	\mean{\fA^{\app}_Y \otimes \fA^{\app}_Y} = \mean{\fA^{\app}_Y \otimes \fA^{\app}_Y}^{\fQ}
\end{equation}
and
\begin{equation}\label{eq:apx_theory_muQ5}
	\mean{\fA^{\app}_Y} \otimes \mean{\fA^{\app}_Y}  = \left(\mean{\fA^{\app}_Y} \otimes \mean{\fA^{\app}_Y} \right)^{\fQ}
\end{equation}
for all elements $\fQ$ of the symmetry group $\mathcal{G}_Y$ and almost every realization $\fA \in \mathcal{A}$. Then, the identity \eqref{eq:apx_theory_muQ1} is a direct consequence.\\
For any two vectors $\bar{\fxi}, \bar{\feta} \in \R^d$, we have
\begin{equation}\label{eq:apx_theory_muQ6}
	\bar{\fxi}\otimes_s \bar{\feta} : \left( \fA^{\app}_Y \otimes \fA^{\app}_Y \right) : \bar{\fxi}\otimes_s \bar{\feta} = \left( \bar{\fxi}^T \fA^{\app}_Y \bar{\feta} \right)^2.
\end{equation}
Defining the integrable function
\begin{equation}\label{eq:apx_theory_muQ7}
	F(\fA) = \mean{ \bar{\fxi}\otimes_s \bar{\feta} : \left( \fA^{\app}_Y \otimes \fA^{\app}_Y \right) : \bar{\fxi}\otimes_s \bar{\feta} },
\end{equation}
the invariance statement \eqref{eq:theory_setup_invariance_assumption}  implies the formula
\begin{equation}\label{eq:apx_theory_muQ8}
	\mean{ \bar{\fxi}\otimes_s \bar{\feta} : \left( \fA^{\app}_Y \otimes \fA^{\app}_Y \right) : \bar{\fxi}\otimes_s \bar{\feta} } = F(\fA) = F(\fA^{\fQ}) = \mean{\left( \bar{\fxi}^T \left(\fA^{\fQ} \right)^{\app}_Y \bar{\feta} \right)^2}.
\end{equation}
The valid statement \eqref{eq:apx_theory_eff6}
\begin{equation}\label{eq:apx_theory_muQ9}
	\bar{\fxi}^T \left(\fA^{\fQ} \right)^{\app}_Y \bar{\feta} = \bar{\fxi}^T  \fQ \fA^{\app}_Y \fQ^T \bar{\feta} \equiv \left( \fQ^T \bar{\fxi}\right)^T  \fA^{\app}_Y \fQ^T \bar{\feta} , \quad \bar{\fxi}, \bar{\feta} \in \R^d,
\end{equation}
implies directly
\begin{equation}\label{eq:apx_theory_muQ10}
	\begin{split}
		\mean{\left( \bar{\fxi}^T \left(\fA^{\fQ} \right)^{\app}_Y \bar{\feta} \right)^2} &= \mean{ \left( \left( \fQ^T \bar{\fxi}\right)^T  \fA^{\app}_Y \fQ^T \bar{\feta}\right)^2}\\
		&= \mean{ \left( \fQ^T \bar{\fxi} \otimes_s \fQ^T \bar{\feta} \right) : \left( \fA^{\app}_Y \otimes \fA^{\app}_Y \right) :   \left( \fQ^T \bar{\fxi} \otimes_s \fQ^T \bar{\feta} \right)}\\
		&= \mean{ \left( \bar{\fxi} \otimes_s \bar{\feta} \right) : \left( \fA^{\app}_Y \otimes \fA^{\app}_Y \right)^{\fQ^T} :   \left( \bar{\fxi} \otimes_s \bar{\feta} \right)}.
	\end{split}
\end{equation}
Thus, the identity \eqref{eq:apx_theory_muQ4} holds with the element $\fQ$ replaced by $\fQ^T \equiv \fQ^{-1}$.\\
A completely analogous argument applies to the integrable function
\begin{equation}\label{eq:apx_theory_muQ11}
	F(\fA) =  \bar{\fxi}\otimes_s \bar{\feta} : \left( \mean{\fA^{\app}_Y} \otimes \mean{\fA^{\app}_Y} \right) : \bar{\fxi}\otimes_s \bar{\feta}
\end{equation}
for fixed vectors $\bar{\fxi}, \bar{\feta} \in \R^d$ and invoke the invariance statement \eqref{eq:theory_setup_invariance_assumption}
\begin{equation}
	\begin{split}
	\label{eq:apx_theory_muQ12}
	 \bar{\fxi}\otimes_s \bar{\feta} : \left( \mean{\fA^{\app}_Y} \otimes \mean{\fA^{\app}_Y} \right) : \bar{\fxi}\otimes_s \bar{\feta} = F(\fA) & =  F(\fA^{\fQ}) \\
	 &=  \bar{\fxi}\otimes_s \bar{\feta} : \left( \mean{\qty(\fA^{\fQ})^{\app}_Y} \otimes \mean{\qty(\fA^{\fQ})^{\app}_Y} \right) : \bar{\fxi}\otimes_s \bar{\feta},
	 	\end{split}
\end{equation}
yielding the formula \eqref{eq:apx_theory_muQ5} from
\begin{equation}\label{eq:apx_theory_muQ13}
	\begin{split}
		\qty(\bar{\fxi}\otimes_s\bar{\feta}) : \qty(\mean{\fA^{\app}_Y} \otimes \mean{\fA^{\app}_Y} )^{\fQ^T} : \qty(\bar{\fxi}\otimes_s\bar{\feta}) &= \qty( \fQ^T \bar{\fxi} \otimes_s \fQ^T \bar{\feta}) : \qty(\mean{\fA^{\app}_Y} \otimes \mean{\fA^{\app}_Y} ) : \qty( \fQ^T \bar{\fxi} \otimes_s \fQ^T \bar{\feta}) \\
		&= \qty( \qty(\fQ^T \bar{\fxi})^T \fA^{\app}_Y \fQ^T \bar{\feta} )^2\\
		&= \qty(  \bar{\fxi} ^T \qty(\fA^{\fQ})^{\app}_Y \bar{\feta} )^2\\
		&= \qty(\bar{\fxi}\otimes_s\bar{\feta}) : \qty(\mean{ \qty(\fA^{\fQ})^{\app}_Y} \otimes \mean{ \qty(\fA^{\fQ})^{\app}_Y} )  : \qty(\bar{\fxi}\otimes_s\bar{\feta}).
	\end{split}
\end{equation}
%\begin{equation}\label{eq:apx_theory_muQ12}
%	\bar{\fxi}\otimes_s \bar{\feta} : \left( \mean{\fA^{\app}_Y} \otimes \mean{\fA^{\app}_Y} \right) : \bar{\fxi}\otimes_s \bar{\feta} = 
%	\bar{\fxi}\otimes_s \bar{\feta} : \left( \mean{\fA^{\app}_Y} \otimes \mean{\fA^{\app}_Y} \right)^{\fQ^T} : \bar{\fxi}\otimes_s \bar{\feta}.
%\end{equation}
%\section{Numerical results of apparent conductivities}
%\label{apx:numerical_apparent_conductivities}
%This section presents the collection of mean and variances of the isotropic, transversely isotropic and orthotropic apparent conductivity tensors presented from Section \ref{sec:computations}.

\section{Averaging the action over the group provides an orthogonal projector}
\label{sec:apx_theory_projector}

The purpose of this section is to show that the bounded linear operator 
\begin{equation}\label{eq:apx_theory_projector_defn}
	P \in L(V), \quad \fv\mapsto\int_{\mathcal{G}} \fL_{\fQ}\fv \, d\mu_{\mathcal{G}}(\fQ),
\end{equation}
defines an orthogonal projector onto the closed subspace of invariant vectors \eqref{eq:SymmetryInformedStrats_projectors_invariant_set}
\begin{equation}\label{eq:apx_theory_projector_invariantVector}
	V^{\mathcal{G}} = \left\{ \fv \in V \, \middle| \, \fL_{\fQ}\fv = \fv \quad \text{for all} \quad \fQ \in \mathcal{G} \right\},
\end{equation}
i.e., the identity \eqref{eq:SymmetryInformedStrats_projectors_projection_via_averaging} holds. To show this orthogonal projector property, we follow Fulton-Harris's proof~\cite{fulton2013representation} for the case of a finite group and consider three separate properties:
\begin{enumerate}
	\item For every element $\fw \in V$, the action of the operator $P$ defined in eq.~\eqref{eq:apx_theory_projector_defn} produces an element of the invariant set \eqref{eq:apx_theory_projector_invariantVector}, i.e.,
	\begin{equation}\label{eq:apx_theory_projector_statement1}
		P\fw \in V^{\mathcal{G}} \quad \text{for all} \quad \fw \in V.
	\end{equation}
	\item Every invariant element $\fv \in V^{\mathcal{G}}$ is preserved by the operator $P$, i.e.,
	\begin{equation}\label{eq:apx_theory_projector_statement2}
		P\fv = \fv \quad \text{for all} \quad \fv \in V^{\mathcal{G}}.
	\end{equation}
	\item The operator \eqref{eq:apx_theory_projector_defn} is self-adjoint \eqref{eq:SymmetryInformedStrats_projectors_self-adjoint_projector}, i.e., the condition
	\begin{equation}\label{eq:apx_theory_projector_statement3}
		\langle P \fv, \fw \rangle_V = \langle \fv, P \fw \rangle_V
	\end{equation}
	holds for all $\fv, \fw \in V$.
\end{enumerate}
Once these three properties are shown, the claimed projection property \eqref{eq:SymmetryInformedStrats_projectors_projection_via_averaging} follows. Indeed, by the first two properties \eqref{eq:apx_theory_projector_statement1} and \eqref{eq:apx_theory_projector_statement2}, the operator \eqref{eq:apx_theory_projector_defn} is a projector with image \eqref{eq:apx_theory_projector_invariantVector}. By the third property \eqref{eq:apx_theory_projector_statement2}, the operator \eqref{eq:apx_theory_projector_defn} is orthogonal \eqref{eq:SymmetryInformedStrats_projectors_self-adjoint_projector}.\\
So, let us consider each of the properties individually.\\
Suppose an element $\fw \in V$ is given. We wish to show the identity
\begin{equation}\label{eq:apx_theory_projector_argument1}
	\fL_{\tilde{\fQ}}\fv = \fv \quad \text{for} \quad \fv := \int_{\mathcal{G}} \fL_{\fQ}\fw \, d\mu_{\mathcal{G}}(\fQ) \quad \text{and all} \quad \tilde{\fQ} \in \mathcal{G}.
\end{equation}
We notice
\begin{equation}\label{eq:apx_theory_projector_argument2}
	\fL_{\tilde{\fQ}}\fv = \fL_{\tilde{\fQ}} \int_{\mathcal{G}} \fL_{\fQ}\fw \, d\mu_{\mathcal{G}}(\fQ) = \int_{\mathcal{G}} \fL_{\tilde{\fQ}}\fL_{\fQ}\fw \, d\mu_{\mathcal{G}}(\fQ) = \int_{\mathcal{G}} \fL_{\tilde{\fQ} \fQ}\fw \, d\mu_{\mathcal{G}}(\fQ).
\end{equation}
By the invariance of the Haar measure under left multiplication \eqref{eq:SymmetryInformedStrats_projectors_left_invariant} by group elements, a change of variables $\fQ' = \tilde{\fQ} \fQ$ leads to the transformations
\begin{equation}\label{eq:apx_theory_projector_argument3}
	\int_{\mathcal{G}} \fL_{\tilde{\fQ} \fQ}\fw \, d\mu_{\mathcal{G}}(\fQ) = \int_{\mathcal{G}} \fL_{\fQ'}\fw \, d\mu_{\mathcal{G}}(\tilde{\fQ}^{-1}\fQ')  = \int_{\mathcal{G}} \fL_{\fQ'}\fw \, d\mu_{\mathcal{G}}(\fQ')  = \fv,
\end{equation}
showing the claimed statement \eqref{eq:apx_theory_projector_statement1}.\\
Concerning the next item \eqref{eq:apx_theory_projector_statement2}, we notice that any element $\fv$ of the invariant set \eqref{eq:apx_theory_projector_invariantVector} satisfies the equation
\begin{equation}\label{eq:apx_theory_projector_argument4}
	\fL_{\fQ}\fv = \fv
\end{equation}
by definition. Thus, averaging over the group \eqref{eq:apx_theory_projector_defn} yields
\begin{equation}\label{eq:apx_theory_projector_argument5}
	P \fv \equiv \int_{\mathcal{G}} \fL_{\fQ}\fv \, d\mu_{\mathcal{G}}(\fQ) = \int_{\mathcal{G}} \fv \, d\mu_{\mathcal{G}}(\fQ) = \fv,
\end{equation}
where we used the normalization of the Haar measure $d\mu$. Last but not least, we consider the self-adjointness property \eqref{eq:apx_theory_projector_statement3}. We notice
\begin{equation}\label{eq:apx_theory_projector_argument6}
	\langle P \fv, \fw \rangle_V = \left\langle \int_{\mathcal{G}} \fL_{\fQ}\fv \, d\mu_{\mathcal{G}}(\fQ), \fw \right\rangle_V = \int_{\mathcal{G}} \langle  \fL_{\fQ}\fv , \fw \rangle_V \, d\mu_{\mathcal{G}}(\fQ).
\end{equation}
Using the group homomorphism property of the representation \eqref{eq:SymmetryInformedStrats_projectors_representation_general} and the orthogonality \eqref{eq:SymmetryInformedStrats_projectors_representation_norm-preserving}, we observe
\begin{equation}\label{eq:apx_theory_projector_argument7}
	\begin{split}
		\langle P \fv, \fw \rangle_V &=	\int_{\mathcal{G}} \langle  \fL_{\fQ}\fv , \fw \rangle_V \, d\mu_{\mathcal{G}}(\fQ)\\
		&= \int_{\mathcal{G}} \langle  \fL_{\fQ}\fv , \fL_{\fQ}\fL_{\fQ^{-1}}\fw \rangle_V \, d\mu_{\mathcal{G}}(\fQ)\\
		&= \int_{\mathcal{G}} \langle  \fv , \fL_{\fQ^{-1}}\fw \rangle_V \, d\mu_{\mathcal{G}}(\fQ)\\
		&= \left\langle  \fv , \int_{\mathcal{G}}  \fL_{\fQ^{-1}}\fw \, d\mu_{\mathcal{G}}(\fQ) \right\rangle_V\\
		&= \left\langle  \fv , \int_{\mathcal{G}}  \fL_{\fQ}\fw \, d\mu_{\mathcal{G}}(\fQ) \right\rangle_V\\
		&= \langle \fv, P\fw \rangle_V,
	\end{split}
\end{equation}
where we used the inversion invariance \eqref{eq:SymmetryInformedStrats_projectors_inversion_invariant} of the Haar measure. Thus, the property \eqref{eq:apx_theory_projector_statement3} is established.

\bibliographystyle{ieeetr}
{\footnotesize \bibliography{literature}}

\end{document}